\documentclass[letterpaper,11pt]{article}
\usepackage[letterpaper,top=2cm,bottom=2cm,left=3cm,right=3cm,marginparwidth=1.75cm]{geometry}
\usepackage{array}
\usepackage{float}
\usepackage{amsmath}
\usepackage{authblk}

\usepackage{amssymb}
\usepackage{graphicx}
\usepackage{subcaption}
\usepackage{xcolor}
\usepackage[numbers,sort&compress]{natbib}
\usepackage{hyperref}
\usepackage{comment}
\definecolor{darkgreen}{RGB}{0, 150, 0}

\newcommand \Pomeron {I\!\!P}
\newcommand \Reggeon {I\!\!R}

\title{Diffractive open charm photoproduction in ultraperipheral lead-lead and proton-lead collisions at the LHC}

\author[1,2]{Vadim Guzey}
\author[3]{Gian Michele Innocenti}
\author[4]{Anna M. Sta\'sto}
\author[4]{Mark Strikman}

\affil[1]{\small \it University of Jyv\"askyl\"a, Department of Physics, P.O. Box 35, FI-40014 University of Jyv\"askyl\"a, Finland}
\affil[2]{\small \it Helsinki Institute of Physics, P.O. Box 64, FI-00014 University of Helsinki, Finland }
\affil[3]{\small \it Massachusetts Institute of Technology, Cambridge, MA 02139,  USA}
\affil[4]{\small \it Department of Physics, Penn State University, University Park, PA 16802, USA}

\begin{document}
\maketitle
\begin{abstract}
We calculate diffractive \(D^{0}\) photoproduction in ultraperipheral lead--lead (Pb--Pb) collisions at the Large Hadron Collider (LHC) within the recently developed G\(\gamma\)A--FONLL framework, where  photon--lead diffraction is modeled using nuclear diffractive parton distributions obtained in the leading twist shadowing approach
and the photon fluxes 
include corrections for independent electromagnetic dissociation accompanying the hard photoproduction process. We then use the predicted diffractive cross section to quantify the coherent diffractive contribution rejected by the \(Xn0n\) neutron-tagged event selection adopted in the first measurement of \(D^0\) photoproduction in Pb-Pb collisions at the LHC, which requires neutron emission from only one of the two lead nuclei. In this work, we also extend the G\(\gamma\)A--FONLL framework to proton--lead (\(p\)--Pb) UPCs and present predictions for inclusive and diffractive \(D^{0}\) photoproduction at the LHC. In this case, the dominant configuration is photon emission from the lead ion followed by photon--proton scattering, and the diffractive contribution is evaluated using proton diffractive parton distributions constrained by HERA data.
\end{abstract}
\clearpage
\tableofcontents
\clearpage


\section{Introduction}
\label{sec:introduction}
In recent years, ultraperipheral collisions (UPCs) of highly relativistic ions have become an important laboratory for studying QCD with photon-induced processes  at collider energies; for reviews, see, for example, Refs.~\cite{Bertulani:1987tz,Baltz:2007kq,Klein:2020fmr}. In UPCs, 
the average impact parameter exceeds the sum of the radii of the colliding projectiles, which strongly suppresses hadronic interactions and makes electromagnetic interactions dominant. As a result, nucleus--nucleus and proton--nucleus collisions can be used as effective photon--photon and photon--hadron colliders. In particular, photonuclear interactions in high-energy UPCs provide direct access to the gluon density in nuclei at very small values of $x$~\cite{Strikman:2005yv},
 in a kinematic regime that is complementary to and, at present, only partially covered by dedicated electron--ion facilities.\\[4pt]
A broad program of photonuclear measurements in UPCs has been carried out at Relativistic Heavy Ion Collider (RHIC) and the Large Hadron Collider (LHC), with a  benchmark  channel being the exclusive photoproduction of vector mesons. When a hard scale is present, as in the production of heavy vector mesons such as $J/\psi$, this process becomes a sensitive probe of gluon dynamics in nucleons and nuclei, while its exclusive final state provides a particularly clean experimental signature. Measurements of $J/\psi$ photoproduction in UPCs have been performed at RHIC~\cite{STAR:2021wwq,STAR:2023nos,PHENIX:2009xtn} and at the LHC~\cite{ALICE:2013wjo,ALICE:2012yye,CMS:2023snh,ALICE:2021gpt,ALICE:2023jgu,LHCb:2022ahs}. Another important hard photonuclear process is dijet photoproduction, in which the jet transverse momenta set the perturbative scale and the measured jet kinematics provide differential access to the nuclear partonic structure and to photon--parton scattering dynamics. Dijet photoproduction in nuclear UPCs has been studied extensively in theory~\cite{Guzey:2016tek,Guzey:2018dlm,Eskola:2024fhf} and has recently been measured at the LHC~\cite{ATLAS:2024mvt}. \\[4pt]
Open-charm photoproduction provides another important hard photonuclear probe in UPCs. Since heavy-quark pairs are produced predominantly through photon--gluon fusion, $D^0$ production directly probes the nuclear gluon density at small $x$. In contrast to exclusive $J/\psi$ photoproduction, which probes a relatively narrow range of scales, and dijet photoproduction, which is restricted to sufficiently hard jets, open charm can be measured over a broad range of transverse momenta and rapidities, allowing one to scan both the gluon momentum fraction $x$ and the hard scale $Q^2 \simeq m_c^2 + p_T^2$,
where $m_c$ is the charm quark mass and $p_T$ is the  $D^0$ transverse momentum.
The experimental relevance of this channel has recently been underscored by the first CMS measurement of $D^0$ photoproduction in ultraperipheral Pb--Pb collisions at $\sqrt{s_{\scriptscriptstyle NN}}=5.36$~TeV in the neutron-tagged $Xn0n$ event class~\cite{CMS:2025jjx}. \\[4pt]
Open heavy-flavor photoproduction in UPCs was first studied at leading order in Ref.~\cite{Klein:2002wm}. More recently, inclusive open-charm photoproduction in Pb--Pb UPCs was revisited within the G$\gamma$A--FONLL framework~\cite{Cacciari:2025tgr}, which combines the FONLL description of heavy-flavor photoproduction with nuclear parton distributions, UPC photon fluxes, and corrections for electromagnetic dissociation. The neutron-tagged event selection used in measurements, however, introduces an additional complication. The analysis requires no neutrons in one zero-degree calorimeter and at least one neutron in the opposite one, so the measured sample is not fully inclusive. This is particularly important for coherent diffraction, where the target nucleus remains intact at the hard-scattering level and therefore does not satisfy the $Xn$ requirement unless it undergoes an additional independent electromagnetic breakup. Although Ref.~\cite{Cacciari:2025tgr} accounted for electromagnetic dissociation accompanying the hard process, it did not include a dedicated correction for the fraction of diffractive events removed by the $Xn$ condition. Quantifying this bias requires a theoretical prediction for diffractive open-charm production in Pb--Pb UPCs and thus motivates the present extension of the G$\gamma$A--FONLL framework.
\\[4pt]
In this work, we present a detailed study of diffractive open-charm photoproduction in ultraperipheral Pb--Pb and p--Pb collisions at the LHC. Diffractive open-charm photoproduction is a particularly valuable channel because it combines a perturbative hard scale, set by $m_c$ and $p_T$, with a large-rapidity-gap final state. It is therefore directly sensitive to the gluon content of the diffractive exchange and, in nuclei, to the interplay between hard diffraction, coherent multiple scattering, and nuclear shadowing. The analogous process was studied experimentally in $ep$ collisions at HERA, where diffractive open-charm production was measured in both deep-inelastic scattering and photoproduction~\cite{H1:1998csb,ZEUS:1998wxs,H1:2011myz}. 
By contrast, while inclusive photonuclear $D^0$ production in Pb--Pb UPCs has now been measured by CMS~\cite{CMS:2025jjx}, a dedicated measurement of diffractive open-charm photoproduction in photon--nucleus collisions is still missing. UPCs at the LHC therefore provide a unique opportunity to access this channel and to test nuclear diffractive parton distributions. \\[4pt]
For Pb--Pb UPCs, we extend the G$\gamma$A--FONLL framework to include coherent diffraction using nuclear diffractive parton distributions from the leading-twist shadowing approach (LTA)~\cite{Frankfurt:2003gx,Frankfurt:2011cs}. We compute the diffractive $D^0$ cross section and the ratio of diffractive to inclusive production as functions of rapidity and transverse momentum. We then use this diffractive component to quantify how the neutron-tagged \(Xn0n\) event selection modifies the inclusive prediction for the CMS \(D^0\) measurement. We also improve the treatment of the UPC photon flux by using a realistic nuclear charge distribution. For \(p\)--Pb UPCs, where the dominant contribution comes from \(\gamma p\) scattering with the photon emitted by the lead nucleus, we present predictions for both inclusive and diffractive \(D^0\) production using proton diffractive parton distributions constrained by HERA data. 
\\[4pt]
The paper is organized as follows. 
Section~2 introduces diffractive open-charm photoproduction, the relevant diffractive kinematics,  the proton and nuclear diffractive PDFs used in the calculation.
In Sec.~3 we review the main ingredients of the G$\gamma$A--FONLL framework and discuss the UPC photon fluxes and electromagnetic-dissociation effects relevant for Pb--Pb collisions. 
In Sec.~4 we present predictions for coherent diffractive \(D^0\) photoproduction in Pb--Pb UPCs, including  outline of the  experimental prospects  at the LHC. 
Section~5 quantifies 
the correction induced by the neutron-tagged \(Xn0n\) selection in the CMS \(D^0\) measurement. 
In Sec.~6 we turn to \(p\)--Pb UPCs and present predictions for inclusive and diffractive \(D^0\) photoproduction. 
We summarize our conclusions in Sec.~7. In Appendix A we include additional plots on typical \(x\)  probed in the nuclear gluon density. 
\newpage
\section{Charm diffraction in UPCs}
\label{sec:charm_in_diffraction}

Diffractive processes are characterized by the presence of a large rapidity gap between the central hadronic system $X$, known as the photon dissociative system, and the forward-going hadronic system $Y$, which constitutes in coherent diffraction 
an intact or coherently scattered proton or nucleus. In charm photoproduction, the system $X$ contains at least a charm quark, possible hadronic remnants from the diffractive exchange, usually modeled by the Pomeron $\Pomeron$ and, in the resolved-photon case, from the photon, when the latter fluctuates into its partonic constituents before interacting with the proton or nucleus. \\[4pt]
In collinear factorization the hard subprocess for diffractive charm production is the same as in inclusive open-charm electroproduction or photoproduction: a virtual or quasi-real photon interacts with a parton to produce a heavy-quark pair, one of whose members may subsequently fragment into the observed charmed meson. The distinctive feature of the diffractive process is the final state which includes a large rapidity gap. For the gap to exist the exchange between the system \(X\) and diffractively scattered target \(Y\) needs to be mediated by the color singlet.   In  the collinear factorization framework the hard diffractive process is   described in terms of the  diffractive parton distributions.

\begin{figure}[H]
    \centering
    \includegraphics[width=0.45\linewidth]{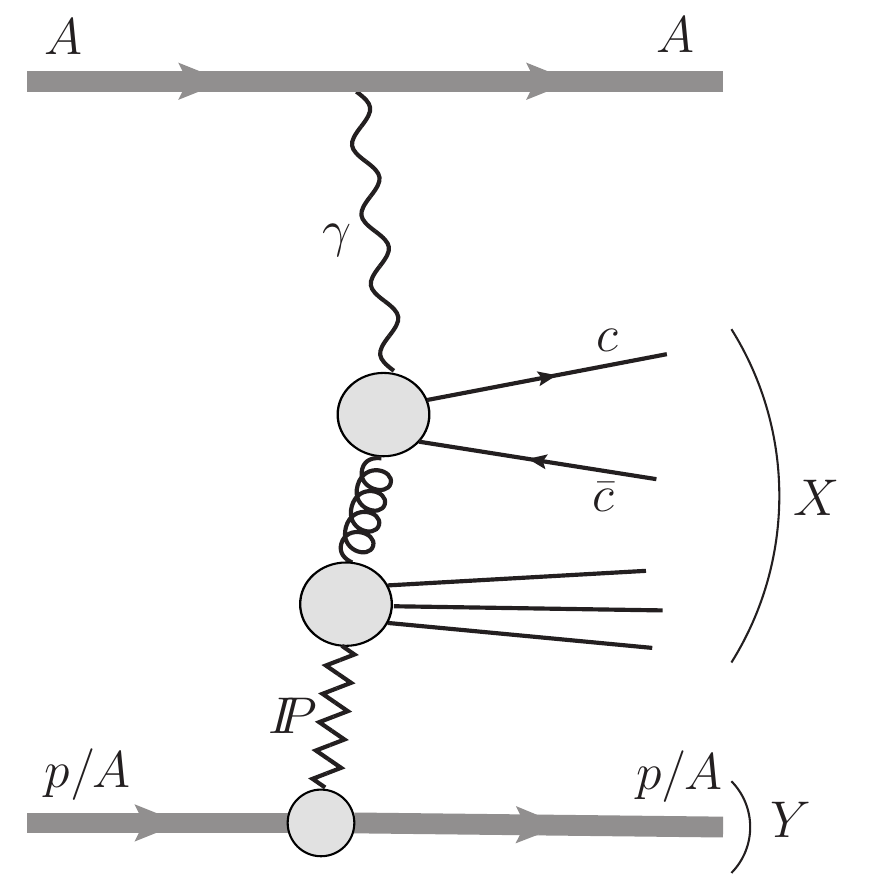}\hfill
      \includegraphics[width=0.45\linewidth]{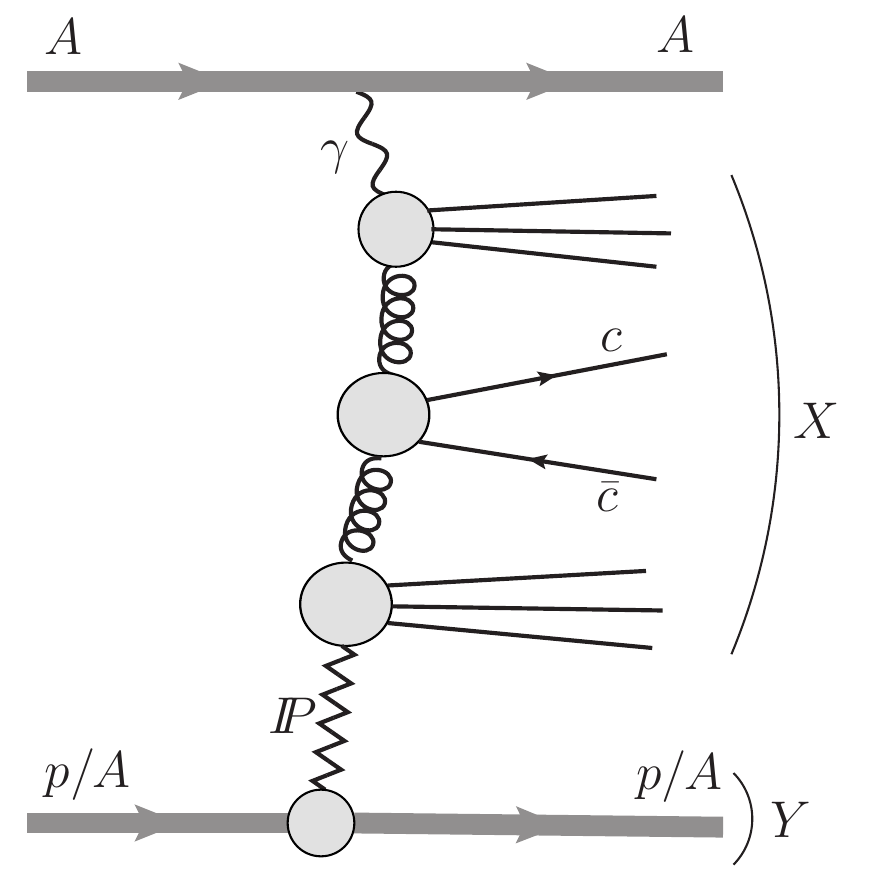}
    \caption{Diagram for the coherent diffractive charm production in $AA$ and $pA$ UPCs in the direct-photon (left panel) and resolved-photon (right panel) cases, where $X$ is the diffractively produced hadronic system, 
    $Y$ is the forward-going nucleus (proton) or its low-mass excitations, and 
    $\Pomeron$ denotes the diffractive exchange (Pomeron). 
    }
    \label{fig:upc_diff}
\end{figure}
\noindent

For the photon--hadron interaction in UPCs, the diffractive process can be written as
\begin{equation}
\gamma(q) + H(P) \longrightarrow X(p_X) + H(p_Y) \; .
\end{equation}
Here, $H(P)$ denotes the incoming hadron, proton or nucleus, with four-momentum $P$. In coherent diffraction, the hadron scatters elastically and remains intact in the final state, with four-momentum $p_Y$. The diffractively produced hadronic system $X$ is characterized by the invariant mass squared $M_X^2 = p_X^2$. A large rapidity gap, characterized by the absence of associated hadron production, separates the hadronic system $X$ from the elastically scattered hadron. 
The coherent diffractive process in nucleus--nucleus and  
proton--nucleus 
UPCs is illustrated in Fig.~\ref{fig:upc_diff}. 
In \(p\)--Pb case, the dominant photonuclear configuration is the one in which the lead ion emits the quasi-real photon and the proton acts as the hadronic target; 
the reverse configuration, in which the proton emits the photon and the lead ion acts as the target, is also possible, but is strongly suppressed.
\\[4pt]
In the above process, the hard partonic cross sections  for the \(\gamma\)-parton scattering to produce the charm quark in the final state is the same as in the inclusive open charm photoproduction. The distinguishing difference of the process  stems from the presence of the rapidity gap due to the colorless diffractive exchange, also known as Pomeron, denoted by \(\Pomeron\). 
In the case of the diffractive process one introduces additional variables
\begin{equation}
 t=(P-p_Y)^2, \; \; \; x_{\Pomeron} = \frac{q\cdot(P-p_Y)}{q\cdot P} \; .
\end{equation}
The variable $t$ is the squared four-momentum transfer at the target vertex. Physically, it characterizes the recoil of the target hadron in the diffractive scattering: small $|t|$ corresponds to a small deflection and is typical of coherent diffraction, where the proton or nucleus remains intact. The variable $x_{\Pomeron}$ is the longitudinal momentum fraction of the incoming hadron carried by the color-singlet diffractive exchange, with 
smaller values of $x_{\Pomeron}$ generally corresponding to more forward scattered intact targets and, approximately, to larger rapidity gaps. 
The  invariant mass of the diffractively produced system is 
$M_X^2=(q+P-p_Y)^2$.
In the literature one can also find the following notation  for the longitudinal momentum fraction of the proton carried by the Pomeron: 
\(
\xi = x_{\Pomeron} 
\).
In the case of the photoproduction $Q^2 \sim 0$, the $x_{\Pomeron}$ fraction can be written as
\begin{equation}
x_{\Pomeron} = \frac{M_X^2+Q^2-t}{W^2+Q^2} \simeq \frac{M_X^2-t} {W^2} \; ,
\end{equation}
where \(W^2=(q+p)^2\) is the  center-of-mass energy squared. 
For fixed photon–hadron energy $W$, larger values of $M_X$ imply larger values of $x_{\Pomeron}$. Physically, this means that the diffractively produced system extends closer in rapidity to the outgoing target, and the rapidity gap is therefore reduced. Although the rapidity-gap size is not introduced as an independent variable in this formalism, it is approximately related to the diffractive momentum fraction through $\Delta y_{\rm gap} \sim \ln(1/x_{\Pomeron})$. 

\subsection{Diffractive PDFs for the proton}
\label{sec:diffraction_proton}

For diffractive hard scattering, collinear factorization provides a framework in which the nonperturbative proton structure is encoded in diffractive parton distributions (PDFs)~\cite{Collins:1997sr,Berera:1995fj}. These distributions obey Dokshitzer--Gribov--Lipatov--Altarelli--Parisi (DGLAP) evolution equations, analogously to ordinary PDFs (see also Ref.~\cite{Trentadue:1993ka}). In this framework, diffractive parton densities describe the partonic content of the proton under the additional requirement that the final state satisfies a diffractive condition, such as the presence of an elastically scattered forward proton. Diffractive proton PDFs have been constrained from QCD fits to diffractive structure-function measurements at HERA~\cite{H1:2006zyl,ZEUS:2009uxs}. In their most differential form, they depend on four variables
\begin{equation}
    f_j^{D(4)}(z,\mu^2,x_{\Pomeron},t) \; ,
\label{eq:fdiffpart}
\end{equation}
where \(j\) denotes the parton species, i.e., gluon or quark flavor, and the superscript \(D(4)\) indicates that the diffractive PDF is differential in four variables. In Eq.~\eqref{eq:fdiffpart}, \(z\) is the longitudinal momentum fraction carried by the parton entering the hard subprocess, measured with respect to the diffractive exchange, \(\mu^2\) is the factorization scale that sets the resolution scale of the hard process, and \(t\) is the squared momentum transfer at the proton vertex. If the recoil momentum transfer \(t\) is not measured, one integrates over \(t\), thereby summing over all allowed proton recoil configurations consistent with diffraction. This defines the corresponding \(D(3)\) diffractive PDFs
\begin{equation}
   f_j^{D(3)}(z,\mu^2,x_{\Pomeron}) \; = \; \int_{|t|_{\rm min}}^{|t|_{\rm max}} d|t|   \,f_j^{D(4)}(z,\mu^2,x_{\Pomeron},t) \; ,
\end{equation}
where 
\(|t|_{\rm min}= x_{\Pomeron}^2 m_p^2/(1-x_{\Pomeron})\) 
is constrained by kinematics. Indeed, for fixed \(x_{\Pomeron}\), the minimum momentum transfer is reached when the final-state proton is scattered with zero transverse momentum. In the high-energy small-angle limit, \(t=-(p_T^2+x_{\Pomeron}^2 m_p^2)/(1-x_{\Pomeron})\), which gives \(|t|_{\rm min}=x_{\Pomeron}^2 m_p^2/(1-x_{\Pomeron})\) for \(p_T=0\). The upper limit \(|t|_{\rm max}\) is set by the experimental acceptance for the recoil kinematics of the diffractively scattered proton. In the H1~\cite{H1:2006uea} and ZEUS~\cite{ZEUS:2009uxs} diffractive measurements at HERA, a typical value was \(|t|_{\rm max}\sim 1~\mathrm{GeV}^2\). The proton diffractive PDFs were extracted from these data by fitting the measured diffractive reduced cross sections, integrated over the accepted proton-vertex phase space. 
\\[4pt] The H1 and ZEUS parametrizations rely on similar assumptions. In particular, they use the following two-component form
\begin{equation}
 f_j^{D(3)}(z,\mu^2,x_{\Pomeron}) \; = \; \Phi_{\Pomeron}(x_{\Pomeron}) f_j^{\Pomeron}(z,\mu^2) \; + \; \Phi_{\Reggeon}(x_{\Pomeron}) f_j^{\Reggeon}(z,\mu^2) \;,
    \label{eq:param}
\end{equation}
where the two terms correspond to the Pomeron (\(\Pomeron\)) and Reggeon exchange (\(\Reggeon\)), respectively. The first component, the Pomeron, is a color-singlet exchange with vacuum quantum
numbers and dominates at low values of \(x_{\Pomeron}\), which are reached in high-energy collisions. The second component, the  Reggeon,  is a subleading  exchange. It becomes important at larger values of \(x_{\Pomeron}\), which are explored in lower-energy collisions. In Eq.\eqref{eq:param}, the functions \(\Phi_{\Pomeron}(x_{\Pomeron})\) and \(\Phi_{\Reggeon}(x_{\Pomeron})\) are flux factors describing the emission of the corresponding color-singlet exchange from the proton. The functions \(f_j^{\Pomeron}(z,\mu^2)\) and \(f_j^{\Reggeon}(z,\mu^2)\) describe the partonic structure of the Pomeron and Reggeon, respectively. We note that, the variable \(x_{\Pomeron}\) denotes the fraction of the incoming proton longitudinal momentum carried by the diffractive exchange, whether this exchange is modeled as a Pomeron or as a Reggeon. \\[4pt] The separation of the \((t,x_{\Pomeron})\) dependence from the \((z,\mu^2)\) dependence is the Regge-factorization assumption used in the HERA fits. Its validity is not guaranteed in QCD, and possible limitations have recently been discussed in the framework of Soft-Collinear Effective Theory \cite{Lee:2025fml}. 
\\[4pt] 
The flux factors introduced above, \(\Phi_{\Pomeron}(x_{\Pomeron})\) and \(\Phi_{\Reggeon}(x_{\Pomeron})\), are \(t\)-integrated quantities, which are obtained by integrating the following \(t\)-dependent fluxes over the momentum transfer \(t\)
\begin{equation}
     \phi_{\Pomeron,\Reggeon}(x_{\Pomeron},t) \; = \; A_{\Pomeron,\Reggeon} \, \frac{e^{B_{\Pomeron,\Reggeon} t}}{x_{{\Pomeron}}^{2\alpha_{{\Pomeron,\Reggeon}}(t)-1}} \; .
     \label{eq:phi4tdep}
\end{equation}
Only the product of the flux factors and the parton distributions enters the diffractive PDFs, so their relative normalization is conventional. The convention chosen in the H1 and ZEUS fits~\cite{H1:2006uea,ZEUS:2009uxs} was such that \(x_{\Pomeron}\Phi_{\Pomeron}(x_{\Pomeron})=1\) at \(x_{\Pomeron}=0.003\). The parameters \(B_{\Pomeron,\Reggeon}\), \(\alpha_{\Pomeron,\Reggeon}(0)\), and \(\alpha'_{\Pomeron,\Reggeon}\) were determined from fits to the data~\cite{H1:2006uea,ZEUS:2009uxs}. The Pomeron parton densities \(f_j^{\Pomeron}(z,\mu^2)\) were parametrized at the starting scale, fitted to the HERA diffractive data, and then evolved with DGLAP evolution~\cite{H1:2006zyl,ZEUS:2009uxs}. By contrast, the Reggeon parton densities \(f_j^{\Reggeon}(z,\mu^2)\) were not fitted as an independent set of PDFs. They were taken from the GRV parametrization of the pion structure function \cite{Gluck:1991ey}. 
\\[4pt] 
A recent study for the Electron-Ion Collider (EIC) showed that this facility could significantly improve the existing constraints on the Reggeon contribution~\cite{Armesto:2024dhc}. This is due to its high luminosity, variable beam energies, and forward proton instrumentation, which enable multidifferential diffractive measurements over a broad range of \(x_{\Pomeron}\) and \(t\), including the larger-\(x_{\Pomeron}\) region where the Reggeon term is enhanced.

\subsection{Diffractive nuclear PDFs in the LTA approximation}
\label{sec:diffraction_nucleus}

For the calculation of the diffractive contribution to charm production in nucleus--nucleus UPCs, one needs diffractive PDFs for a nucleus, generalizing those for the proton discussed in Sec.~\ref{sec:diffraction_proton}. In our analysis, we use the leading-twist approach (LTA), which allows one to express nuclear diffractive PDFs at small \(x\) as a Gribov--Glauber series of multiple coherent interactions with \(i=1,2,\dots,A\) nucleons of the nuclear target. Each term in this series can be written in terms of the nucleon (proton) diffractive PDFs
times the factor accounting for nuclear attenuation~\cite{Frankfurt:2003gx,Frankfurt:2011cs}.
\begin{figure}[t!]
    \centering
    \includegraphics[width=1.0\linewidth]{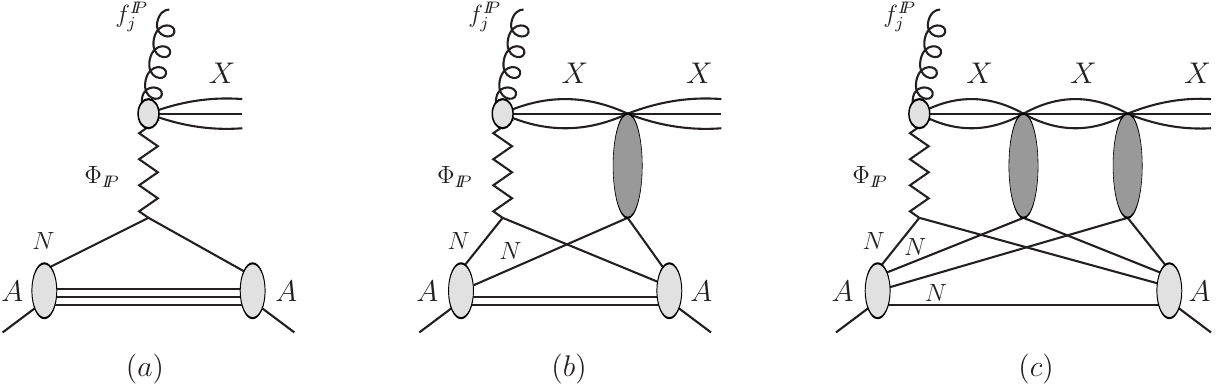}
    \caption{Nuclear gluon diffractive PDF \(f_{j/A}^{D(3)}\) represented as a Gribov--Glauber series of interactions with one (\(a\)), two (\(b\)), and three (\(c\)) nucleons of the nuclear target \(A\); higher-order terms are not shown. Diffractive exchanges with target nucleons \(N\) are factorized into the Pomeron flux \(\Phi_{\Pomeron}\) and Pomeron PDFs \(f_{j}^{\Pomeron}\). The state \(X\) is diffractively produced and subsequently scatters elastically from the remaining nucleons, as indicated by the dark shaded ovals. The lower blobs denote the nuclear wave function.}

    \label{fig:lta_diff_2026_gluon}
\end{figure}

It is shown schematically in Fig.~\ref{fig:lta_diff_2026_gluon}, which illustrates how the nuclear gluon diffractive distribution is built as a series of interactions with one (graph \(a\)), two (graph \(b\)), and three (graph \(c\)) nucleons \(N\) of the nucleus \(A\); the terms with \(i \geq 4\) are not shown. Each diagram contains diffractive scattering on a target nucleon, \(\gamma^{\ast}+N \to X+N\), and \(i-1\) elastic scatterings of the diffractively produced state \(X\) on the remaining nucleons. These rescatterings are represented by the dark shaded ovals.\footnote{For a term involving \(i\) nucleons, one nucleon participates in the diffractive transition \(\gamma^\ast+N\to X+N\), while the diffractively produced state \(X\) undergoes \(i-1\) elastic rescatterings on the remaining participating nucleons. Thus, graph \(a\) has no rescattering, graph \(b\) has one rescattering, and graph \(c\) has two rescatterings.}
The diffractive exchanges are factorized into the product of the Pomeron flux $\Phi_{\Pomeron}$ 
(denoted by the vertical zigzag lines) and the Pomeron  PDFs $f_j^{\Pomeron}$. 
As explained in Ref.~\cite{Frankfurt:2011cs}, the subleading Reggeon contribution gives a negligibly small contribution to nuclear shadowing and is therefore neglected in LTA calculations of nuclear diffractive PDFs.
The lower shaded blobs denote the ground-state nuclear wave function. Graph \(a\) represents the interaction with a single target nucleon and corresponds to the impulse approximation for \(f_{j/A}^{D(3)}\). Graphs \(b\) and \(c\) include interactions with two and three target nucleons, respectively, and therefore generate shadowing corrections to \(f_{j/A}^{D(3)}\) due to nuclear absorption of the state \(X\). In this treatment, the diffractively produced state \(X\) is assumed to scatter elastically from the target nucleons without transitions to other diffractive states. \\[4pt]
Summing the series illustrated in Fig.~\ref{fig:lta_diff_2026_gluon}, including the higher-order terms with \(4\leq i\leq A\), one obtains the following expression for the nuclear diffractive PDFs \(f_{j/A}^{D(3)}\)~\cite{Frankfurt:2003gx,Guzey:2024xpa}
\begin{eqnarray}
f_{j/A}^{D(3)}(z,\mu^2,x_{\Pomeron})  &=&
 4\pi f_{j/p}^{D(4)}(z,\mu^2,x_{\Pomeron},t_{\rm min}) \nonumber\\ 
&\times& \int d^2 \mathbf{s} \, \bigg|\int_{-\infty}^{+\infty} dz^{\prime} \rho_A(s,z^{\prime}) e^{i z^{\prime} x_{\Pomeron}m_N} e^{-\frac{(1-i\eta)}{2}\sigma_{\rm soft}^j(x,\mu^2) \int_{z^{\prime}}^{\infty} dz^{\prime \prime}\rho_A(s,z^{\prime \prime})}\bigg|^2 .
    \label{eq:dpdf_lta}
\end{eqnarray}
In Eq.\eqref{eq:dpdf_lta}, $f_{j/p}^{D(4)}$ is the 
nucleon (proton) diffractive PDF evaluated at $t=t_{\rm min}=-|t|_{\rm min}=-x_{\Pomeron}^2 m_p^2/(1-x_{\Pomeron}) \approx 0$; 
$\rho_A(s,z^{\prime})$ is the nuclear density~\cite{DeVries:1987atn}, where $\mathbf{s}$ and $z^{\prime}$ are the transverse and longitudinal coordinates of the nucleons involved in 
the interaction with a hard probe. Note that, since the $t$-dependence of the nuclear form factor is much faster than that of the $\gamma^{\ast}+N \to X+N$ amplitude, $f_{j/p}^{D(4)}$ is evaluated at 
$t \approx 0$ and all the active nucleons are located at the same transverse distance 
$s=|\mathbf{s}|$. 
Further, the effective cross section $\sigma_{\rm soft}^j(x,\mu^2)$, 
where $x=z x_{\Pomeron}$, 
is a model-dependent parameter controlling the strength of interaction of the state $X$ with target nucleons (denoted by dark shaded ovals in Fig.~\ref{fig:lta_diff_2026_gluon}).
Finally, $\eta=\pi/2(\alpha_{\Pomeron}(0)-1) = 0.17$ is the ratio of the real to imaginary parts of the $\gamma^{\ast}+N \to X+N$ amplitude, which is estimated using 
the dispersion relation and the value of the Pomeron intercept determined from  the QCD fits to HERA data on inclusive diffraction in electron-proton DIS,
$\alpha_{\Pomeron}(0)=1.111$~\cite{H1:2006zyl}.
\\[4pt]
It is instructive to consider the small-\(x_{\Pomeron}\) limit of Eq.~\eqref{eq:dpdf_lta}, corresponding to diffractive masses that are not large compared with the photon--nucleon energy $W$,
\begin{eqnarray}
f_{j/A}^{D(3)}(z,\mu^2,x_{\Pomeron})  &=&
 f_{j/p}^{D(4)}(z,\mu^2,x_{\Pomeron},t_{\rm min}) \nonumber \\
 &\times &\frac{16 \pi}{(1+\eta^2) (\sigma^j_{\rm soft}(x,\mu^2))^2} \int d^2 \mathbf{s} \; \bigg|1- e^{-\frac{(1-i\eta)}{2}\sigma_{\rm soft}^j(x,\mu^2) T_A(s)}\bigg|^2 =\nonumber\\
 &=&
 f_{j/p}^{D(3)}(z,\mu^2,x_{\Pomeron}) \frac{1}{\sigma_{\rm el}^j(x,\mu^2)} \int d^2 \mathbf{s} \; \bigg|1- e^{-\frac{(1-i\eta)}{2}\sigma_{\rm soft}^j(x,\mu^2) T_A(s)}\bigg|^2 
 \,,
    \label{eq:dpdf_lta_2}
\end{eqnarray}
where $T_A(s)=\int^{\infty}_{-\infty} dz \rho_A(s,z)$ is the nuclear density (profile) in the transverse plane normalized to the number of nucleons, $\int d^2 \textbf{s}\,T_A(s)=A$.
In the second line, we used that $f_{j/p}^{D(4)}(z,\mu^2,x_{\Pomeron},t_{\rm min})
= B_{\rm diff} f_{j/p}^{D(3)}(z,\mu^2,x_{\Pomeron})$, where 
 \(B_{\rm diff}\approx 6~\mathrm{GeV}^{-2}\) is the slope parameter of the exponential \(t\)-dependence of the proton diffractive structure function
 measured at HERA in diffractive DIS events with a leading proton, i.e., events in which the final-state proton remains intact and is detected in the forward proton spectrometer~\cite{H1:2006uea}. We also introduced the elastic cross section
$\sigma_{\rm el}^j(x,\mu^2)$ by relating it to the total cross section $\sigma_{\rm soft}^j(x)$ through the optical theorem
\begin{equation}
\sigma_{\rm el}^j(x,\mu^2)=(1+\eta^2) \frac{(\sigma^j_{\rm soft}(x,\mu^2))^2}{16 \pi B_{\rm diff}} \,.
\label{eq:sigma_el}    
\end{equation}
Equation~(\ref{eq:dpdf_lta_2}) offers a straightforward physical interpretation of Fig.~\ref{fig:lta_diff_2026_gluon} and the mechanism of nuclear shadowing in the LTA model. It states that 
the ratio of nuclear and proton diffractive PDFs, $f_{j/A}^{D(3)}/f_{j/p}^{D(3)}$, is shadowed in proportion to the ratio of the nucleus and proton elastic cross sections, which are calculated using the effective total nucleon cross section $\sigma^j_{\rm soft}(x,\mu^2)$, with larger values of \(\sigma_{\rm soft}^j(x,\mu^2)\) corresponding to stronger absorption of \(X\) as it propagates through the nucleus.
\\[4pt]
The key observation of the LTA approach is that nuclear shadowing is expressed in terms of diffraction on the proton. To make this connection more transparent, one notices that the cross section $\sigma_{\rm soft}^j(x,\mu^2)$,
which determines the amount of nuclear absorption of state $X$ and the magnitude of nuclear shadowing, is proportional to probability of diffraction for a given parton flavor $j$. More precisely, 
$\sigma_{\rm soft}^j(x,\mu^2)$
is bounded from below by the cross section $\sigma_2^j(x,\mu^2)$
\begin{equation}
\sigma_2^j(x,\mu^2)=\frac{16 \pi B_{\rm diff}}{(1+\eta^2)f_{j/p}(x,\mu^2)} \int_x^{0.1} \frac{dx_{\Pomeron}}{x_{\Pomeron}} f_{j/p}^{D(3)}(z,\mu^2,x_{\Pomeron}) \,,
\label{eq:sigma_2}
\end{equation}

where $f_{j/p}$ is the usual proton PDF. In this equation, the lower integration limit comes from the kinematic constraint $x_{\Pomeron} \geq x$, and the upper integration limit is given by the usual condition on the produced diffractive masses $M_X^2/W^2 \leq 0.1$,
leading to $x_{\Pomeron} \leq 0.1$~\cite{H1:2006uea}. \\[4pt]
The experimental observation of significant, leading twist diffraction in electron-proton DIS at HERA
translates into relatively sizable diffractive PDFs of the proton, resulting in large  $\sigma_2^j(x,\mu^2)$ and $\sigma^j_{\rm soft}(x,\mu^2)$. 
To estimate a theoretical uncertainty of the LTA predictions, one can vary $\sigma_{\rm soft}^j(x,\mu^2)$ 
in the range between the $q{\bar q}$ dipole-nucleon cross section and the total pion-nucleon cross section, which correspond to the ``high shadowing'' and ``low shadowing'' scenarios, respectively~\cite{Frankfurt:2011cs}.
This gives $\sigma^j_{\rm soft}(x,\mu^2) \approx 40-50$ mb for $x=10^{-4}-10^{-3}$ at $\mu^2=4$ GeV$^2$, leading to 
$f_{j/A}^{D(3)}/[Af_{j/p}^{D(3)}] \approx 0.5$ both for quark and gluon distributions.
\\[4pt]
Note that the terminology ``high'' and ``low'' shadowing refers to LTA predictions for usual nuclear PDFs.
In the case of diffractive scattering, the trend is opposite: the ``high shadowing'' scenario with a lower  $\sigma^j_{\rm soft}(x,\mu^2)$ results in a weaker nuclear suppression for nuclear diffractive PDFs $f_{j/A}^{D(3)}$, and ``low shadowing'' with larger $\sigma^j_{\rm soft}(x,\mu^2)$ gives a stronger suppression to $f_{j/A}^{D(3)}$, see Eq.~(\ref{eq:dpdf_lta_2}). Thus, the LTA model naturally predicts a strong suppression of the nuclear diffractive PDFs at small $x$ due to nuclear shadowing. To appreciate the size of this suppression, it is useful to compare it with the impulse-approximation estimate,
\(f_{j/A}^{D(3)}/[A f_{j/p}^{D(3)}]\approx 4\)~\cite{Guzey:2024xpa}. Since the LTA prediction for 
this ratio is \(f_{j/A}^{D(3)}/[A f_{j/p}^{D(3)}]\approx 0.5\),  
the suppression relative to the impulse approximation is \(4/0.5\simeq 8\). In summary, Eqs.~(\ref{eq:dpdf_lta}) -- (\ref{eq:sigma_2}) are used to calculate nuclear diffractive PDFs at a certain initial scale, $\mu^2=Q_0^2=4$ GeV$^2$~\cite{Frankfurt:2011cs,Guzey:2024xpa}. Their subsequent $\mu^2$ dependence is given by 
Dokshitzer-Gribov-Lipatov-Altarelli-Parisi (DGLAP) scale evolution equations at fixed $x_{\Pomeron}$.\\[4pt]
It is worth noting that, in the context of the future Electron--Ion Collider (EIC), the ratio of diffractive to total cross sections in electron--nucleus and electron--proton DIS is a particularly sensitive observable for discriminating among different mechanisms of small-\(x\) QCD dynamics~\cite{Accardi:2012qut}. In this respect, the strong suppression of nuclear diffractive PDFs predicted by LTA should be contrasted with the enhancement predicted in saturation-based frameworks~\cite{Kowalski:2007rw,Lappi:2023frf}.

\clearpage
\section{Open-charm photoproduction in AA collisions with G\texorpdfstring{$\gamma$}{gamma}A--FONLL: overview of the framework and recent updates}
\label{sec:theoryGammaFONLL}

In G$\gamma$A--FONLL framework~\cite{Cacciari:2025tgr}, the observed $D^0$ spectrum in ultraperipheral collisions is built from four ingredients: the effective UPC photon flux emitted by the projectile ion, the parton distribution of the target nucleus, the FONLL description of the short-distance heavy-quark production cross section, and the fragmentation of the charm quark into the observed meson. To ensure consistency with UPC measurements, the photon flux is adapted to the event selections used experimentally to isolate clean photonuclear interactions, in particular the requirement of no forward neutrons in the 
zero-degree calorimeter (ZDC)
on the photon-emitting side, which is sensitive to additional electromagnetic dissociation of the emitting nucleus that can lead to nuclear breakup.

\begin{figure}[ht]
    \centering
    \includegraphics[width=0.5\textwidth]{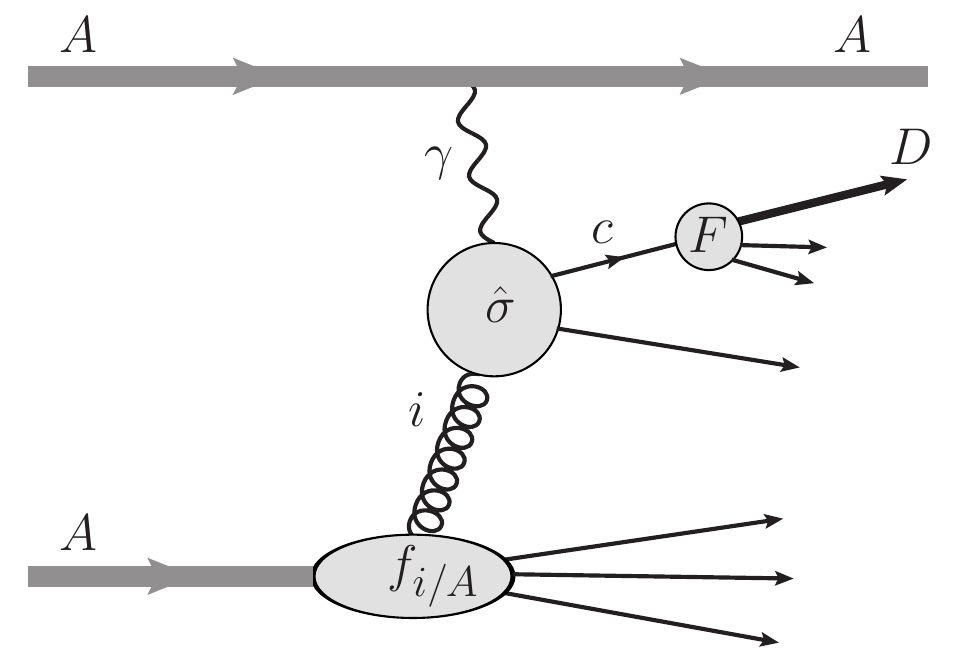}
    \caption{Schematic description of the inclusive $D$  meson production in ultraperipheral heavy-ion collisions,
    where  
    $f_{i/A}$ denotes PDF in the 
    nucleus, $F$ fragmentation function and $\hat{\sigma}$ hard scattering process. 
    }
    \label{fig:charm_upc}
\end{figure}
A schematic representation of the inclusive process is shown in Fig.~\ref{fig:charm_upc}. One nucleus emits a quasi-real photon, which interacts with a parton from the other nucleus and produces a charm quark that subsequently fragments into the observed meson. The hard partonic cross section is denoted by $\hat{\sigma}$, the fragmentation function by $F$, and the parton distribution of parton $i$ in nucleus $A$ by $f_{i/A}$. Because the final state of the target nucleus is left unspecified, the inclusive cross section is understood as summed over all target final states compatible with the measured $D^0$ meson. It therefore receives contributions from both non-diffractive interactions, which involve color exchange and generally lead to target dissociation, and diffractive interactions, which proceed through color-singlet exchange and typically leave the target intact. 
Below we briefly review the main ingredients of G$\gamma$A--FONLL~\cite{Cacciari:2025tgr} used to describe inclusive open-charm photoproduction in ultraperipheral collisions.

\subsection{Charm quark production and hadronization}

G$\gamma$A--FONLL uses the FONLL formalism for the short-distance heavy-quark production cross section. FONLL provides a perturbative QCD description of heavy-quark production over a broad kinematic range, from transverse momenta of the order of the heavy-quark mass to the asymptotic high-$p_T$ regime. Originally developed for heavy-quark production in hadron--hadron collisions \cite{Cacciari:1998it} and later extended to photoproduction in electron--proton scattering \cite{Cacciari:2001td}, FONLL is based on collinear factorization and includes the resummation of logarithmically enhanced terms. This construction enables a smooth interpolation between the low- and high-transverse-momentum regions of heavy-quark production. The matching formula for the resummed FONLL cross section  can be schematically expressed as \cite{Cacciari:2001td,Frixione:2002zv}
\begin{equation}
   {\rm  FONLL=FO+(RS-FOM0)} \times G(m,p_T) \; .
    \label{eq:fonllmatching}
\end{equation} 
In this formula, the  FO  represents the exact fixed NLO calculation~\cite{Ellis:1988sb,Smith:1991pw}. In this term the contributions at order $\alpha_{\rm em} \alpha_s$ and $\alpha_{\rm em} \alpha_s^2$  are included exactly, with mass effects. This term is dominant at low transverse momenta \(p_T\) of the heavy quark.
 RS is the resummed result, which includes  logarithmically enhanced terms $\alpha_{\rm em} \alpha_s (\alpha_s \ln(p_T/m))^k$ and $\alpha_{\rm em} \alpha_s^2 (\alpha_s \ln(p_T/m))^k$, up to  terms suppressed by powers of $m/p_T$. FOM0 is the massless limit of fixed order FO without  terms suppressed by powers of $m$, while logarithms of the mass are retained. Its subtraction removes the terms already common to FO and RS, thereby avoiding double counting. Finally, $G(m,p_T)$ is  an arbitrary interpolating function, that must be regular in $p_T$, and must approach unity up to terms suppressed by powers of $m/p_T$ at large transverse momenta. The produced charm quark subsequently fragments into the observed $D^0$ meson, therefore one needs to include   the corresponding fragmentation function.  As a baseline choice, we use the Braaten-Cheung-Fleming-Yuan (BCFY) fragmentation function~\cite{Braaten:1994bz}. This fragmentation function depends on a single non-perturbative parameter, $r$, which can be interpreted as the ratio of the constituent mass of the light quark to the meson mass. Following Refs.~\cite{Cacciari:2003zu,Cacciari:2012ny}, and consistently with the recent analysis of Ref.~\cite{Cacciari:2025tgr}, we take $r=0.1$.

\subsection{PDFs of the real photon}

A further ingredient of the calculation is the partonic structure of the incoming photon. Since the process under consideration is a photoproduction cross section, one must include both {\it direct} (or {\it pointlike}) and {\it resolved} (or {\it hadronic}) contributions. In the direct contribution, the photon directly enters the hard cross section, and this contribution can be expressed in the form
\begin{equation}
\frac{d\sigma}{dydp_T}\bigg|_{\rm dir} \; = \; \sum_j \int dx_p \,f_{j/H}(x_p) \, \frac{d\hat{\sigma}_{j\gamma}}{dydp_T}(P_{\gamma},x_p P_H) \; , 
    \label{eq:direct}
\end{equation}
where $f_{j/H}(x_p)$ is the distribution of parton $j$ in the hadron $H$, which can be either a proton or a nucleus, 
$d\hat{\sigma}_{j\gamma}(P_{\gamma},x_p P_H)/(dy dp_T)$
is the partonic cross section for the scattering of parton $j$ with a photon $\gamma$, producing a heavy quark in the final state, $P_{\gamma}$ and $P_H$ are the four-momenta of the photon and hadron, respectively, and $x_p$ is the momentum fraction of the hadron $H$ carried by parton $j$. In the above notation, we suppress the dependence on the renormalization and factorization scales. \\[4pt]
However, it has long been known that the photon can have a hadronic structure, see, for example, the reviews in Refs.~\cite{Nisius:1999cv,Krawczyk:2000mf,Buras:2005nj}. Therefore, in addition to the pointlike component, one must also include the {resolved}, or {hadronic}, contribution. The formula for the cross section including the resolved photon is given in the following collinearly factorized form
\begin{equation}
\frac{d\sigma}{dydp_T}\bigg|_{\rm res} \; = \; \sum_{kj} \int dx_{\gamma} \, dx_p \, f_{k/\gamma}(x_\gamma) \, f_{j/H}(x_p) \, \frac{d\hat{\sigma}_{kj}}{dydp_T}(x_\gamma P_{\gamma},x_p P_H) \; , 
    \label{eq:resolved}
\end{equation}
where $f_{k/\gamma}$ is the parton density in the photon and 
$d\hat{\sigma}_{kj}(x_\gamma P_{\gamma},x_p P_H)/(dydp_T)$ 
is the partonic cross section for the scattering of two partons, one from the hadron ({\it j/H}) and the other from the photon ($k/\gamma$), producing a heavy quark in the final state. Here, $x_{\gamma}$ denotes the fraction of the momentum of the photon carried by parton labeled $k$. The separation between the pointlike and resolved components is clear at leading order, but at higher orders in perturbation theory it is not well defined, and the two contributions are related, see the discussion in Refs.~\cite{Cacciari:2001td,Cacciari:2025tgr}. In the present calculation, G$\gamma$A--FONLL adopts the GRV~\cite{GRVPhotonPDF} parametrizations for the photon PDFs.

\subsection{Photon flux parametrization in \texorpdfstring{$AA$}{AA} UPCs}
\label{sec:photon_fluxes}

\begin{figure}[t!]
    \centering
    \includegraphics[width=0.65\linewidth]{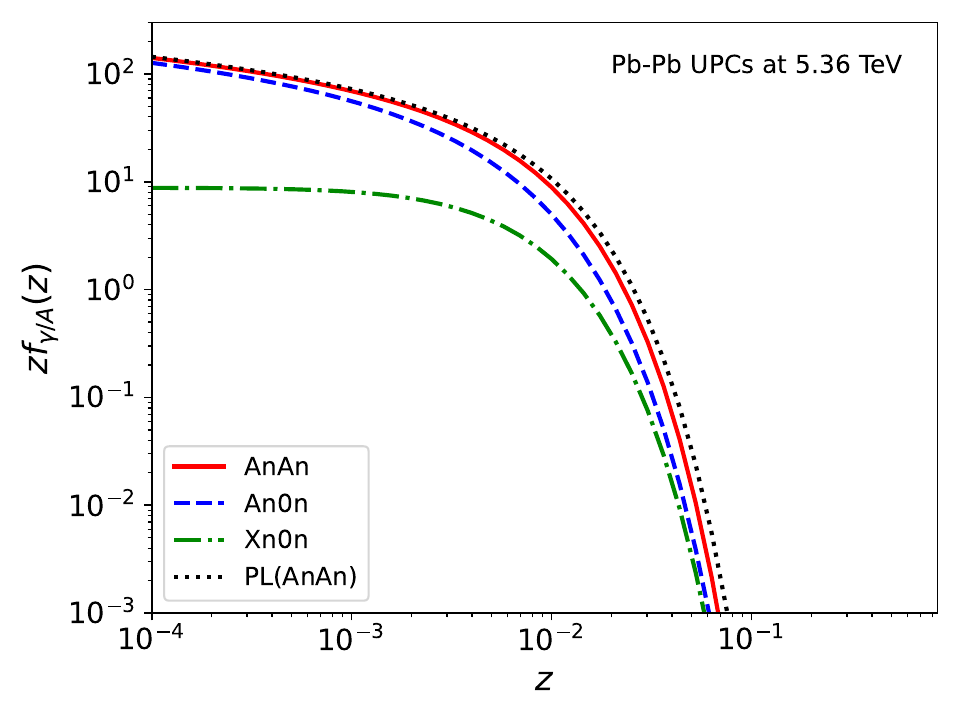}
    \caption{The photon fluxes for $AnAn$ (red solid), $An0n$ (blue dashed), and $Xn0n$ (green dot-dashed) neutron classes as a function of the energy fraction $z$ for Pb--Pb UPCs at 5.36 TeV. The black dotten curve is the PL approximation.}
    \label{fig:flux_AA_classes}
\end{figure}

\begin{figure}[h]
    \centering
    \includegraphics[width=0.65\linewidth]{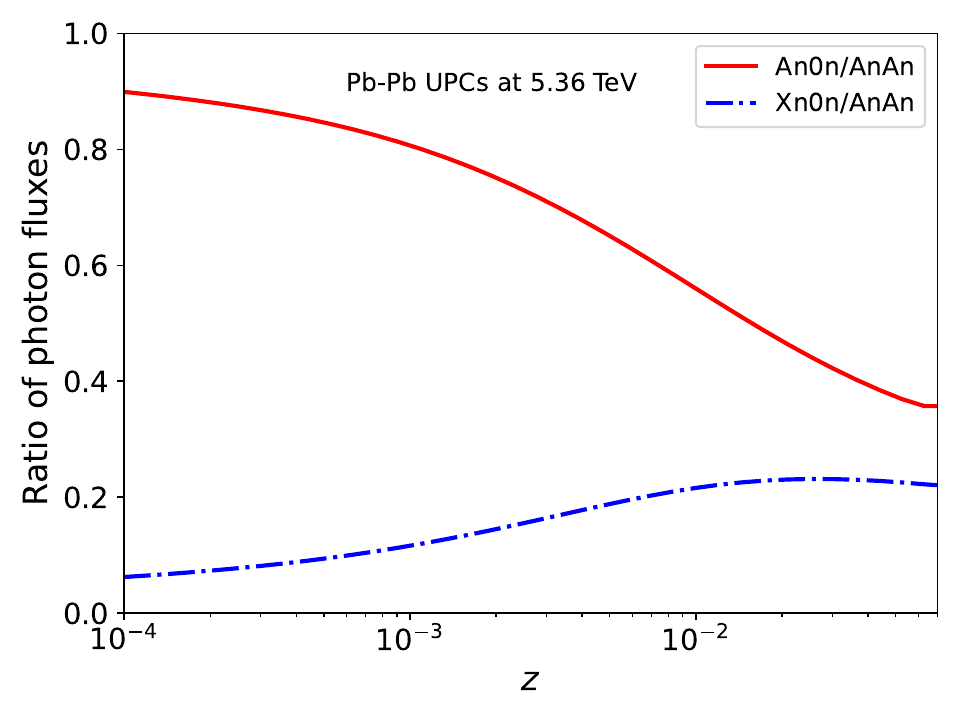}
\caption{The ratio of the photon fluxes for different neutron classes as function of $z$ for Pb--Pb UPCs at 5.36 TeV: $An0n/AnAn$ (red solid) and $Xn0n/AnAn$ (blue dot-dashed).}
    \label{fig:ratio_An0n_Xn0n}
\end{figure}

Using the Weizs\"acker--Williams method of equivalent photons, or equivalent photon approximation (EPA)~\cite{Fermi:1924tc,Fermi:1925fq,vonWeizsacker:1934nji,Williams:1934ad}, UPC cross sections are written as the convolution of the underlying photoproduction cross section with a photon flux.  
This flux is obtained by integrating the impact-parameter-dependent photon density over the transverse geometry of the collision, while excluding configurations in which the nuclei overlap and therefore interact hadronically as well imposing conditions on the number of forward neutrons from possible mutual electromagnetic dissociation of the colliding ions~\cite{Vidovic:1992ik,Nystrand:2004vn,Guzey:2018dlm,Eskola:2024fhf}. 
\\[4pt]
Specifically, 
experimental measurements typically impose neutron-tagging requirements using the 
ZDC.
In the existing photonuclear \(D^0\) measurement, a \(0n\) condition is imposed on the photon-emitting side to suppress hadronic contamination. In addition, an \(Xn\) condition is applied on the target side to identify the direction of the incoming photon and suppress photon--photon background. This selection is also required by the online-triggering conditions typically used by LHC experiments to increase the rate of collected photonuclear events~\cite{CMS:2025jjx}. 
The presence of a \(0n\) requirement can lead to a suppression of the visible photonuclear cross section as a result of the presence of uncorrelated soft-photon exchanges, which can induce electromagnetic dissociation of the photon-emitting nucleus~\cite{Baltz:2002pp}. As a result, not all photonuclear events contribute to the \(0n\)-tagged sample. 
\\[4pt]
A separate effect might arise from the \(Xn\) requirement on the photon-receiving nucleus since it also implies that the measured cross section is not fully inclusive. The largest potential bias is expected for coherent diffraction, where the diffractively scattered lead nucleus remains intact at the hard-scattering level. Such events contribute to the inclusive photonuclear cross section, but can fail the \(Xn\) selection unless the target nucleus is subsequently broken up by an additional independent electromagnetic interaction. Previous versions of G\(\gamma\)A--FONLL did not explicitly account for this condition. In particular, in the results presented in Ref.~\cite{Cacciari:2025tgr}
this effect was neglected, based on a crude estimate using the leading twist shadowing model for nuclear diffraction~\cite{Frankfurt:2011cs}, which suggested that the corresponding correction would be small. In Sec.~\ref{sec:impactonXn0n}, we will revisit this assumption by explicitly evaluating the coherent diffractive contribution and by constructing \(Xn0n\)-corrected cross sections that can be directly compared with the neutron-tagged experimental measurement.\\[4pt]
To enable an accurate comparison with data, theoretical calculations must therefore be corrected to account for these neutron-tagging criteria. Below we consider the photon fluxes corresponding to the following three experimental situations: no conditions are imposed on the final states of the outgoing nuclei and the number of forward neutrons (\(AnAn\)), zero forward neutrons in the ZDC on the photon-emitting side and no conditional on the nuclear 
target side (\(An0n\)), and zero forward neutrons in the ZDC on one side and at least one neutron in the ZDC on the other side (\(Xn0n\)).
\\[4pt]
In the \(AnAn\) channel, 
the photon flux $f_{\gamma/A}(z)$ as a function of the photon energy fraction $z=k/E_A$, where $E_A$ is the nuclear beam energy,
is given the following expression
\begin{equation}
f_{\gamma/A}^{(AnAn)}(z)=\int d^2  \textbf{b} \,f_{\gamma/A}(z,\textbf{b}) \, \Gamma_{AA}(\textbf{b}) \,,
\label{eq:flux}
\end{equation} 
where the integration runs over the transverse distance $\textbf{b}$ (impact parameter) between 
the centers of the two nuclei in the transverse plane of the collision.
One can in principle refine Eq.~(\ref{eq:flux}) by taking into account details of the transverse-plane geometry of UPC  events and the spatial dependence nuclear PDFs~\cite{Eskola:2024fhf}.
While it somewhat increases $f_{\gamma/A}(z)$ for large $z$, the photon flux is suppressed by several orders of magnitude in this region.   
\\[4pt]
In Eq.~(\ref{eq:flux}), $f_{\gamma/A}(z,\textbf{b})$ is the photon flux emitted by an ultrarelativistic charge distribution at the transverse distance $\textbf{b}$ from its center~\cite{Vidovic:1992ik}
\begin{equation}
f_{\gamma/A}(z,\textbf{b})=\frac{\alpha_{\rm em}Z^2}{\pi^2}\frac{1}{z} \left|\int_0^{\infty} dk_{\perp}
\frac{k_{\perp}^2 F_A(k_{\perp}^2+z^2 m_N^2)}{k_{\perp}^2+z^2 m_N^2}J_1(k_{\perp}|\textbf{b}|)\right|^2 \,,
\label{eq:flux_vidovic}    
\end{equation}
where $\alpha_{\rm em}$ is the fine-structure constant, $Z$ is the nucleus electric charge, $F_A(t)$ is the nucleus (charge) form factor normalized to unity, $F_A(t=0)=1$, 
and $J_1$ is the Bessel function of the first kind. 
One can see from Eq.~(\ref{eq:flux_vidovic}) that the maximal values of the photon transverse 
momentum $k_{\perp}$ and the photon energy $k=z E_A$
are controlled by the nuclear form factor $F_A(t)$. Therefore, they can be estimated as $k_{\perp} \sim 1/R_A$ and $k \sim \gamma_L/R_A$, where 
$R_A$ is the nucleus radius and $\gamma_L=E_A/m_N=\sqrt{s_{NN}}/(2 m_N)$ is the nucleus Lorentz factor.
\\[4pt]
The factor of $\Gamma_{AA}(\textbf{b})$ in Eq.~(\ref{eq:flux}) suppresses hadronic (non-UPC) interactions between the colliding ions and represents the probability of the absence of strong nucleus-nucleus interactions at the distance $\textbf{b}$. One commonly calculates it using the optical limit of the Glauber model for high-energy $AA$ scattering
\begin{equation}
\Gamma_{AA}(\textbf{b})=e^{-\sigma_{NN} \int d^2 \textbf{b}^{\prime} T_A(|\textbf{b}^{\prime}|) T_A(|\textbf{b}-\textbf{b}^{\prime}|)} \,,
\label{eq:Gamma_AA}
\end{equation}
where $\sigma_{NN}$ is the energy-dependent total nucleon-nucleon cross section.
At $\sqrt{s_{NN}}=5.36$ TeV, we use $\sigma_{NN}=92$ mb~\cite{ParticleDataGroup:2020ssz}.
\\[4pt]
In the case, when one is interested in $f_{\gamma/A}(z)$ for small $z$, the photon flux can be reliably estimated using the point-like (PL) approximation for the charge distribution in the photon-emitting nucleus. Substituting $F_A(t)=F_A^{\rm PL}(t)=1$ in Eq.~(\ref{eq:flux_vidovic}), one obtains
\begin{equation}
f_{\gamma/A}^{{\rm PL}}(z,\textbf{b})=\frac{\alpha_{\rm em}Z^2}{\pi^2}\frac{1}{z} \left(z m_N K_1(z m_N |\textbf{b}|)\right)^2\,,
\label{eq:flux_PL}    
\end{equation}
where $K_1$ is the modified Bessel function of the second kind.
Then, introducing the minimal impact parameter $b_{\rm min}$ (the expression in Eq.~(\ref{eq:flux_PL}) logarithmically diverges in the $|\textbf{b}| \to 0$ limit), 
one obtains the final expression for the photon flux in the PL approximation
\begin{equation}
f_{\gamma/A}^{{\rm PL}({AnAn})}(z)=\frac{\alpha_{\rm em}Z^2}{\pi} \frac{1}{z} \left[\zeta^2 \left(K_0^2(\zeta)-K_1^2(\zeta)\right)+2\zeta  K_0(\zeta) K_1(\zeta) \right]\,,
\label{eq:flux_PL_2}    
\end{equation}
where $\zeta=z m_N b_{\rm min}$.
Following~\cite{Nystrand:2004vn,Guzey:2018dlm,Eskola:2024fhf},  
we take $b_{\min} = 2 R = 14.2$ fm, with $R = 7.1$ fm  being the nuclear hard-sphere radius, to account for the finite size of the colliding nuclei.
\\[4pt]
As we discussed above, the photonuclear events in the case of the $D^0$ production are selected by requiring that there are no neutrons (\(0n\)) in a
ZDC
in one direction, and at least one neutron (\(Xn\)) in the opposite ZDC.  This is interpreted as the UPC event in which the \(0n\)-nucleus is the photon emitting one, and the \(Xn\)-nucleus is the parton emitting one, playing the role of a nuclear target.
\footnote{This is supplemented by the additional requirement of the rapidity gap 
in the \(0n\)-direction, 
suppressing
the hadronic and two-photon dominated processes.}
\\[4pt]
Assuming that the soft electromagnetic interactions factorize from the hard interaction~\cite{Baltz:2002pp}, one can
introduce the effective flux corresponding to the \(An0n\) neutron class 
\begin{equation}
f_{\gamma/A}^{({An0n})}(z)=\int d^2 \textbf{b} \,f_{\gamma/A}(z,\textbf{b}) \,\Gamma_{AA}(\textbf{b}) \,P_{\text{no-EM}}(\textbf{b}) \,,
\label{eq:nucfluxEMD_An0n}
\end{equation} 
where the factor of $P_{\text{no-EM}}$ is the probability for not having the electromagnetic breakup of the photon-emitting nucleus. 
It can be parametrized as  
\begin{equation}
  P_{\text{no-EM}}(\textbf{b})=\exp(-S/|\textbf{b}|^2) \,, 
  \label{eq:PnoEM}
\end{equation}
where  the area parameter \(S\) has been taken to be  $(17.4 \, \rm fm)^2$, see \cite{Gimeno-Estivill:2025rbw}.  
\\[4pt]
Similarly, the effective 
flux for \(Xn0n\) neutron class can be evaluated using the following formula (compare with Eq.~(\ref{eq:nucfluxEMD_An0n}))
\begin{equation}
f_{\gamma/A}^{(Xn0n)}(z)=\int d^2 \textbf{b} \, f_{\gamma/A}(z,\textbf{b}) \, \Gamma_{AA}(\textbf{b}) \, P_{\text{no-EM}}(\textbf{b})  \,\left(1-P_{\text{no-EM}}(\textbf{b}) \right) \,,
\label{eq:nucfluxEMD_Xn0n}
\end{equation} 
where the term $1-P_{\text{no-EM}}$ corresponds to the probability of at least one neutron emission.
\\[4pt]
Figure~\ref{fig:flux_AA_classes} shows the photon fluxes for different neutron classes
discussed in this section as a function of the energy fraction $z$ for Pb--Pb UPCs at 5.36 TeV.
One can see from the figure that in the $AnAn$ case, the PL approximation (black dotted line)
reproduces the exact calculation (red solid line) very well for small $z < 0.01$, 
with 
the PL flux
being slightly higher and the largest differences for $z> 0.01$, where both fluxes are suppressed by 
several orders of magnitude.
At the same time, the $An0n$ (blue dashed) and $Xn0n$ (green dot-dashed) fluxes are suppressed 
with respect to the $AnAn$ case.
\\[4pt]
To better illustrate this suppression, it is also instructive to compare the magnitudes of the two effective fluxes, $An0n$ and $Xn0n$, in more details. This is 
presented
in Fig.~\ref{fig:ratio_An0n_Xn0n}, where both fluxes are shown as ratios to the 
$AnAn$ flux. 
One can see from the figure that in the $An0n$ case (red solid curve), the ratio
is most suppressed in the large $z$ region, and thus it has the largest effect on suppression in the large \(p_T\) and \(y\) region of the produced $D^0$. It was shown \cite{Cacciari:2025tgr} that in the CMS kinematics the suppression on the level of the cross section can range between 0.8 (for low $p_T\sim 2 \;\rm GeV$ and negative rapidity, $y\sim (-2,-1)$ to 0.4 (for higher $p_T\sim 12 \;\rm GeV$ and high rapidity, $y\sim (1,2)$). 
In the $Xn0n$ case (blue dot-dashed curve), the ratio
is strongly suppressed for small values of $z$ and then only comparable to the $An0n$ case for large $z$, 
where the overall flux is 
in any case negligible.

\clearpage
\clearpage
\section{Diffractive \texorpdfstring{$D^0$}{D0} photoproduction in Pb--Pb UPCs with G\texorpdfstring{$\gamma$}{gamma}A-FONLL}
\label{sec:results}

\begin{figure}[H]
\centering
\begin{subfigure}{0.49\textwidth}
    \centering
    \includegraphics[width=\textwidth]{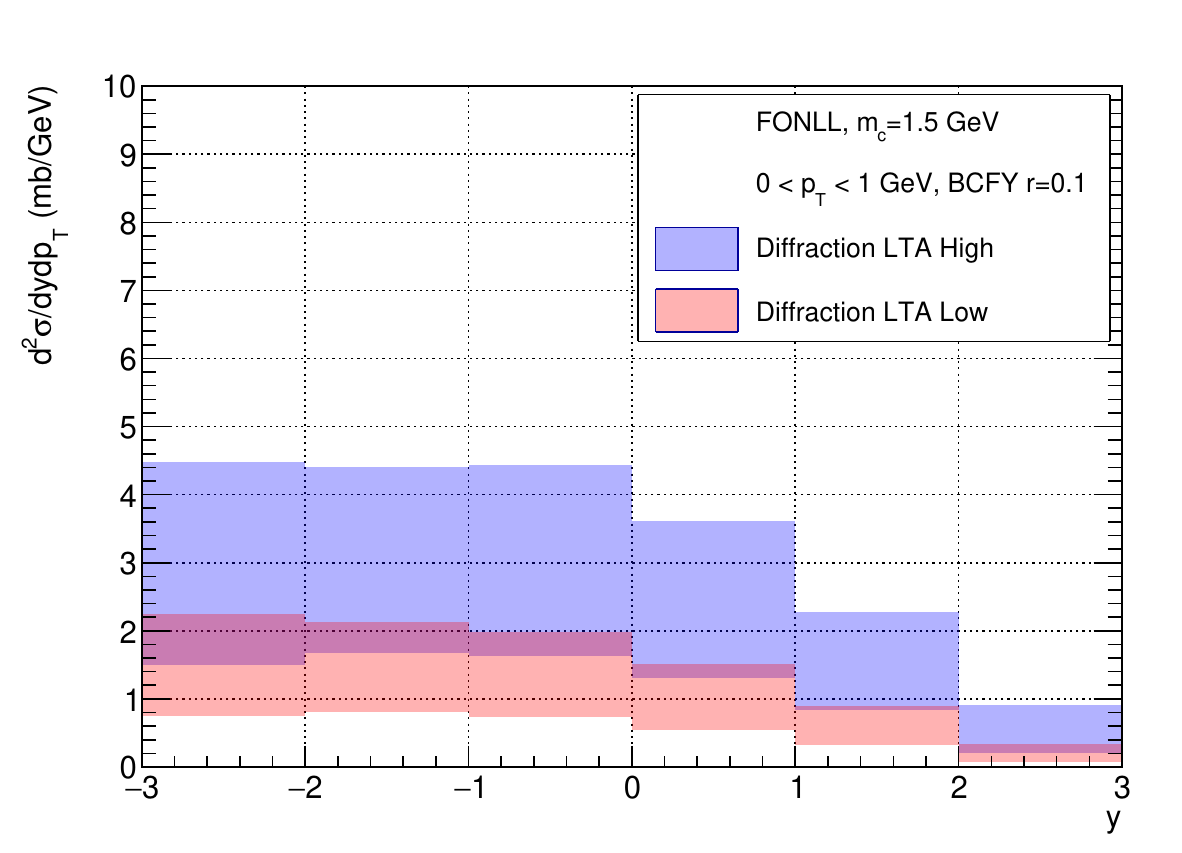}
    \label{fig:ydpt0_diff}
\end{subfigure}
\hfill
\begin{subfigure}{0.49\textwidth}
    \centering
    \includegraphics[width=\textwidth]{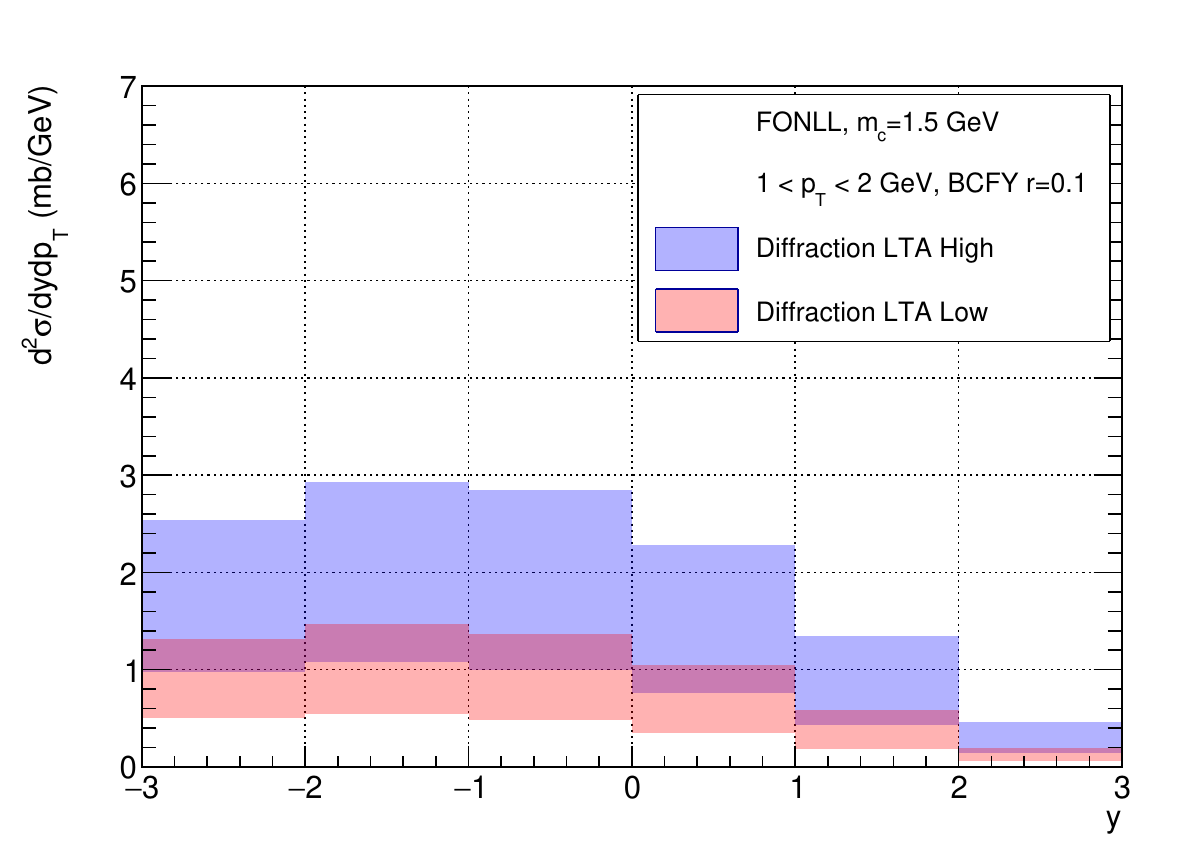}
    \label{fig:ydpt1_diff}
\end{subfigure}
\begin{subfigure}{0.49\textwidth}
    \centering
    \includegraphics[width=\textwidth]{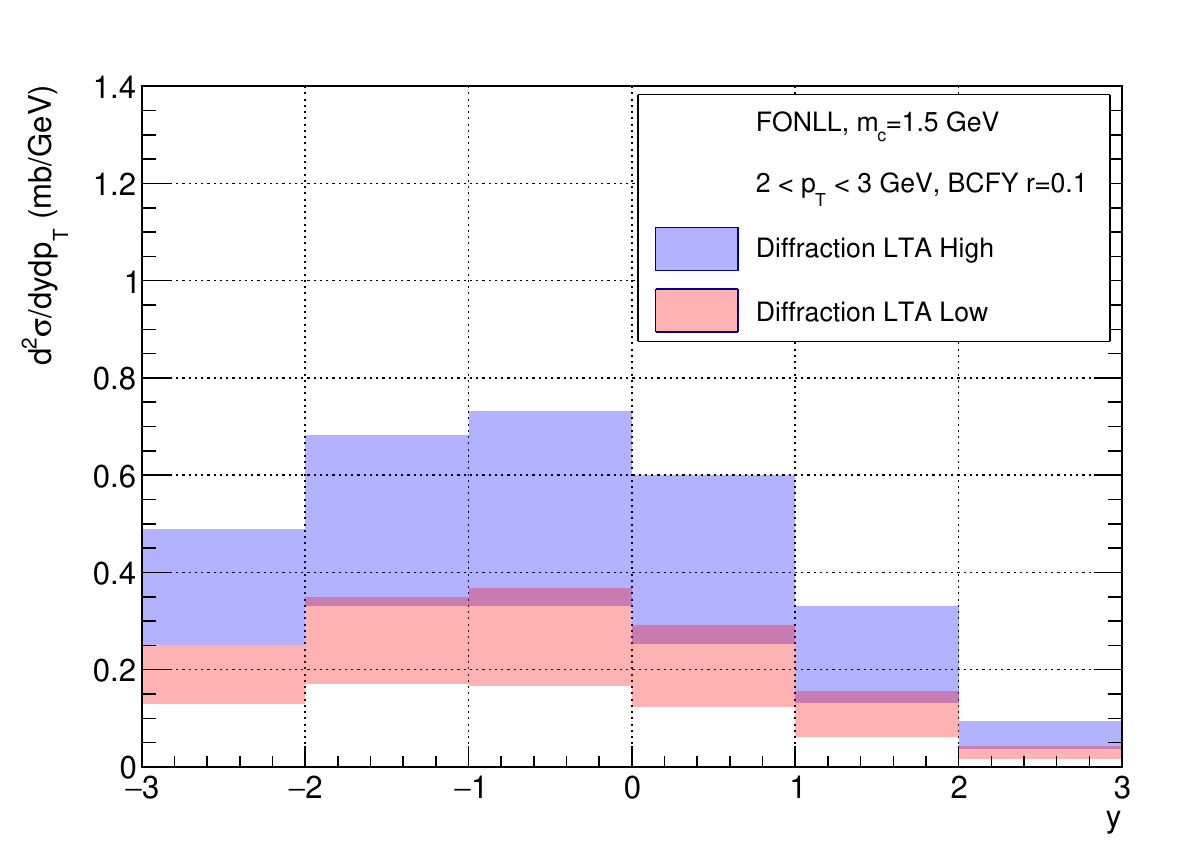}
    \label{fig:ydpt2_diff}
\end{subfigure}
\hfill
\begin{subfigure}{0.49\textwidth}
    \centering
    \includegraphics[width=\textwidth]{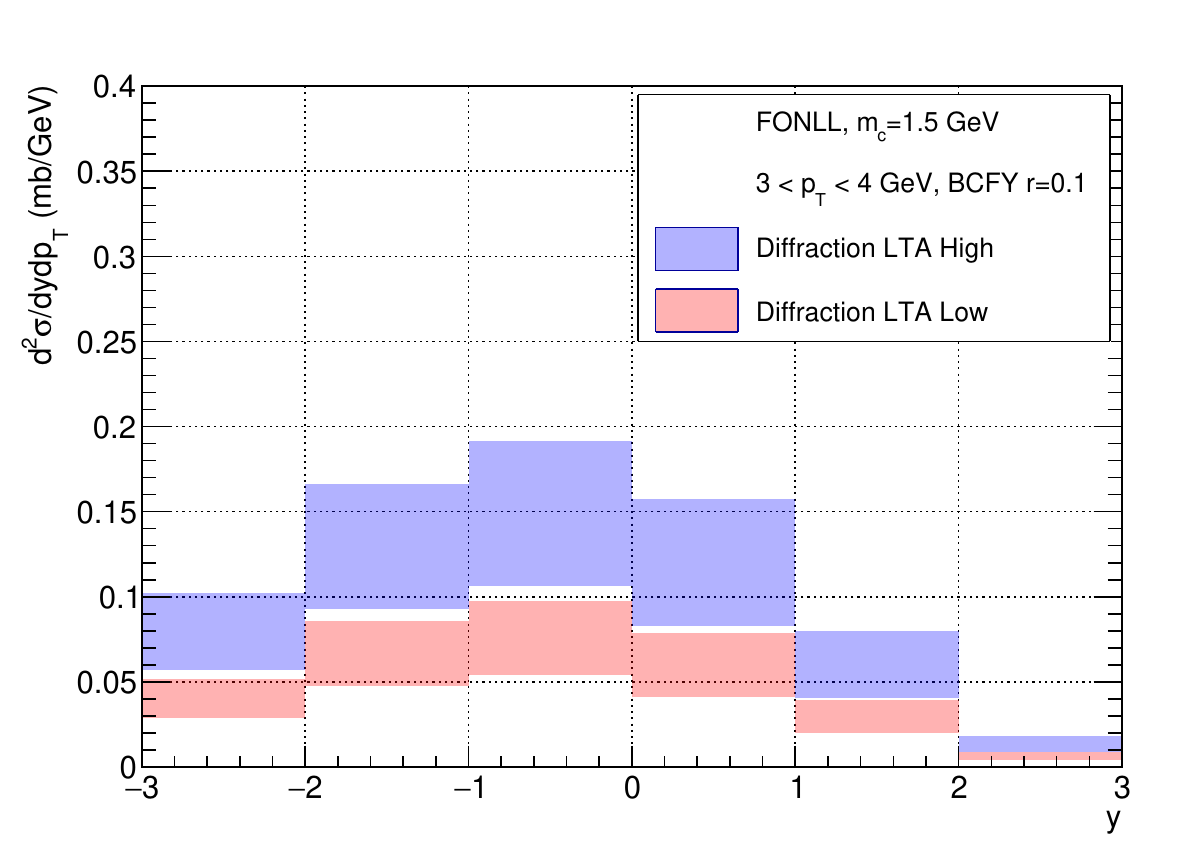}
    \label{fig:ydpt3_diff}
\end{subfigure}
\begin{subfigure}{0.49\textwidth}
    \centering
    \includegraphics[width=\textwidth]{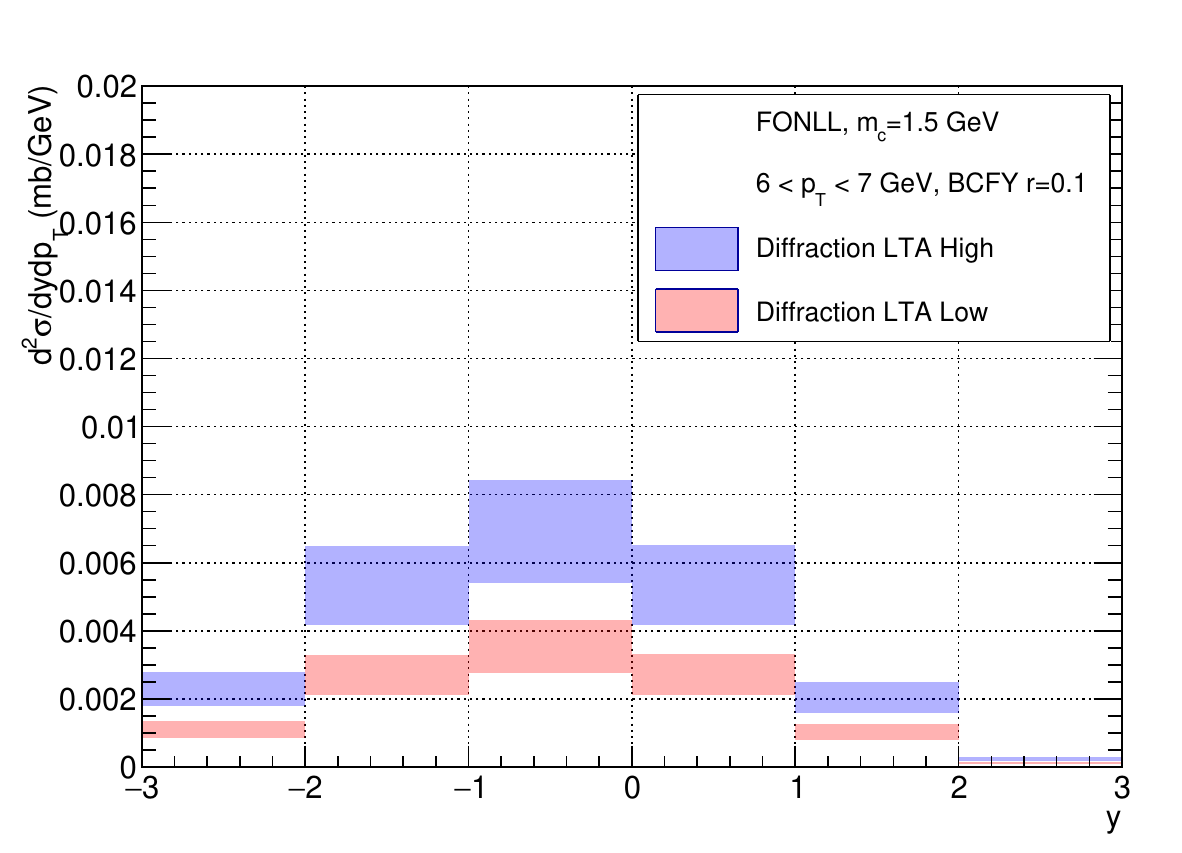}
    \label{fig:ydpt6_diff}
\end{subfigure}
\hfill
\begin{subfigure}{0.49\textwidth}
    \centering
    \includegraphics[width=\textwidth]{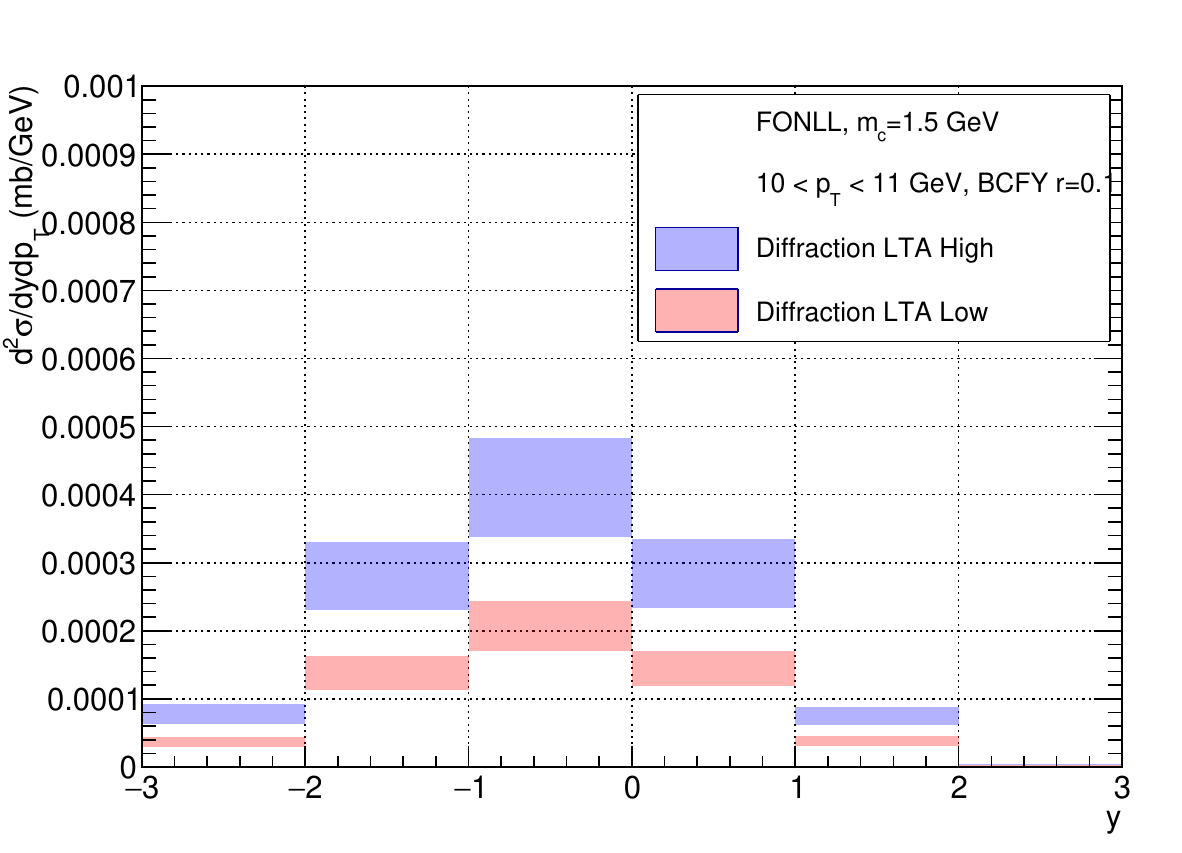}
    \label{fig:ydpt10_diff}
\end{subfigure}
\caption{Diffractive cross section for   the $D^0$ photoproduction in AA UPCs in bins  of \(p_T\) as a function of rapidity \(y\) of the produced charmed meson. Bands indicate renormalization and factorization scale variation. 
Red bands correspond to the `Low' shadowing scenario and the blue bands to the `High' shadowing scenario within the LTA  shadowing model.
 Value of $x_{\Pomeron}$ has been integrated up to $0.1$, charm mass has been set to $m_c=1.5$ GeV, and BCFY fragmentation function was used with parameter $r=0.1$. }
 \label{fig:ydpt_diff_aa_upc}
\end{figure}

\begin{figure}[h]
    \centering
    \begin{subfigure}{0.49\textwidth}
    \centering
    \includegraphics[width=\textwidth]{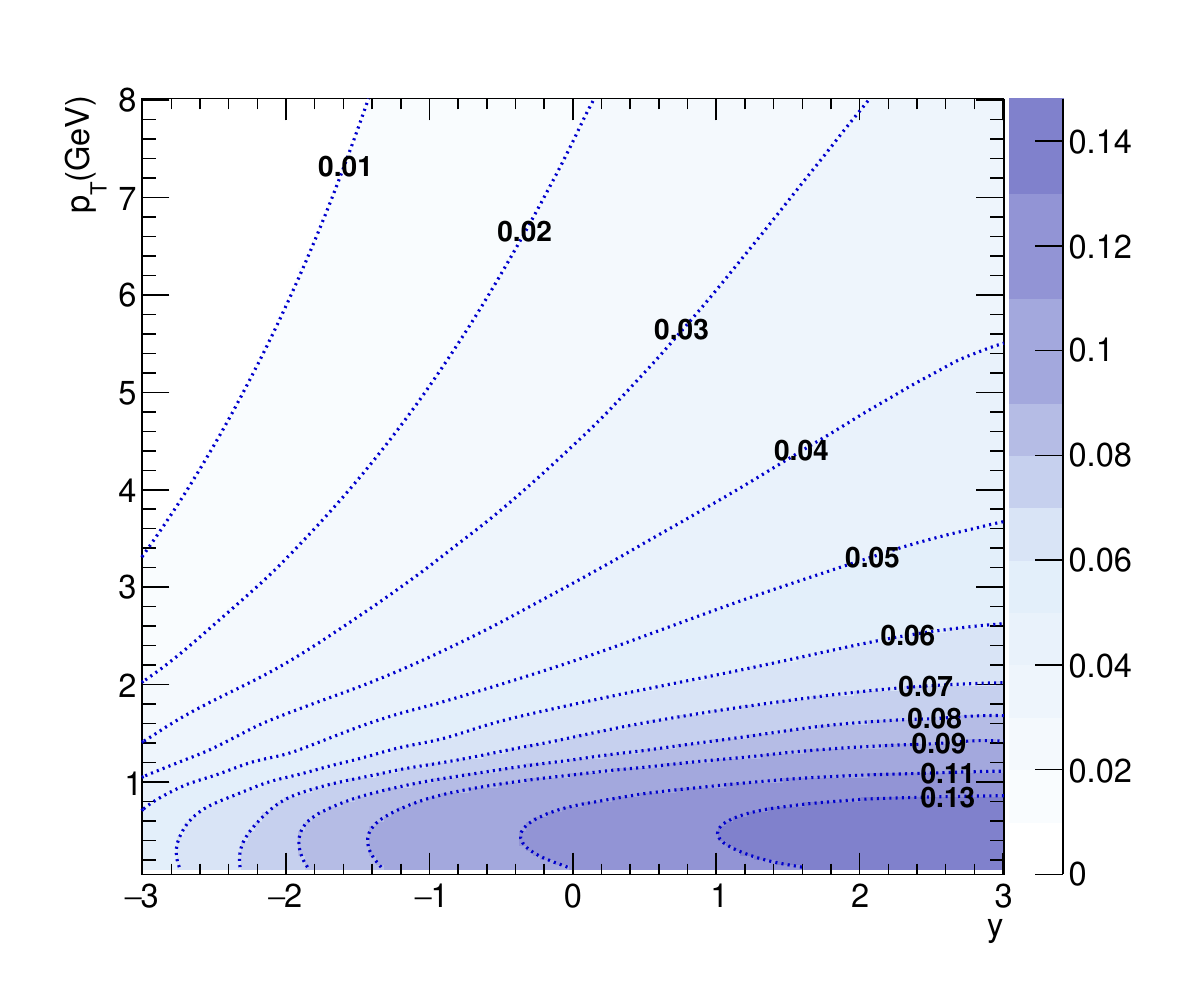}
    \end{subfigure}
   \begin{subfigure}{0.49\textwidth}
    \centering
    \includegraphics[width=\textwidth]{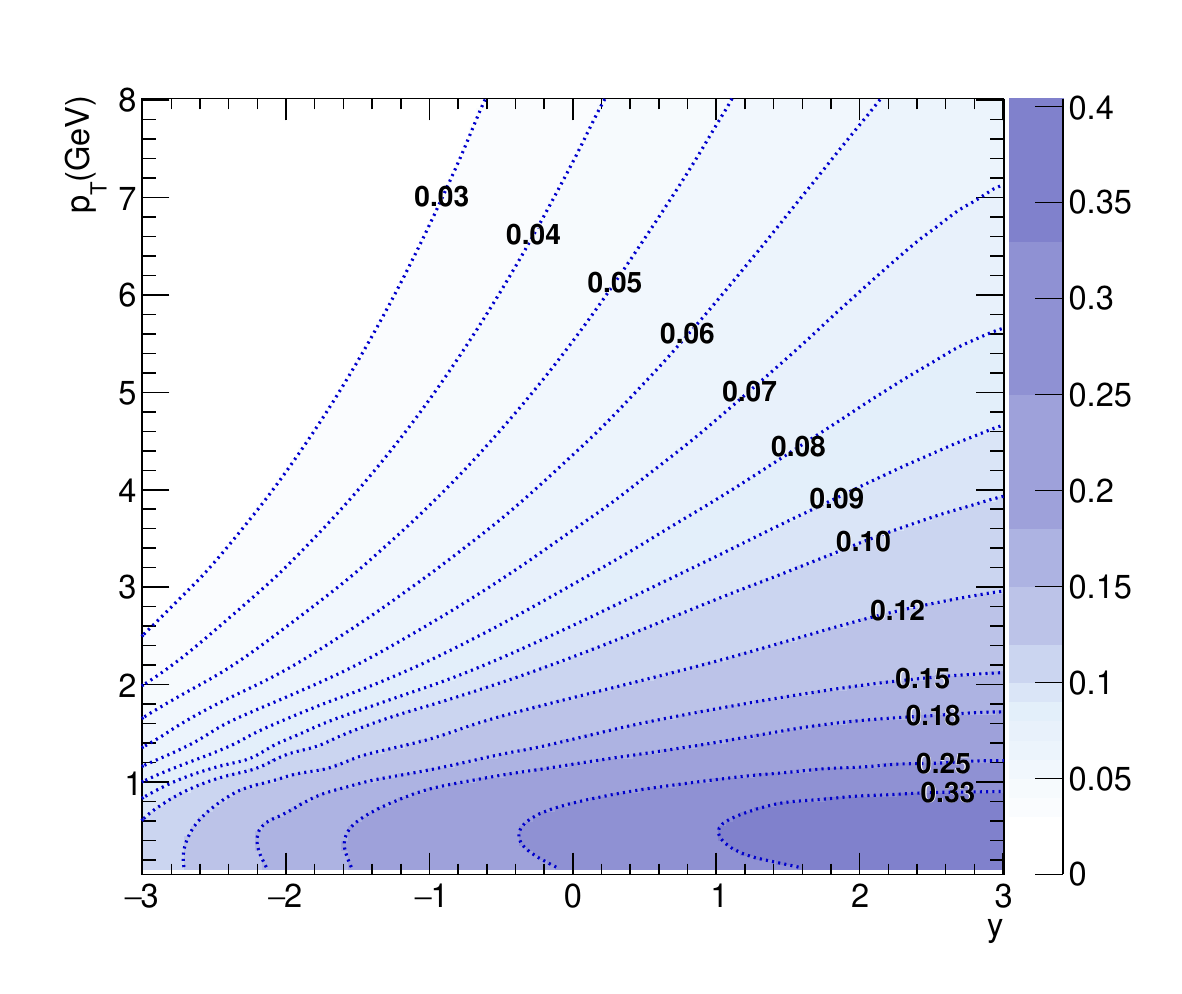}
    \end{subfigure}
    \caption{Contour plots of the diffractive to inclusive ratio for the UPC $D^0$ photoproduction cross section in \(p_T\) and \(y\) plane. Value of $x_{\Pomeron}$ has been integrated up to $0.1$ in the nuclear diffractive PDFs. Both inclusive and diffractive cross sections have been evaluated using LTA   diffractive and inclusive nuclear PDFs respectively. Left: LTA ``low shadowing'', and Right: LTA ''high shadowing''.}
    \label{fig:diff_to_inclusive}
\end{figure}

In this section we apply the diffractive framework to predict coherent diffractive \(D^0\) photoproduction in Pb--Pb UPCs. In this case, the photon emitted by one lead ion scatters diffractively from the other lead ion, which remains intact. 
The corresponding cross section reads
\begin{equation}
   d\sigma^{A+A\rightarrow D^0+A}_{An0n}=f_{\gamma/A}^{({An0n})}\otimes d\sigma_{\rm diff}^{(\gamma+A \rightarrow D^0+A)} \;.
   \label{eq:UPC_diff}
\end{equation}
In Eq.~(\ref{eq:UPC_diff}),
the short-distance charm-production and fragmentation ingredients are the same as in the inclusive G\(\gamma\)A--FONLL calculation. 
Following Ref.~\cite{Cacciari:2003zu}, where the charm-quark mass and the BCFY fragmentation parameter were simultaneously tuned to provide the best description of charm-hadroproduction data, we use \(m_c=1.5\,\mathrm{GeV}\) together with \(r=0.1\).
For the calculation of $d\sigma_{\rm diff}^{(\gamma+A \rightarrow D^0+A)}$, 
we use nuclear diffractive PDFs obtained in the leading twist shadowing (LTA) approach~\cite{Frankfurt:2003gx,Frankfurt:2011cs}.
The calculation is performed for the two standard LTA scenarios, conventionally referred to as ``low shadowing'' and ``high shadowing'', see Sec.~\ref{sec:diffraction_nucleus}.
\\[4pt] 
The LTA calculation provides the diffractive-to-inclusive ratio, while the inclusive \(D^0\) cross section is taken from the G\(\gamma\)A--FONLL setup of Ref.~\cite{Cacciari:2025tgr}, which uses modern inclusive nuclear PDFs. We therefore obtain the absolute diffractive cross section by multiplying the inclusive prediction by the LTA diffractive-to-inclusive ratio.
In Ref.~\cite{Cacciari:2025tgr}, the inclusive cross section was evaluated with both EPPS21~\cite{Eskola:2021nhw} and nNNPDF3.0~\cite{AbdulKhalek:2022fyi}, which gave very similar results. In the following, we use EPPS21 as the inclusive baseline for this reweighting procedure. The resulting diffractive cross sections, differential in the \(D^0\)-meson rapidity \(y\) and shown in representative \(p_T\) bins, are presented in Fig.~\ref{fig:ydpt_diff_aa_upc}. The bands indicate the factorization- and renormalization-scale variation, \(\mu_F,\mu_R\in(0.5\mu_0,2.0\mu_0)\), with \(\mu_0=m_T=\sqrt{p_T^2+m_c^2}\) and \(1/2\leq \mu_F/\mu_R\leq 2\). The LTA ``high shadowing'' scenario gives a larger diffractive cross section than the LTA ``low shadowing'' scenario. The diffractive cross section decreases with rapidity from about \(y\simeq -0.5\) to \(y=2\). This behavior mainly reflects the decrease of the photon flux with increasing photon energy. 
\\[4pt] 
In Fig.~\ref{fig:diff_to_inclusive}, we present the ratio of the diffractive to inclusive contributions, shown as a contour plot in the plane of the \(D^0\)-meson rapidity \(y\) and transverse momentum \(p_T\). In the diffractive calculation, the nuclear diffractive PDFs are integrated over \(x_{\Pomeron}\) up to \(0.1\). For consistency, both the diffractive and inclusive calculations are evaluated using nuclear PDFs obtained in the leading twist shadowing model. 
The left panel of Fig.~\ref{fig:diff_to_inclusive} corresponds to the LTA ``low shadowing'' scenario, while the right panel corresponds to the LTA ``high shadowing'' scenario~\cite{Frankfurt:2011cs}. The diffractive contribution in both scenarios is largest at low \(p_T\) and large rapidity \(y\), which correspond to the smallest values of the gluon momentum fraction \(x_g\) in the nucleus. The largest diffractive contributions are obtained at the highest rapidities of the \(D^0\) meson. The diffractive fraction decreases rapidly with transverse momentum, reaching only a few percent at \(p_T=8\,\mathrm{GeV}\). For \(p_T=2\,\mathrm{GeV}\), the diffractive cross section amounts to about \(5\%\) of the inclusive cross section in the ``low shadowing'' scenario and about \(10\)--\(15\%\) in the ''high shadowing'' scenario. At large negative rapidity, the diffractive contribution is also suppressed. This can be understood from the approximate relation \(x_g \simeq (m_T/\sqrt{s_{NN}})\exp(-y_c)\), discussed in Ref.~\cite{Cacciari:2025tgr}, which shows that negative rapidities probe larger gluon momentum fractions in the nucleus. The exact relation between \(x_g\) and rapidity is more involved, as shown in Appendix~A. Nevertheless, the suppression at large negative rapidity is connected to the larger values of \(x_g\), for which the available rapidity gap becomes small and the diffractive contribution is reduced. 
In summary,
both Figs.~\ref{fig:ydpt_diff_aa_upc} and~\ref{fig:diff_to_inclusive} show a significant dependence on the choice of the LTA shadowing scenario, with the ``high shadowing''  
predicting a diffractive cross section typically about a factor of two larger than in the ''low shadowing'' case.

\noindent

\noindent
\\[4pt] 
Additionally, 
we have also evaluated the diffractive fraction in \(AA\) UPCs using the following rather simple estimate. We use the impulse approximation, with the diffractive proton PDFs taken from the ZEUS SJ parametrization~\cite{ZEUS:2009uxs}, and include nuclear shadowing in these PDFs by rescaling them by a factor \(R_{i/A}^2\). Here, \(R_{i/A}=f_{i/A}/(A f_{i/p})\) is the nuclear ratio for the inclusive PDFs, taken from the EPPS21 parametrization. The modification by the square of this ratio stems from the fact that diffraction is a quasi-elastic contribution, and therefore the shadowing factor enters at the amplitude level and must be squared in the cross section. Using this prescription, we find that the diffractive ratio calculated in this way is much larger than in the LTA model. More precisely, the diffractive contribution in this model reaches about \(30\%\) along a contour extending from \(y \approx -1.5\) and \(p_T\approx 2\,\mathrm{GeV}\) to \(y\approx 2\) and \(p_T\approx 7\,\mathrm{GeV}\). It can even reach \(50\%\) for \(p_T<2\,\mathrm{GeV}\) and \(y>0.5\). Given the oversimplified nature of this model, the resulting estimate should therefore be treated only as an upper bound on the diffractive cross section.

\subsection{Experimental considerations for future measurements of diffractive production of \texorpdfstring{$D^{0}$}{D0} in PbPb collisions at the LHC} 
Future measurements of coherent diffractive open-charm photoproduction in Pb--Pb UPCs would provide a direct test of the nuclear diffractive gluon distribution in a kinematic region that is not accessible in existing lepton--nucleus data. In this subsection, we discuss experimental considerations for such a measurement in the kinematic region accessible by a general-purpose large-acceptance detector, such as CMS at the LHC, taking advantage of the detector capabilities expected after the HL-LHC upgrades~\cite{Contardo:2020886}. \\[4pt]
The experimental strategy for measuring photon-induced diffractive $D^{0}$ production in $\mathrm{A}n0n$ Pb–Pb UPCs would rely on some of the selections developed for the semi-inclusive $D^{0}$ measurement of Ref.~\cite{CMS:2025jjx}, with a few substantial differences. For events in which the incoming photon propagates toward positive pseudorapidity, the signal would be identified by the presence of a $D^{0}$ candidate reconstructed with the tracker detectors, the absence of forward neutrons in the ZDC at positive rapidity, and the presence of a rapidity gap in the same direction. As previously discussed, this rapidity gap is typically implemented by requiring the absence of energy deposits above a given threshold in the corresponding forward hadronic calorimeters (HFs), in the pseudorapidity region $3<\eta<5$. This requirement allows one to select a high-purity UPC sample and reject contamination from hadronic interactions. In contrast to the selection used in Ref.~\cite{CMS:2025jjx}, which focused on semi-inclusive $D^{0}$ production in the $\mathrm{X}n0n$ neutron class, no requirement on the presence of at least one neutron in nucleus-going direction should be imposed for this measurement. Dedicated effort should instead be devoted to designing an experimental strategy to enhance or isolate diffractive events within the inclusive photoproduction sample, exploiting their characteristic large rapidity gap and intact, or coherently scattered, target nucleus while suppressing non-diffractive photonuclear production. The contribution from photon–photon processes should be also quantified with dedicated phenomenological studies and, where possible, constrained with data-driven experimental methods. Isolating the diffractive component from the inclusive photonuclear sample would there require nontrivial experimental techniques, such as template fits to forward-activity or rapidity-gap observables, and would likely introduce sizable experimental uncertainties. Such an effort could also strongly benefit from extended forward-rapidity instrumentation, such as forward shower counters (FSCs), which can extend the rapidity-gap veto beyond the standard calorimeter acceptance by detecting showers from very forward particles~\cite{Albrow:2008pn,Albrow:2014lrm}. FSCs were installed for the heavy-ion data-taking period in Run 3 and are being considered as a dedicated upgrade of the forward CMS instrumentation for Run 4 and 5.\\[4pt]
\begin{figure}[htp!]
\centering
\includegraphics[width=0.9\linewidth]{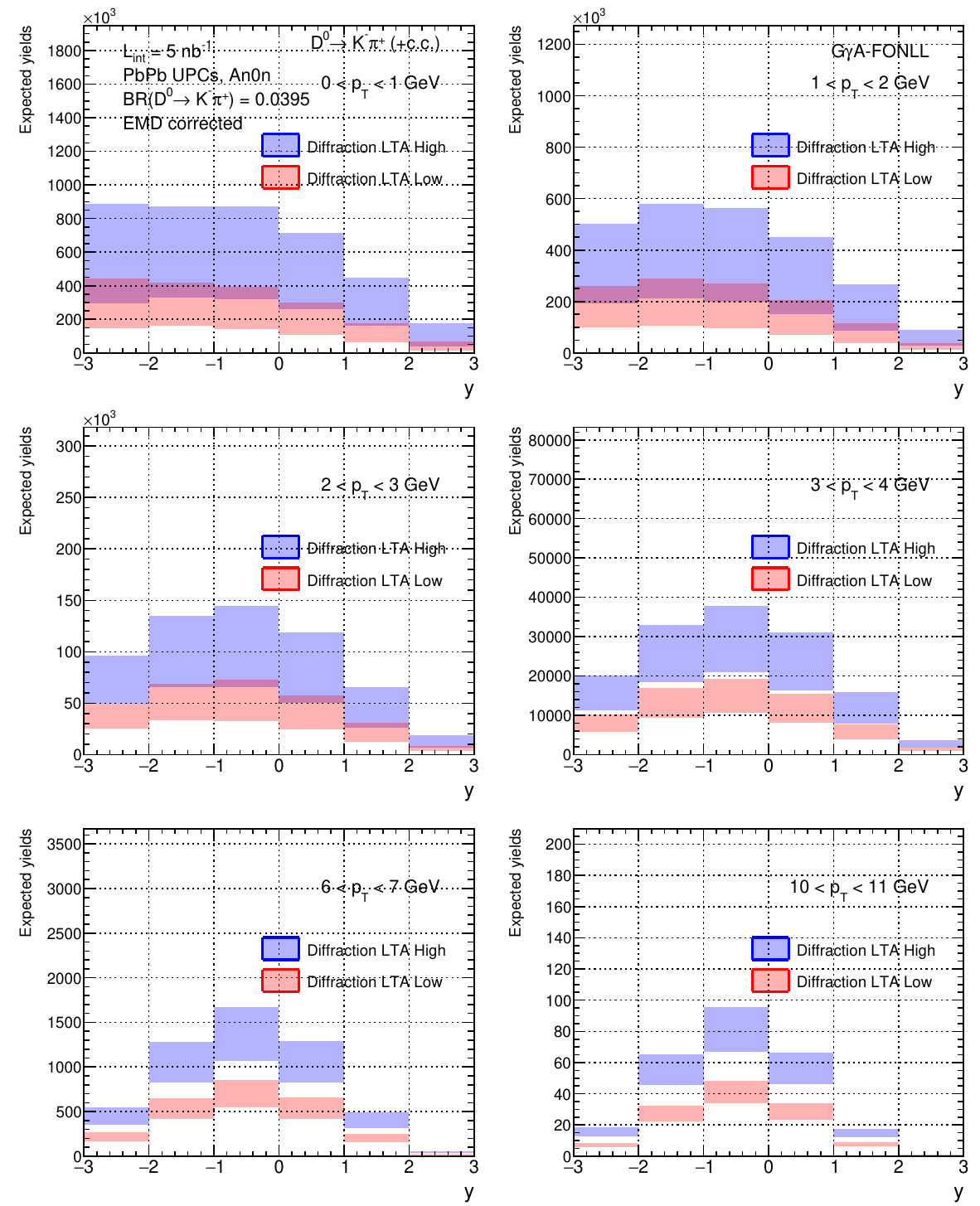}
\caption{Expected yields of $D^{0}\to K^-\pi^+$ (and charge conjugates) for coherent diffractive $D^{0}$ photoproduction in Pb--Pb UPCs at $\sqrt{s_{\scriptscriptstyle NN}}=5.36$ TeV in the $An0n$ neutron class. The yields are obtained from the G$\gamma A$--FONLL calculation with $m_c=1.5~{\rm GeV}$, GRV photon PDFs, EPPS21 PDFs, BCFY fragmentation with $r=0.1$, and the EMD-corrected $An0n$ photon flux, assuming $\mathcal{L}_{\rm int}=5~{\rm nb}^{-1}$ and ${\rm BR}(D^0\to K^-\pi^+)=0.0395$. Detector acceptance, trigger efficiency, and reconstruction efficiency are not included. The blue and red bands show the scale-variation envelopes for the LTA-high and LTA-low nuclear diffractive shadowing scenarios, respectively.}
\label{fig:diff_d0_yields_AA}
\end{figure}
While a detailed assessment of the experimental techniques required for future measurements is beyond the scope of this work, we provide here a simple estimate of the expected yields of diffractive $D^{0}$ signal candidates for a representative Pb--Pb UPC sample with $\mathcal{L}_{\rm int}=5~{\rm nb}^{-1}$ at $\sqrt{s_{\scriptscriptstyle NN}}=5.36~{\rm TeV}$. This estimate does not include detector acceptance, trigger efficiencies, or reconstruction efficiencies, and should therefore be interpreted only as a guide to the available signal statistics rather than as a full experimental projection. \\[4pt]
Figure~\ref{fig:diff_d0_yields_AA} translates the diffractive cross sections presented in Fig.~\ref{fig:ydpt_diff_aa_upc} into expected yields of $D^{0}\to K^\mp\pi^\pm$ candidates (and charge conjugates) for Pb--Pb UPCs at $\sqrt{s_{\scriptscriptstyle NN}}=5.36~{\rm TeV}$ in the $\mathrm{A}n0n$ neutron class. The the G$\gamma A$--FONLL calculation uses $m_c=1.5~{\rm GeV}$, GRV photon PDFs, EPPS21 nuclear PDFs, BCFY fragmentation with $r=0.1$, and the EMD-corrected $An0n$ photon flux discussed in Section~\ref{sec:results}. The plotted yields include the yields of both $D^0+\overline{D}^0$, the charm-to-$D^0$ fragmentation fraction used in the cross-section calculation, and ${\rm BR}(D^0\to K^-\pi^+)=0.0395$. 
\\[4pt]
The six panels show the rapidity dependence of the expected yields in intervals of transverse momentum for An0n events. The expected yields span a wide range across the measured phase space. In the lowest-$p_T$ bins, they reach $\mathcal{O}(10^5)$--$\mathcal{O}(10^6)$ signal candidates, depending on rapidity and on the LTA shadowing scenario. The yields then fall rapidly with increasing transverse momentum, reaching $\mathcal{O}(10^4)$ events in the $3<p_T<4$ GeV bin and $\mathcal{O}(10)$--$\mathcal{O}(10^2)$ events in the $10<p_T<11$ GeV bin, depending on rapidity. These first estimates suggest that high-statistics measurements of diffractive production at low and intermediate $p_T$, differential in rapidity, are likely to be feasible, provided that the main experimental challenges can be addressed. These include, in particular, the extraction of the diffractive fraction and the development of realistic trigger strategies capable of sampling the full delivered luminosity.
\clearpage
\section{Diffraction correction for the 
\texorpdfstring{$Xn0n$}{Xn0n} Pb--Pb \texorpdfstring{$D^0$}{D0} measurement in PbPb UPCs}
\label{sec:impactonXn0n}

\begin{figure}[H]
\centering
\begin{subfigure}{0.49\textwidth}
    \centering
    \includegraphics[width=\textwidth]{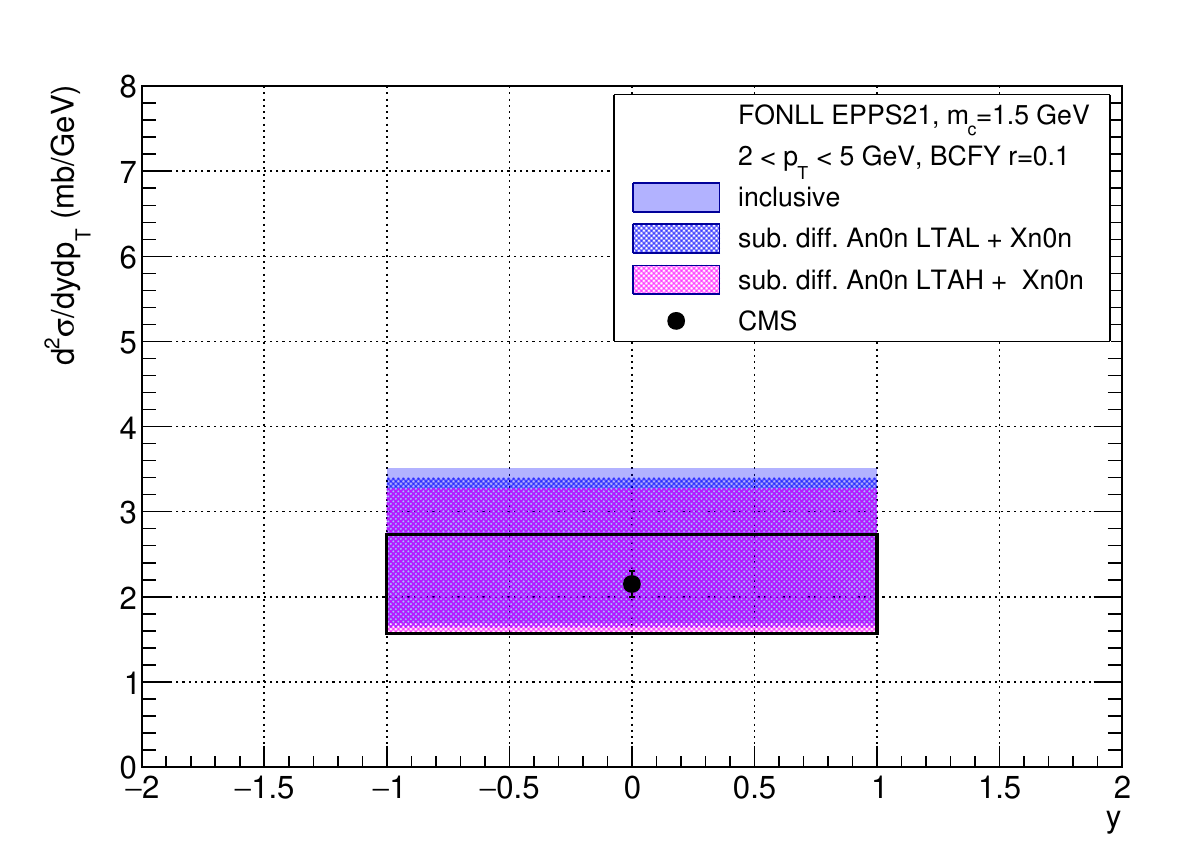}
    \label{fig:ydpt1_1_cms}
\end{subfigure}
\begin{subfigure}{0.49\textwidth}
    \centering
    \includegraphics[width=\textwidth]{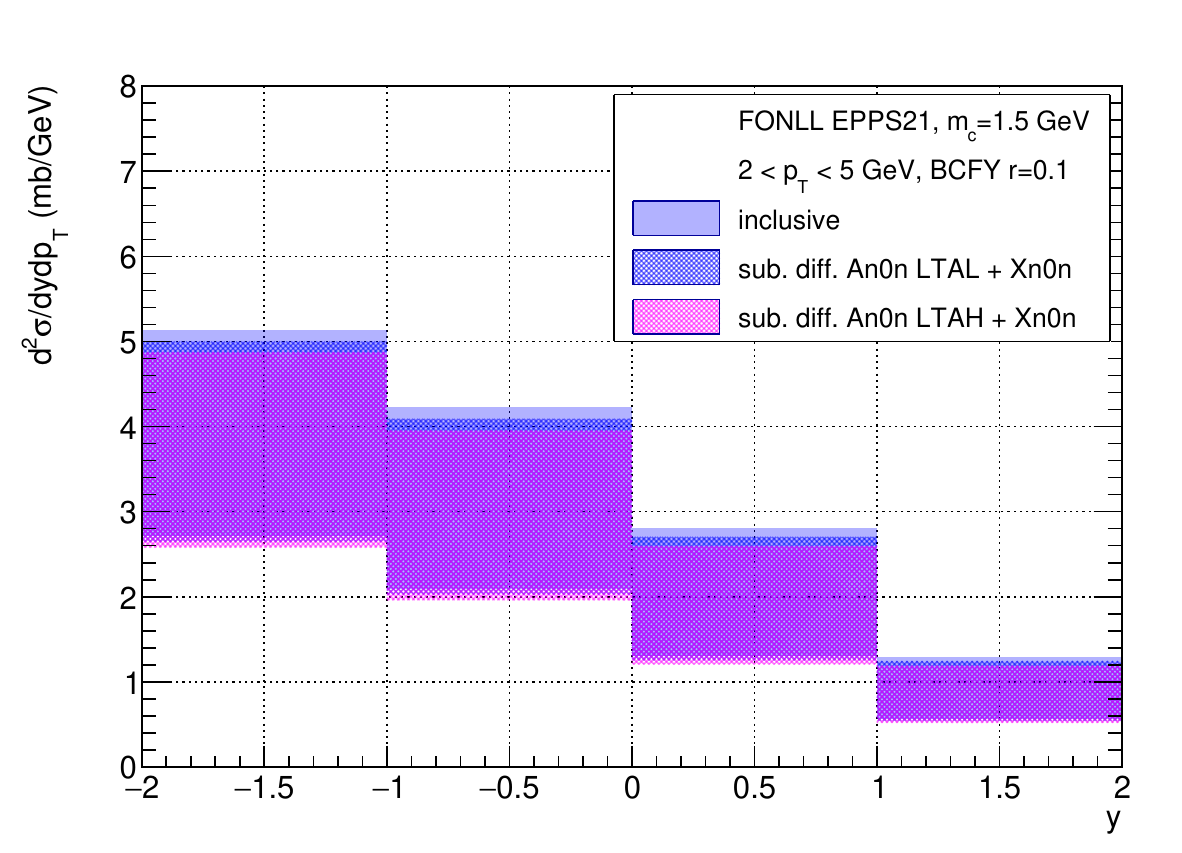}
    \label{fig:ydpt1_cms}
\end{subfigure}
\hfill
\begin{subfigure}{0.49\textwidth}
    \centering
    \includegraphics[width=\textwidth]{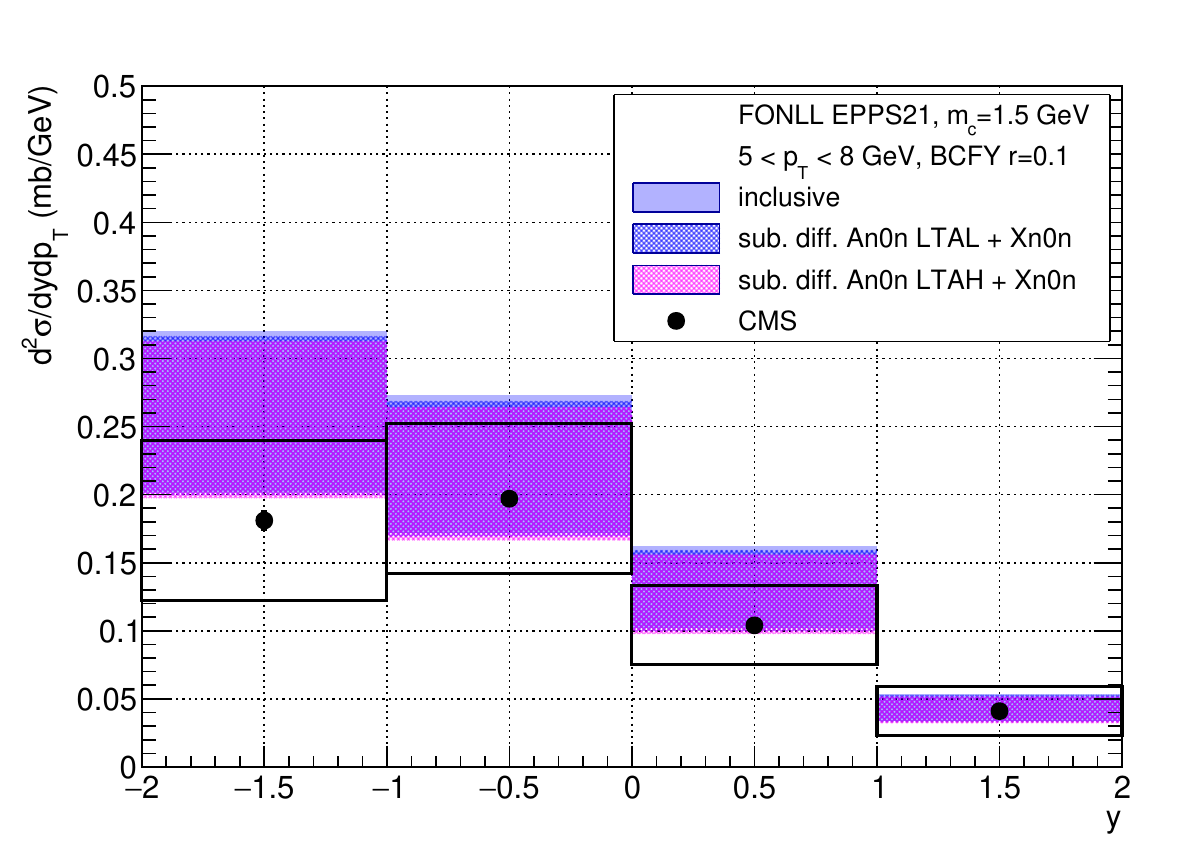}
    \label{fig:ydpt2_cms}
\end{subfigure}
\begin{subfigure}{0.49\textwidth}
    \centering
    \includegraphics[width=\textwidth]{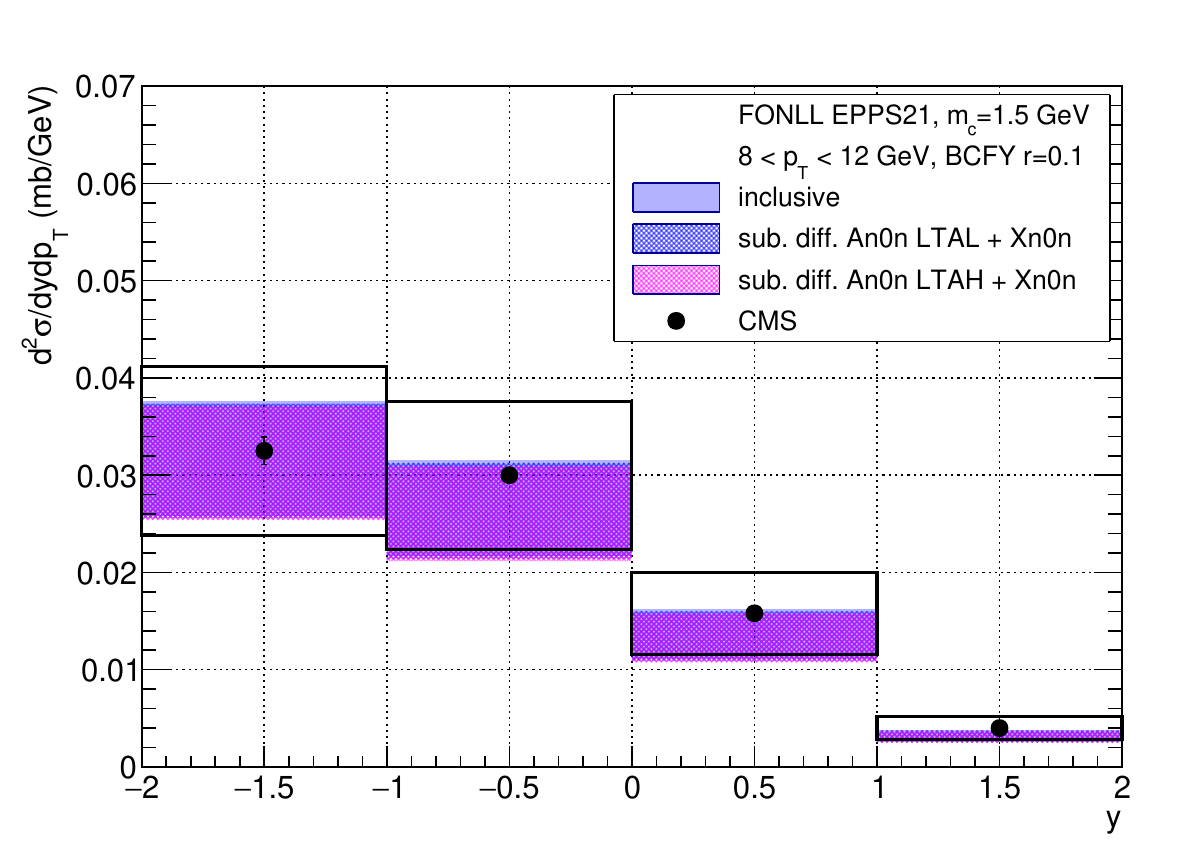}
    \label{fig:ydpt3_cms}
\end{subfigure}
 \caption{Rapidity distribution for the $D^0$  production in UPC Pb--Pb collisions at ${\rm\sqrt{s_{\scriptscriptstyle NN}} }=5.36$ TeV with EPPS21 nuclear PDF in  $p_{T}$ bins $(2-5), (5-8), (8-12)$ GeV. Light blue band: FONLL  calculation with factorization and renormalization scale variation for $m_c=1.5 \, \rm GeV$, dark blue and magenta bands: calculations including  subtraction of the diffractive contribution using the LTA ``low'' (LTAL) and ``high'' (LTAH) shadowing scenarios, 
 according to Eq.~(\ref{eq:crosssection_subtraction}).    
 The data are from CMS~\cite{CMS:2025jjx}.}
 \label{fig:ydpt_epps21bcfy01_subtr2_cms}
\end{figure}

\begin{figure}[H]
\centering
\begin{subfigure}{0.49\textwidth}
    \centering
    \includegraphics[width=\textwidth]{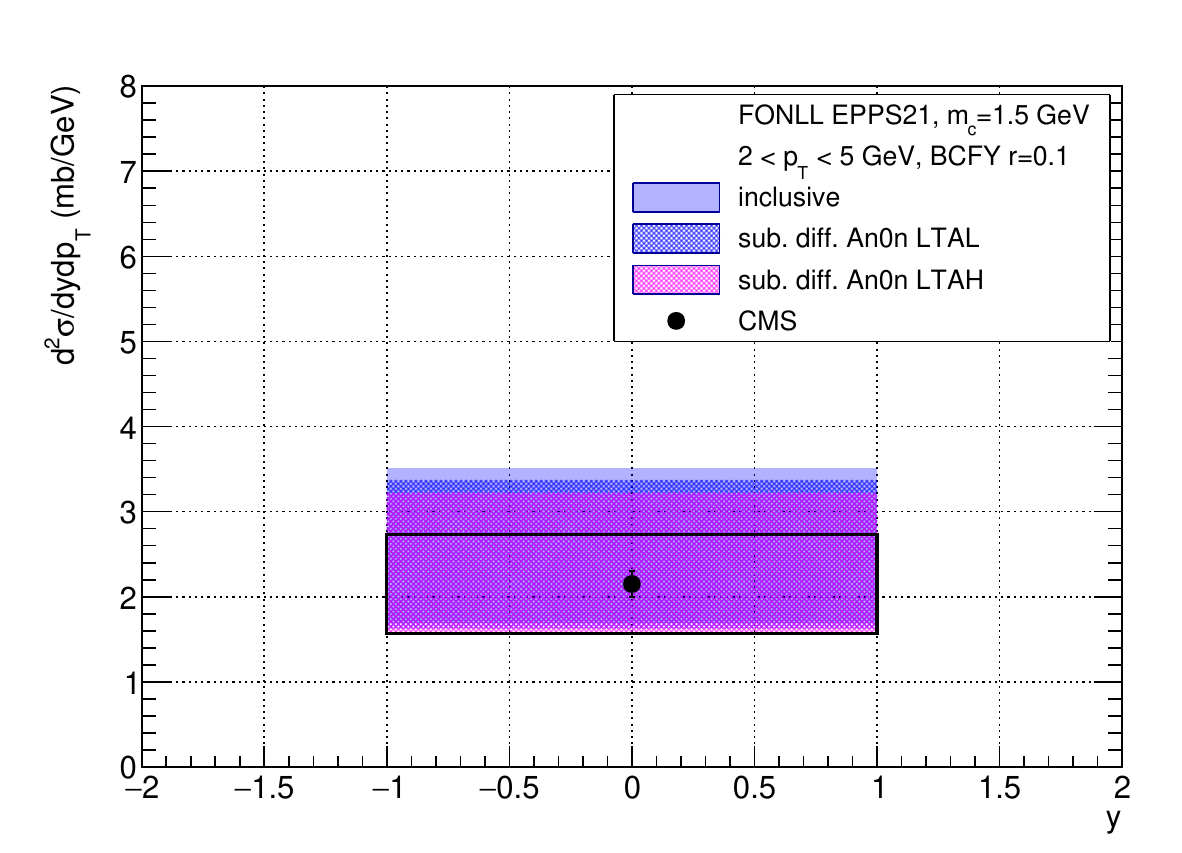}
    \label{fig:ydpt1_1_cms1}
\end{subfigure}
\begin{subfigure}{0.49\textwidth}
    \centering
    \includegraphics[width=\textwidth]{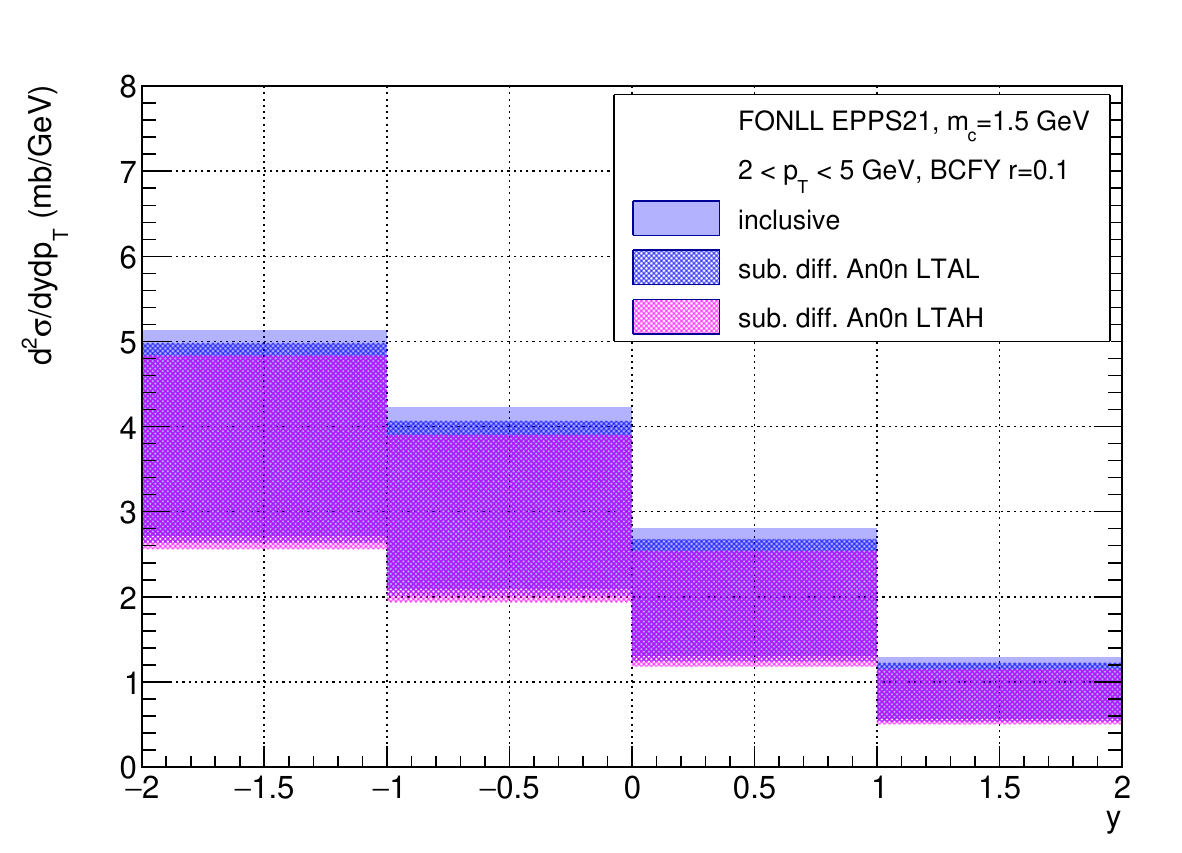}
    \label{fig:ydpt1_cms1}
\end{subfigure}
\hfill
\begin{subfigure}{0.49\textwidth}
    \centering
    \includegraphics[width=\textwidth]{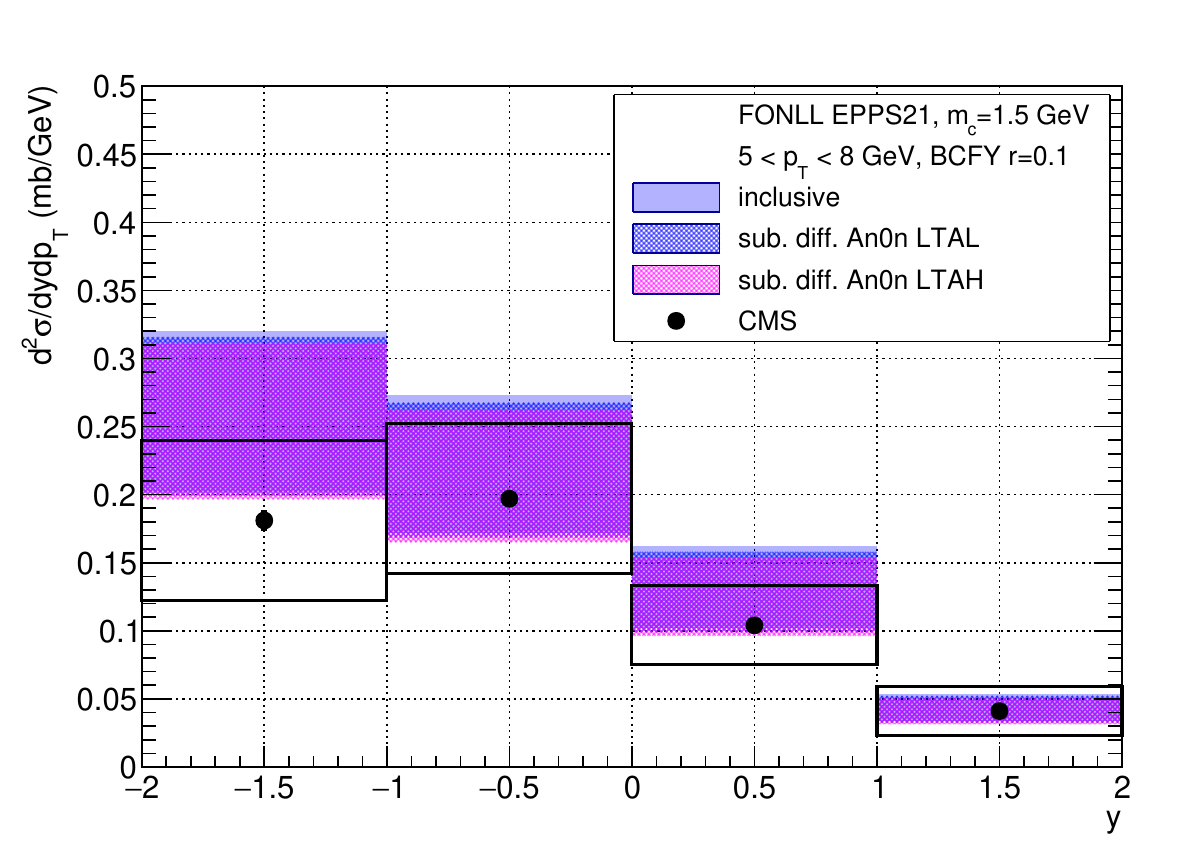}
    \label{fig:ydpt2_cms1}
\end{subfigure}
\begin{subfigure}{0.49\textwidth}
    \centering
    \includegraphics[width=\textwidth]{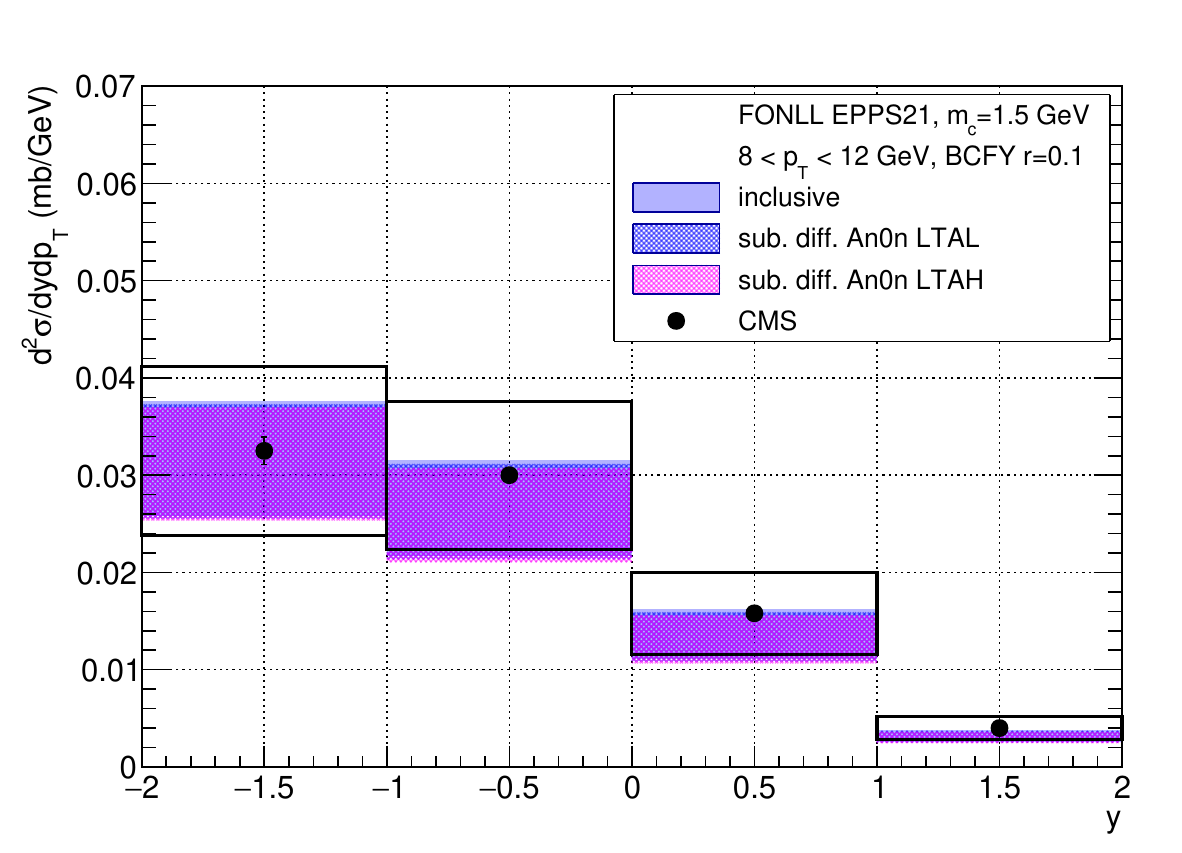}
    \label{fig:ydpt3_cms1}
\end{subfigure}
\caption{Rapidity distribution  for the $D^0$  production in UPC Pb--Pb collisions at ${\rm\sqrt{s_{\scriptscriptstyle NN}} }=5.36$ TeV with the EPPS21 nuclear PDF in  $p_{T}$ bins $(2-5), (5-8), (8-12)$ GeV. Light blue band: FONLL  calculation with factorization and renormalization scale variation for $m_c=1.5 \, \rm GeV$, dark blue and magenta bands: calculations including  subtraction of the diffractive contribution using the LTA ``low'' (LTAL) and ``high'' (LTAH) shadowing 
predictions, respectively, according to formula Eq.~(\ref{eq:crosssection_subtraction}) but without the $Xn0n$ term. The data are from CMS~\cite{CMS:2025jjx}.}
 \label{fig:ydpt_epps21bcfy01_subtr1_cms}
\end{figure}

In this section, we use the coherent diffractive \(D^0\) cross section presented in Sec.~\ref{sec:results} to build a prediction for the double-differential \(D^0\) cross section in Pb--Pb UPCs in the presence of the \(Xn0n\) selection used in the recent CMS measurement~\cite{CMS:2025jjx}. We quantify how this neutron-tagged event selection modifies the inclusive photonuclear prediction for \(D^0\) production. As discussed above, the predictions for inclusive \(D^0\) production in Pb--Pb UPCs reported in Ref.~\cite{Cacciari:2025tgr} explicitly accounted for the \(0n\) condition on the photon-emitting side, which reduces the visible cross section through electromagnetic-dissociation effects. However, no correction was applied for the requirement of neutron emission in the opposite ZDC, associated with nuclear breakup of the photon-receiving target. This requirement can preferentially remove the coherent diffractive component of the inclusive cross section, since in coherent diffraction the target nucleus remains intact at the hard-scattering level. The absence of a dedicated correction was motivated by a qualitative estimate based on the leading twist shadowing model for nuclear diffraction~\cite{Frankfurt:2011cs}, which suggested that the diffractive contribution would be small~\cite{Cacciari:2025tgr}. Using the diffractive G\(\gamma\)A--FONLL predictions obtained in Sec.~\ref{sec:results}, we now provide a quantitative estimate of this correction and construct a prediction that accounts for the \(Xn0n\) selection applied in the measurement. 
\\[4pt]
The corresponding cross section accounting for the Xn0n selection can be written schematically as
\begin{equation}
   d\sigma^{A+A\rightarrow D^0+X}_{Xn0n}=f_{\gamma/A}^{({An0n})}\otimes d\sigma^{(\gamma+A \rightarrow D^0+X)}-f_{\gamma/A}^{({An0n})}\otimes d\sigma_{\rm diff}^{(\gamma+A \rightarrow D^0+A)}+f_{\gamma/A}^{({Xn0n})}\otimes d\sigma_{\rm diff}^{(\gamma+A \rightarrow D^0+A)}.
   \label{eq:crosssection_subtraction}
\end{equation}
The first term,
\(f_{\gamma/A}^{({An0n})}\otimes d\sigma^{(\gamma+A \rightarrow D^0+X)}\),
is the inclusive photonuclear \(D^0\) cross section folded with the effective photon flux for events with zero neutrons on the photon-emitting side and no requirement on the photon-receiving side. This corresponds to the baseline inclusive prediction with the \(0n\) condition applied to the photon-emitting nucleus. The second term,
\(-f_{\gamma/A}^{({An0n})}\otimes d\sigma_{\rm diff}^{(\gamma+A \rightarrow D^0+A)}\),
subtracts the coherent diffractive contribution included in the inclusive cross section, because in this contribution the photon-receiving nucleus remains intact at the hard-scattering level and therefore does not satisfy the \(Xn\) requirement. The third term,
\(f_{\gamma/A}^{({Xn0n})}\otimes d\sigma_{\rm diff}^{(\gamma+A \rightarrow D^0+A)}\),
adds back the subset of coherent diffractive events that pass the \(Xn\) selection due to an additional independent electromagnetic breakup of the diffractively scattered nucleus.
\\[4pt]
Equation~\eqref{eq:crosssection_subtraction} relies on the parametrization of the photon fluxes 
corresponding to the different neutron-emission classes presented in Sec.~\ref{sec:photon_fluxes}.
The inclusive charm-meson photoproduction cross section $d\sigma^{(\gamma+A \rightarrow D^0+X)}$ is calculated using the EPPS21 nuclear PDFs, and for the diffractive charm-meson cross section $d\sigma^{(\gamma+A \rightarrow D^0+A)}$
we used the nuclear diffractive PDFs calculated in the leading twist approximation (LTA) shadowing model~\cite{Frankfurt:2003gx,Frankfurt:2011cs}.
The unsubtracted inclusive calculation is shown in Figs.~\ref{fig:ydpt_epps21bcfy01_subtr2_cms} and~\ref{fig:ydpt_epps21bcfy01_subtr1_cms} by the light-blue bands, which indicate the usual renormalization- and factorization-scale variation. The unsubtracted results correspond to the calculations previously presented in Ref.~\cite{Cacciari:2025tgr}, apart from the treatment of the photon flux. While Ref.~\cite{Cacciari:2025tgr} used the point-like approximation with a sharp impact-parameter cutoff, the present calculation uses the impact-parameter-dependent photon flux of Eq.~\eqref{eq:flux_vidovic}. The dark-blue and magenta bands in Fig.~\ref{fig:ydpt_epps21bcfy01_subtr2_cms} show the inclusive calculation with the diffractive subtraction of Eq.~\eqref{eq:crosssection_subtraction}, using the coherent diffractive component evaluated in the LTA ``low'' and ``high'' shadowing scenarios, respectively.
Overall, the difference between the unsubtracted and difractive-corrected inclusive  calculation is very small.
In addition in Fig.~\ref{fig:ydpt_epps21bcfy01_subtr1_cms}, we show the same diffractive subtraction but without the add-back term proportional to the \(Xn0n\) flux in Eq.~\eqref{eq:crosssection_subtraction}. In other words, the coherent diffractive component is subtracted from the inclusive prediction using the \(An0n\) flux, but no diffractive contribution is restored through the \(Xn0n\) flux. This comparison isolates the numerical impact of the last term in Eq.~\eqref{eq:crosssection_subtraction}. Comparing Figs.~\ref{fig:ydpt_epps21bcfy01_subtr2_cms} and~\ref{fig:ydpt_epps21bcfy01_subtr1_cms}, we see that the impact of the \(Xn0n\) add-back term is very small, as expected from the strong suppression of the \(Xn0n\) flux shown in Fig.~\ref{fig:ratio_An0n_Xn0n}.

\clearpage
\clearpage
\section{Charm production in proton-nucleus UPC with G\texorpdfstring{$\gamma$}{gamma}A-FONLL}
\label{sec:proton_nucleus}

In \(p\)--Pb UPCs, the dominant photonuclear configuration is the one in which the lead ion emits the quasi-real photon and the proton acts as the hadronic target. The reverse configuration, in which the proton emits the photon and the lead ion is the target, is also possible, but it is strongly suppressed by the \(Z^2\) enhancement of the photon flux from the lead ion. We therefore focus on the photon--proton contribution. Future measurements of charm production in \(p\)--Pb UPCs would provide sensitivity to the gluon density in the proton, complementing and extending the constraints obtained from open-charm photoproduction measurements in electron--proton collisions at HERA~\cite{H1:1998csb,ZEUS:1998wxs,H1:2011myz}.
\\[4pt]
We now extend the G$\gamma$A--FONLL calculation of inclusive and diffractive open-charm photoproduction to ultraperipheral proton--lead collisions. For these results, we consider \(p\)--Pb collisions at \(\sqrt{s_{\scriptscriptstyle NN}}=8.16\,\mathrm{TeV}\). This energy corresponds to the highest-energy \(p\)--Pb data set collected at the LHC and has substantially larger luminosity than the earlier \(p\)--Pb run at \(\sqrt{s_{\scriptscriptstyle NN}}=5.02\,\mathrm{TeV}\). We use the same ingredients described in Sec.~\ref{sec:theoryGammaFONLL}: the heavy-quark production cross section is evaluated in FONLL, the produced charm quark fragments into a \(D^0\) meson, and both direct and resolved photon contributions are included. As in the \(AA\) case, the photon PDF is taken from the GRV parametrization, the fragmentation function is BCFY with parameter \(r=0.1\), and the charm-quark mass is set to \(m_c=1.5\,\mathrm{GeV}\). The main differences with respect to the Pb--Pb calculation are the photon flux appropriate to the \(pA\) collision geometry and the replacement of nuclear target PDFs with proton PDFs. For the inclusive calculation, we use the CT18ANLO proton PDF set~\cite{Hou:2019efy}. For calculation of the diffractive cross section we used ZEUS SJ \cite{ZEUS:2009uxs} parametrization of the diffractive PDFs. A detailed description of the effective photon flux parametrization is provided in the following subsection.

\subsection{Photon flux for \texorpdfstring{$pA$}{pA} UPCs}
\begin{figure}[h]
\centering
\includegraphics[width=0.65\linewidth]{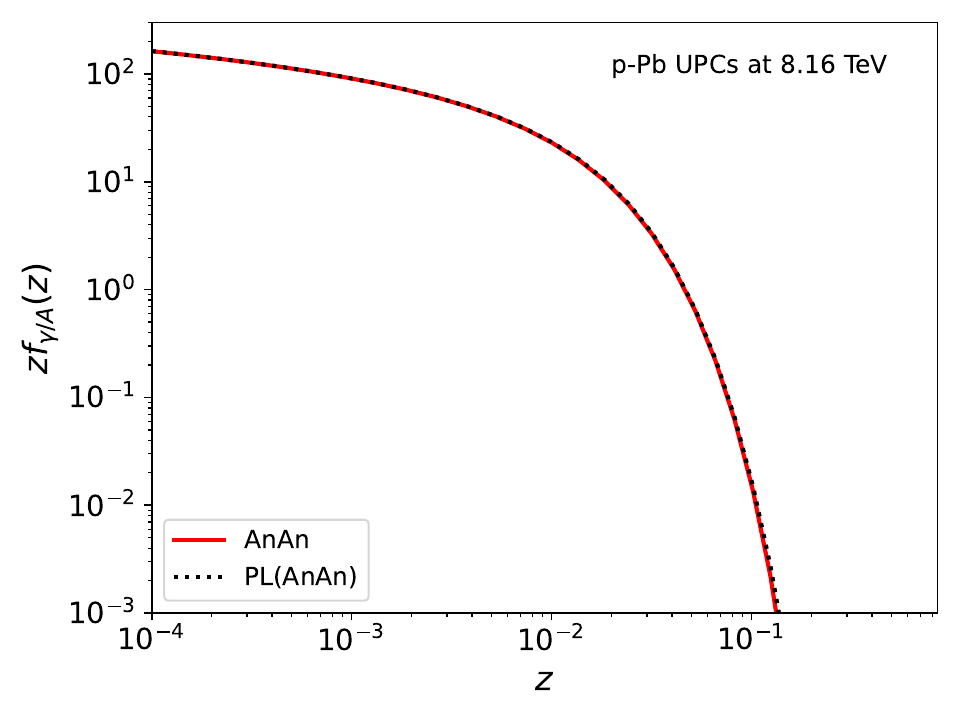}
\caption{Photon flux from the lead ion in \(pA\) UPCs at \(\sqrt{s_{\scriptscriptstyle NN}}=8.16\,\mathrm{TeV}\), shown as a function of the photon energy fraction \(z\). The solid curve shows the default calculation using the Woods--Saxon charge distribution and the \(pA\) hadronic survival factor \(\Gamma_{pA}\), while the dashed curve shows the point-like approximation with \(b_{\rm min}=1.1R_A\).}
    \label{fig:flux_pA}
\end{figure}

In the case of \(pA\) UPCs, the photon flux emitted by an ultrarelativistic nucleus is given by~\cite{Guzey:2013taa}
\begin{equation}
f_{\gamma/A}^{({\rm AnAn})}(z)
=
\int d^2 \mathbf{b} \,
f_{\gamma/A}(z,\mathbf{b}) \,
\Gamma_{pA}(\mathbf{b}) \, ,
\label{eq:flux_pA}
\end{equation}
where \(\Gamma_{pA}\) is the factor suppressing hadronic interactions at small impact parameters. It can be calculated using the optical limit of the Glauber model for high-energy \(pA\) scattering,
\begin{equation}
\Gamma_{pA}(\mathbf{b})
=
e^{-\sigma_{\scriptscriptstyle NN}T_A(\mathbf{b})} \, ,
\label{eq:Gamma_pA}
\end{equation}
where \(\sigma_{\scriptscriptstyle NN}\) is the total nucleon--nucleon cross section. For \(\sqrt{s_{\scriptscriptstyle NN}}=8.16\,\mathrm{TeV}\), we take \(\sigma_{\scriptscriptstyle NN}=99\,\mathrm{mb}\)~\cite{ParticleDataGroup:2020ssz}. Our default calculation uses the photon flux of Eq.~\eqref{eq:flux_pA}, evaluated with the Woods--Saxon charge distribution. This provides the realistic flux used in the numerical predictions below. For reference, we also show the analytic point-like approximation of Eq.~\eqref{eq:flux_PL_2}, in which the suppression of hadronic \(pA\) interactions at small impact parameters is implemented through a sharp cutoff \(b_{\rm min}\). We use \(b_{\rm min}=1.1R_A\) for this comparison. 
\\[4pt]
Figure~\ref{fig:flux_pA} shows the photon flux emitted by the lead ion in \(pA\) UPCs. The solid curve is the default Woods--Saxon result from Eq.~\eqref{eq:flux_pA}, while the dashed black curve shows the point-like sharp-cutoff approximation with \(b_{\rm min}=1.1R_A\).
The two flux calculations are numerically very close.  
Note that no additional electromagnetic-dissociation factor is applied for \(pA\) UPCs, since the probability of independent electromagnetic breakup accompanying the hard photoproduction process is expected to be much smaller than in \(AA\) collisions.

\subsection{Predictions for inclusive \texorpdfstring{$D^0$}{D0} photoproduction in \texorpdfstring{$pA$}{pA} UPCs}

\begin{figure}
    \centering
    \includegraphics[width=0.49\linewidth]{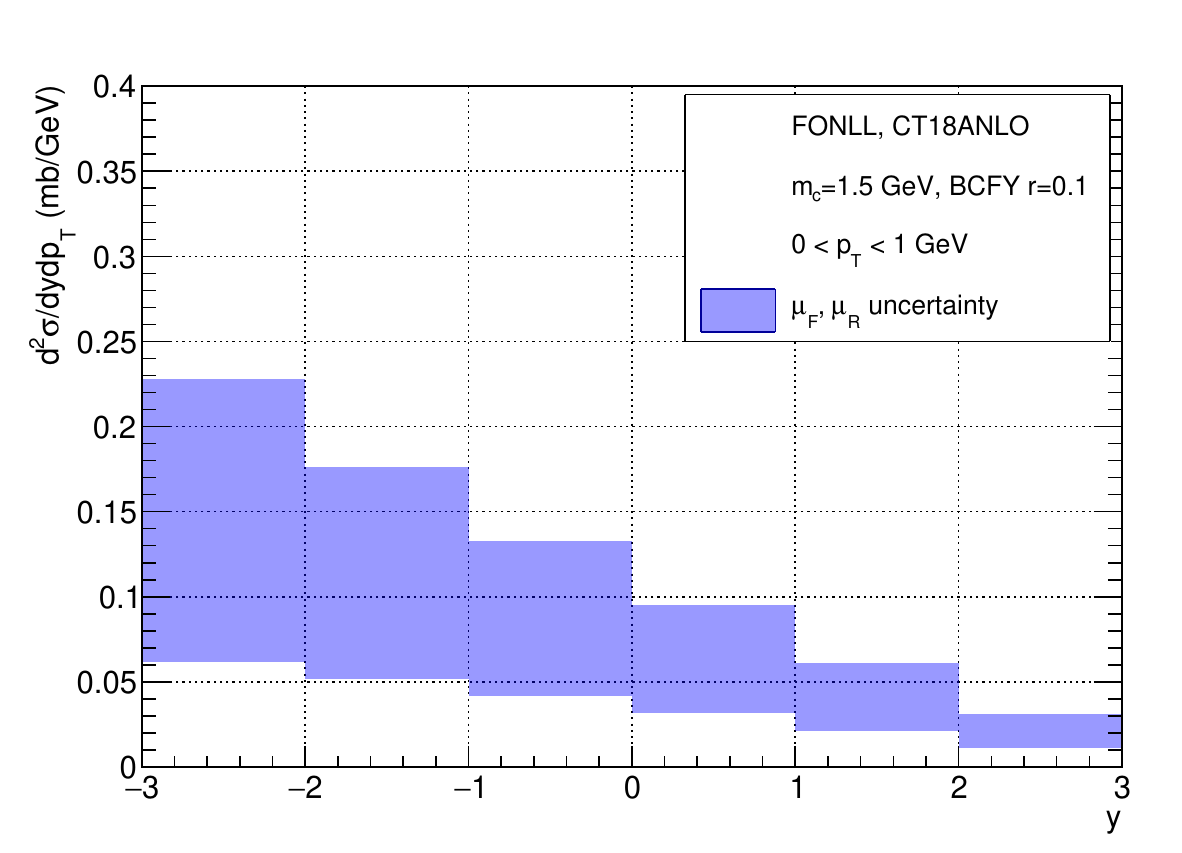}
     \includegraphics[width=0.49\linewidth]{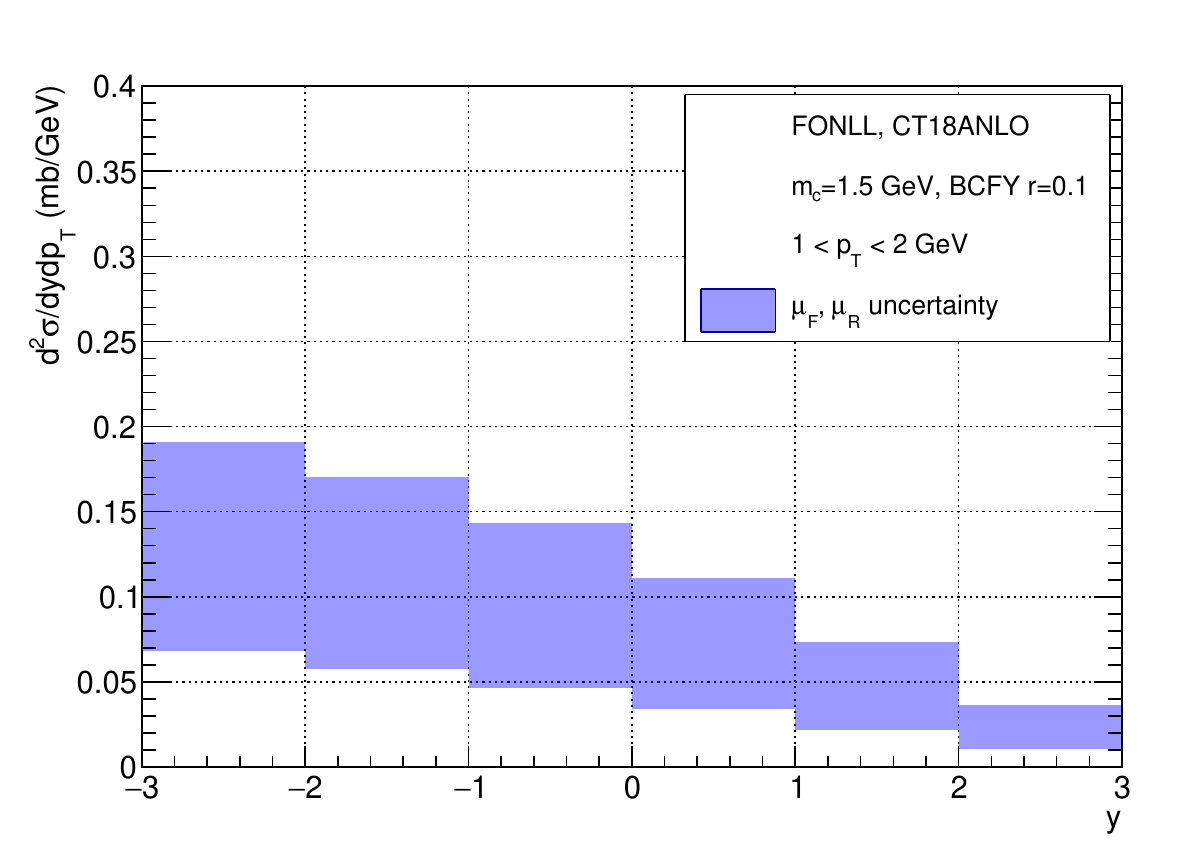}
    \includegraphics[width=0.49\linewidth]{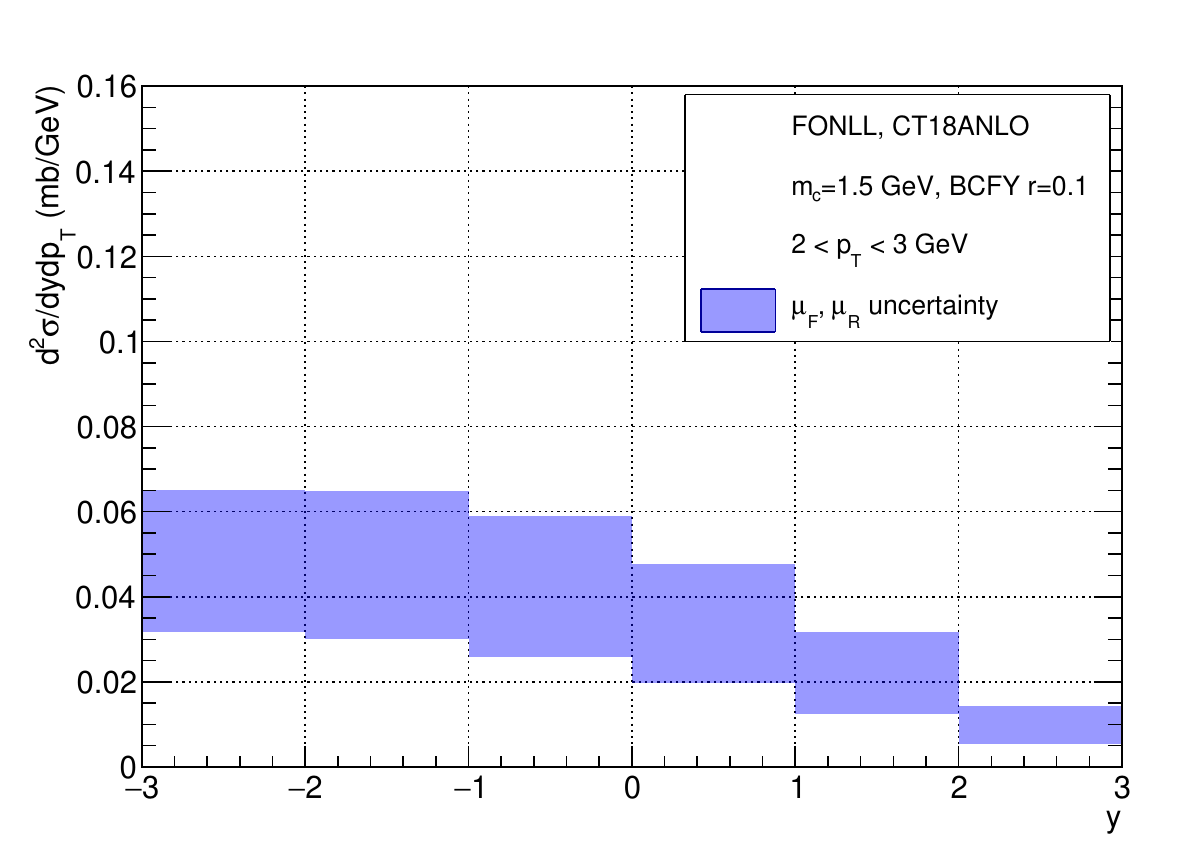}
    \includegraphics[width=0.49\linewidth]{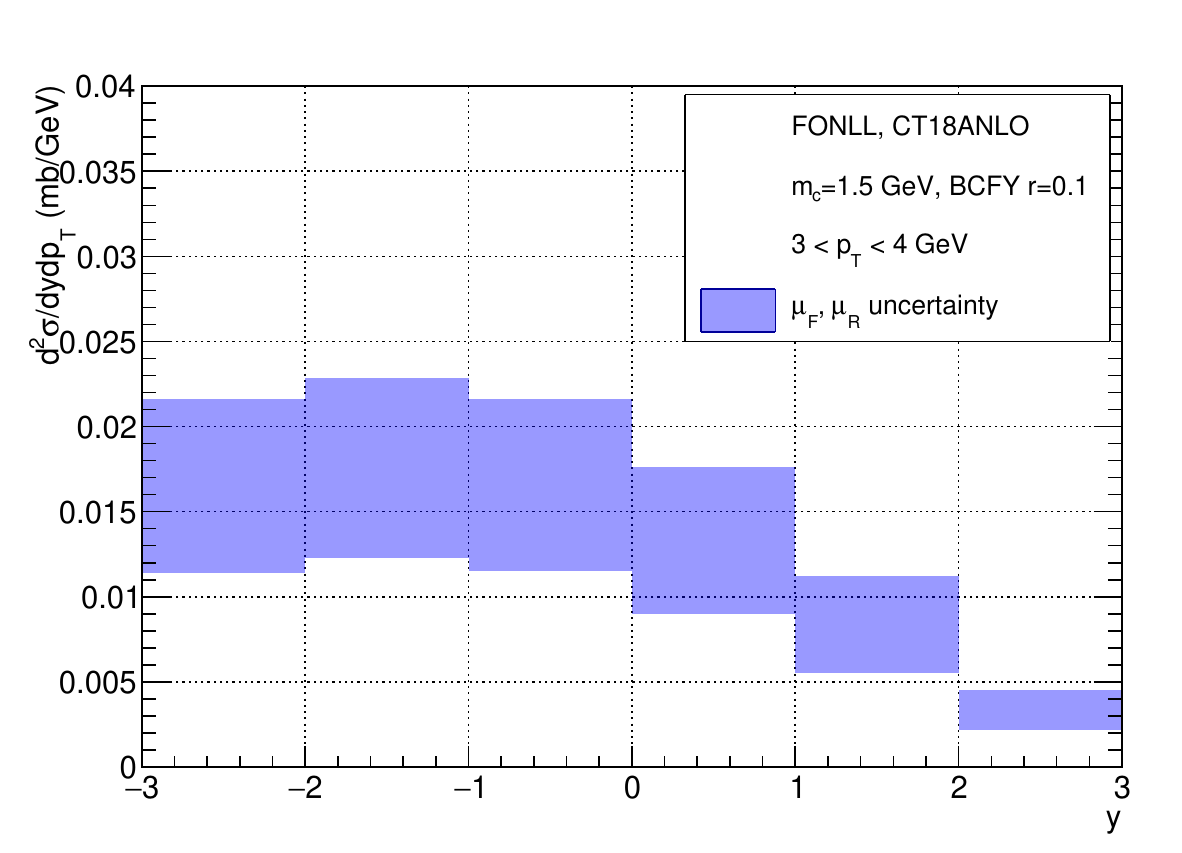}
    \includegraphics[width=0.49\linewidth]{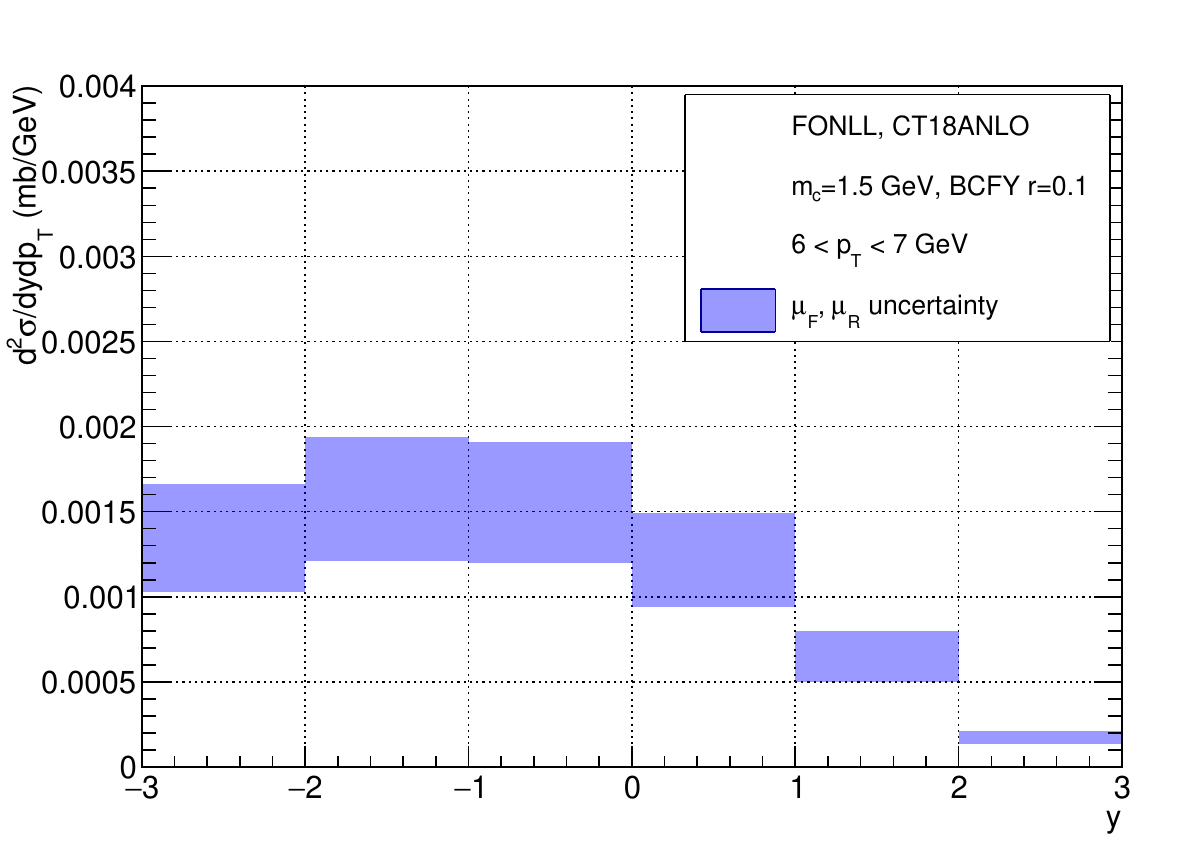}
    \includegraphics[width=0.49\linewidth]{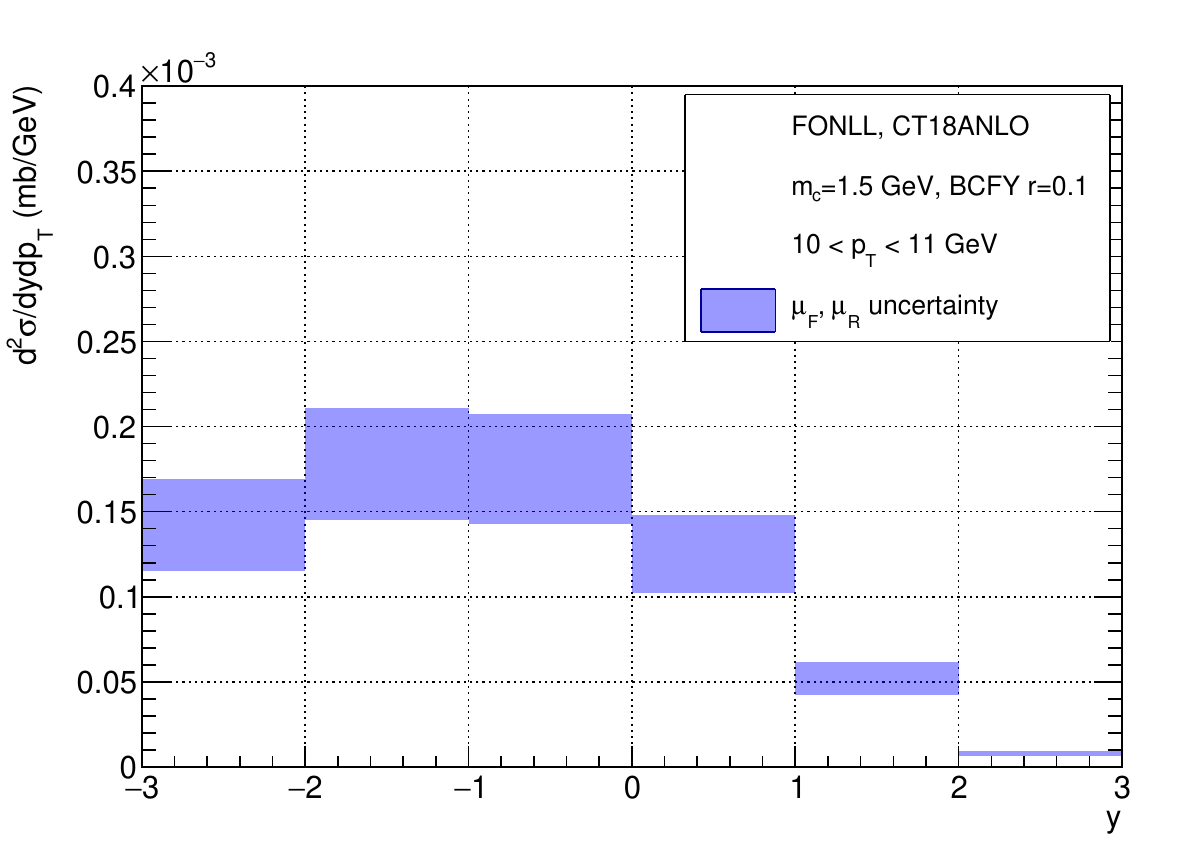}
    \caption{Double differential cross section for the $D^0$ production in $p$Pb UPCs at $\sqrt{\rm s_{\scriptscriptstyle{NN}}}=8.16$ TeV, plotted as a function of rapidity.   Six panels correspond to following $p_{T}$ bins: $(0-1), (1,2), (2,3), (3-4), (6-7), (10-11)$ GeV. Light blue band: FONLL  calculation with factorization and renormalization scale variation. Charm mass in FONLL  is set to  $m_c=1.5 \, \rm GeV$. Proton PDF is CT18ANLO,  fragmentation function is BCFY \cite{Braaten:1994bz,Cacciari:2003zu} with parameter $r=0.1$. Photon-emitting nucleus is moving in the positive rapidity direction, the proton is moving in the negative rapidity direction.}
    \label{fig:ydistr1}
\end{figure}

Figure~\ref{fig:ydistr1} shows the double-differential cross section for $D^0$ photoproduction in 
$pA$ UPCs 
as a function of rapidity in the interval \(-2<y<2\), for six representative \(p_T\) bins of the 
$D^0$ meson: \((0,1)\), \((1,2)\), \((2,3)\), \((3,4)\), \((6,7)\), and \((10,11)\,\mathrm{GeV}\). The bands indicate the renormalization- and factorization-scale dependence, with the constraint \(1/2\le \mu_R/\mu_F \le 2\). The convention is such that the nucleus moves in the positive-rapidity direction and the proton in the negative-rapidity direction. Thus, the smallest values of \(x\) in the proton are probed at the largest rapidities. The features of the double-differential cross section in \(pA\) are very similar to those in the \(AA\) case, with a decreasing cross section as a function of rapidity due to the steeply falling photon flux, as well as a strong \(p_T\) dependence. The \(pA\) cross section, when rescaled by the mass number \(A\), is comparable to the \(AA\) cross section, although some differences remain due to the larger photon flux in \(pA\) than in \(AA\), since smaller impact parameters are involved, the slightly higher collision energy, and the lack of nuclear shadowing.

\subsection{Diffractive \texorpdfstring{$D^0$}{D0} photoproduction and diffractive fractions in \texorpdfstring{$pA$}{pA} UPCs}
\label{sec:diffraction_pA_results}
\begin{figure}[h]
    \centering
    \includegraphics[width=0.49\linewidth]{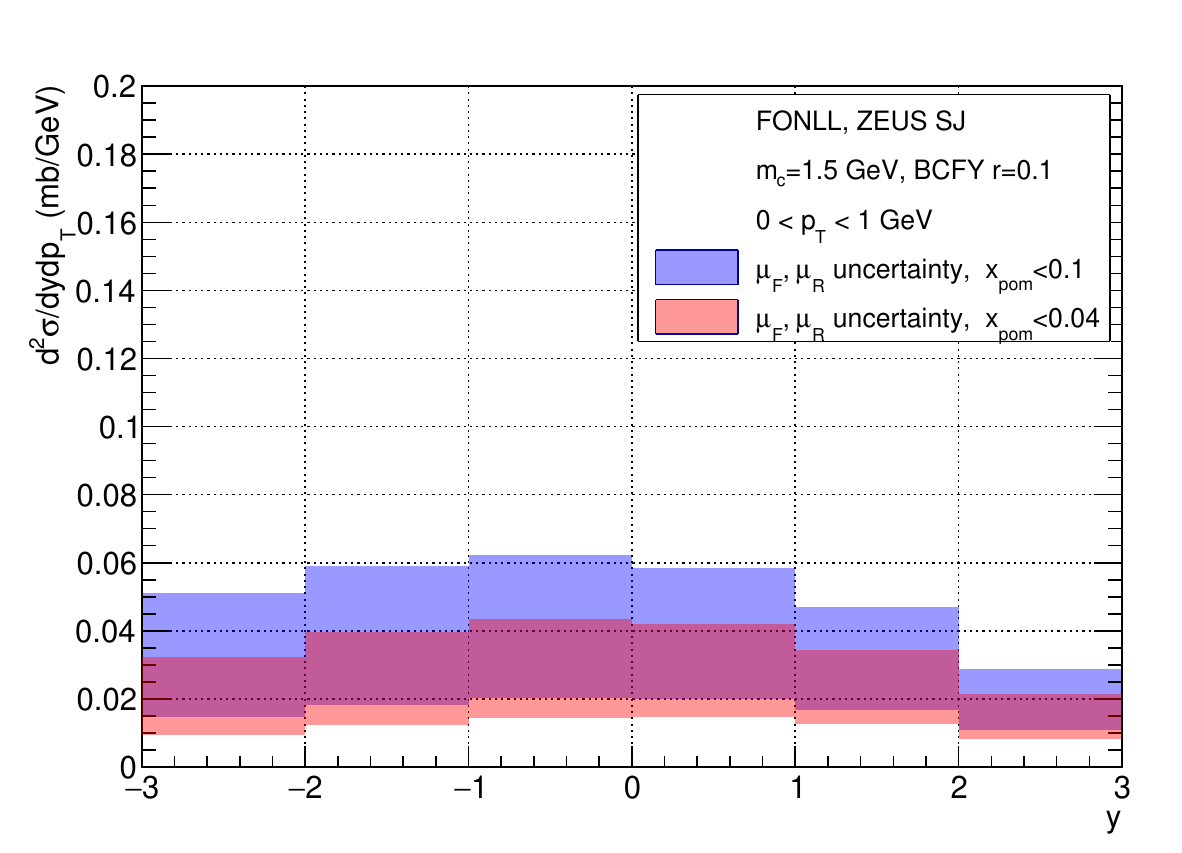}
     \includegraphics[width=0.49\linewidth]{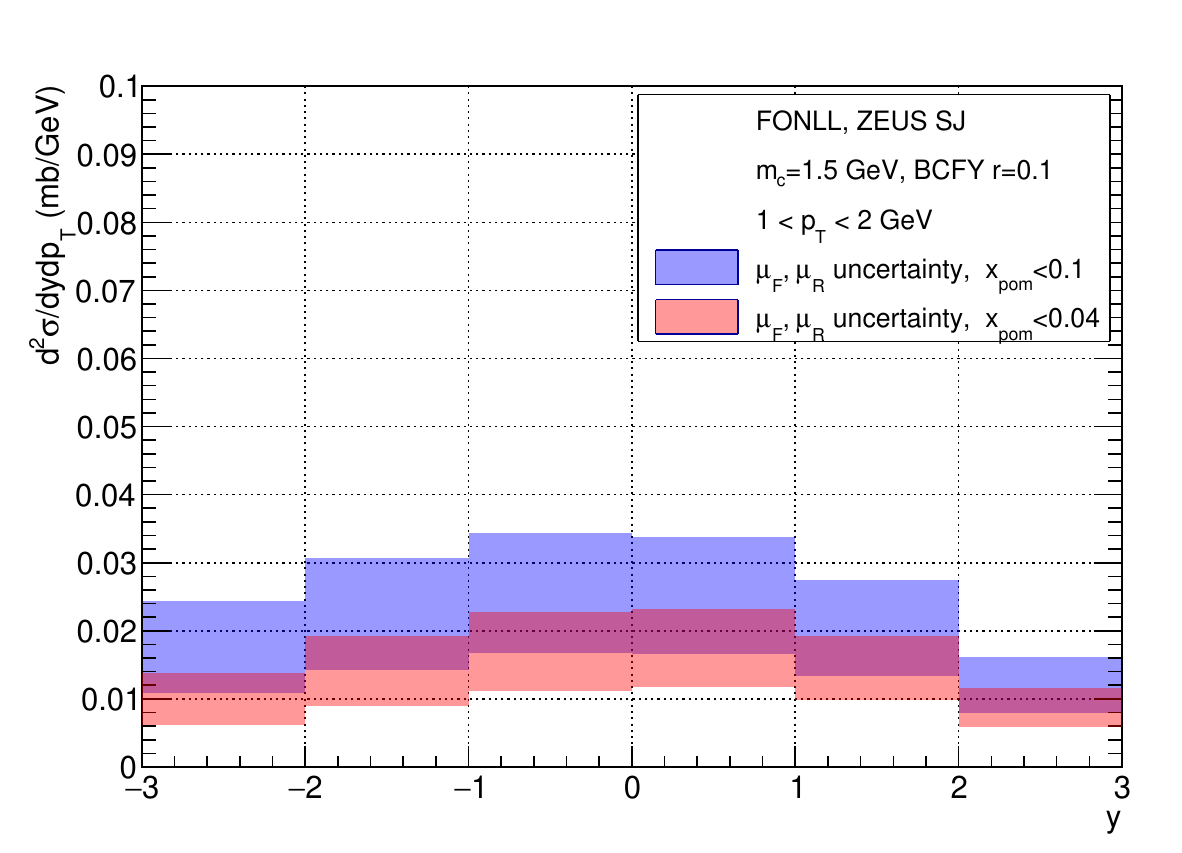}
    \includegraphics[width=0.49\linewidth]{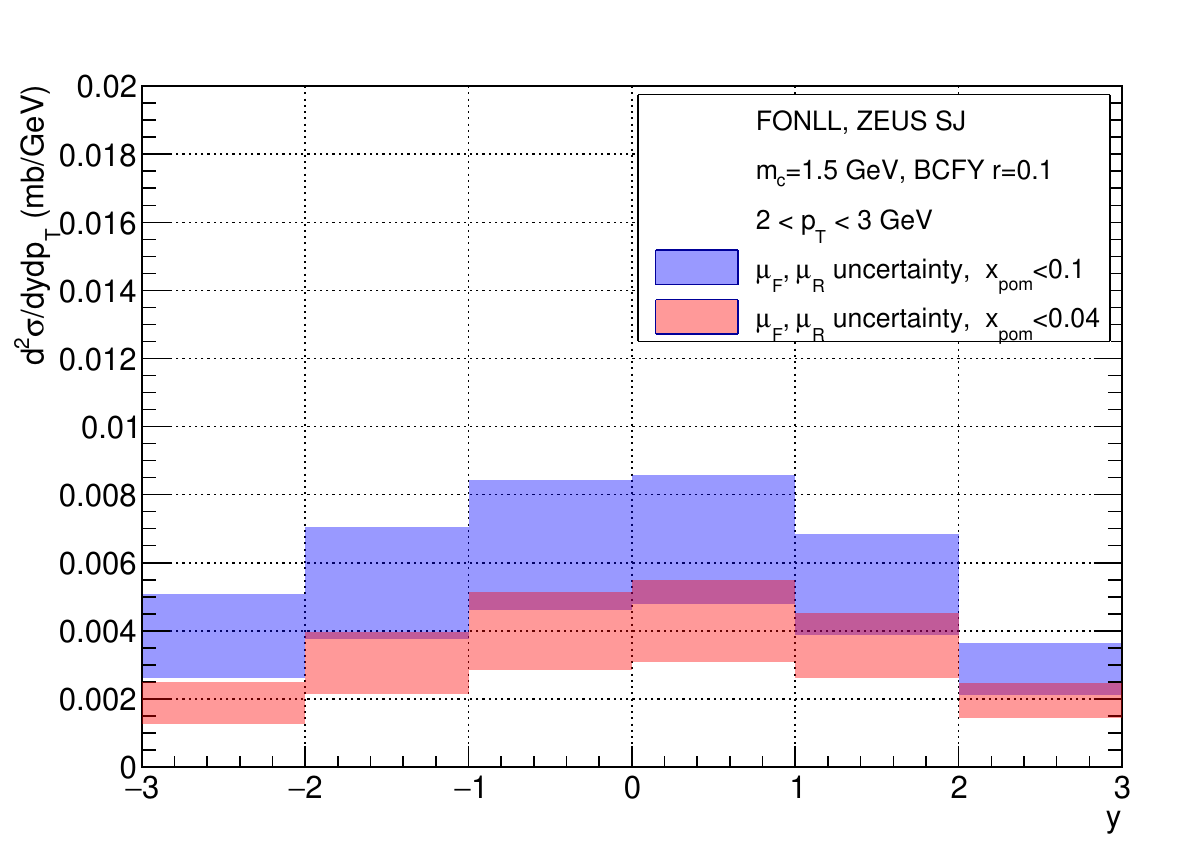}
    \includegraphics[width=0.49\linewidth]{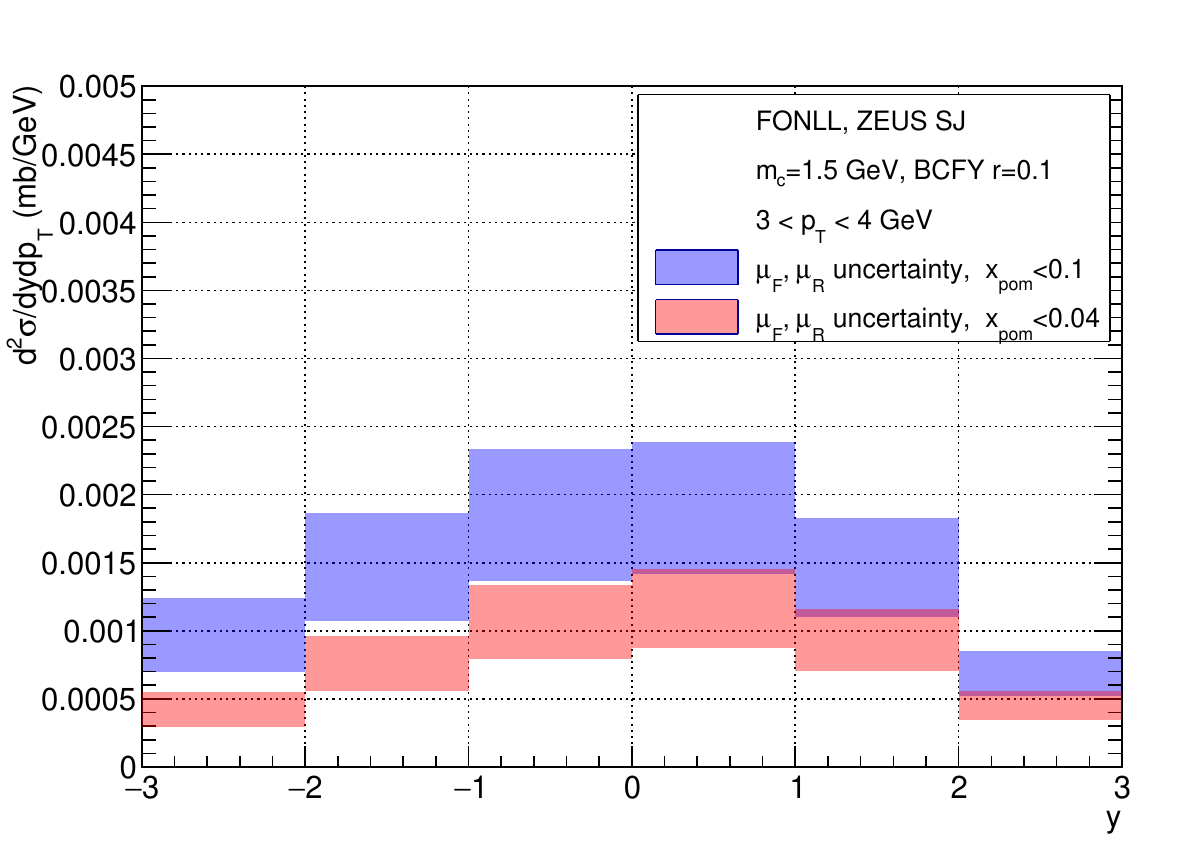}
    \includegraphics[width=0.49\linewidth]{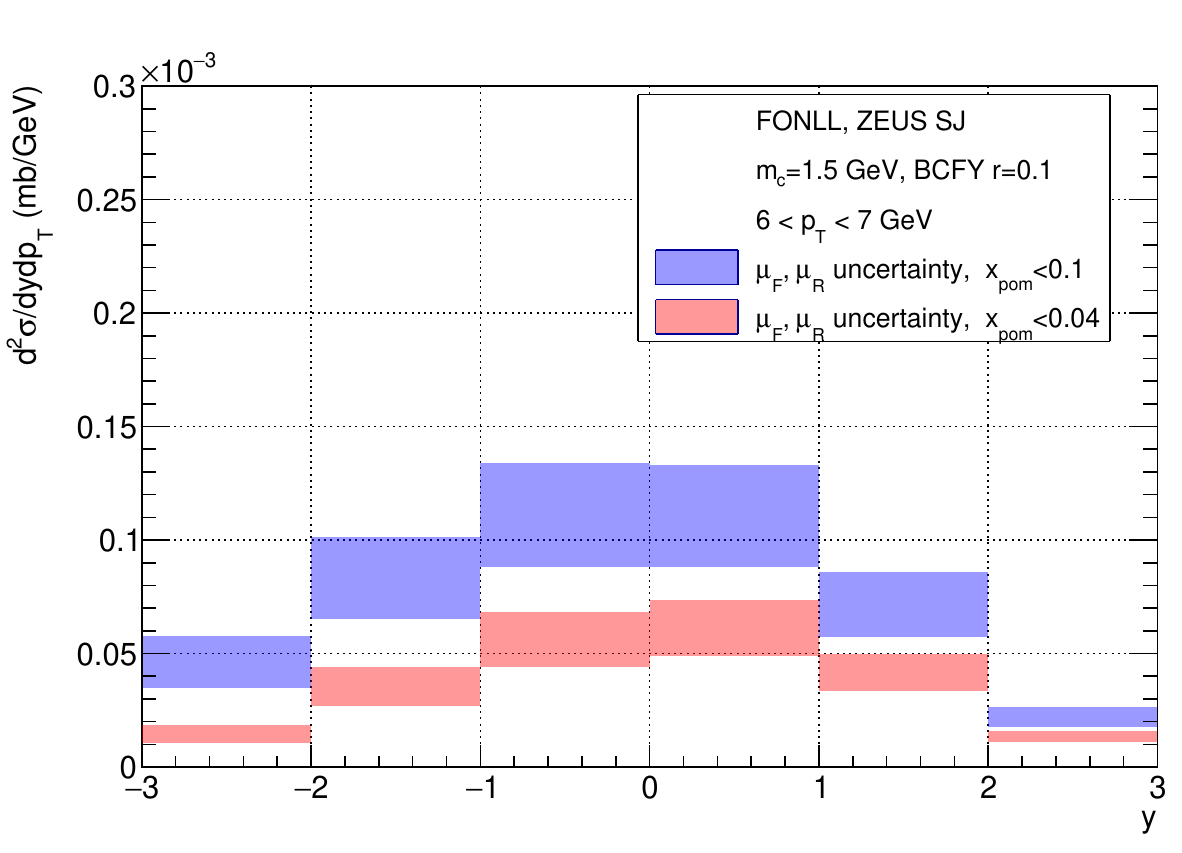}
    \includegraphics[width=0.49\linewidth]{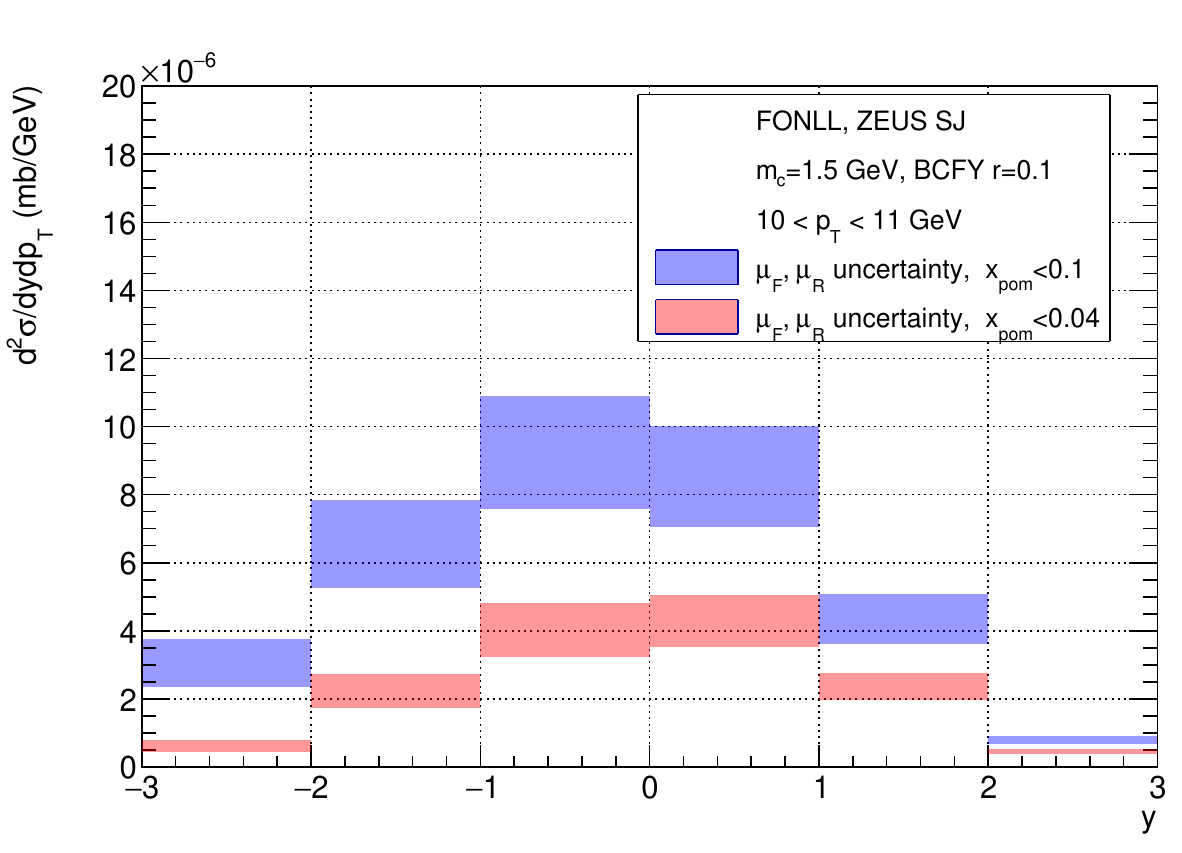}
    \caption{Double-differential cross section for diffractive \(D^0\) production in \(p\)Pb UPCs at \(\sqrt{s_{\scriptscriptstyle NN}}=8.16~\mathrm{TeV}\), shown as a function of rapidity. The six panels correspond to the following \(p_T\) bins: \((0,1)\), \((1,2)\), \((2,3)\), \((3,4)\), \((6,7)\), and \((10,11)~\mathrm{GeV}\). The blue bands correspond to \(x_{\Pomeron}<0.1\), and the red bands to \(x_{\Pomeron}<0.04\). The bands show the FONLL factorization- and renormalization-scale variation. The charm mass is set to \(m_c=1.5~\mathrm{GeV}\). The proton PDF is CT18ANLO, and the fragmentation function is BCFY~\cite{Braaten:1994bz,Cacciari:2003zu} with \(r=0.1\). The photon-emitting nucleus moves in the positive-rapidity direction, and the proton moves in the negative-rapidity direction.}
    \label{fig:ydistr1diff}
\end{figure}

\begin{figure}[h]
    \centering
    \begin{subfigure}{0.49\textwidth}
    \centering
    \includegraphics[width=\textwidth]{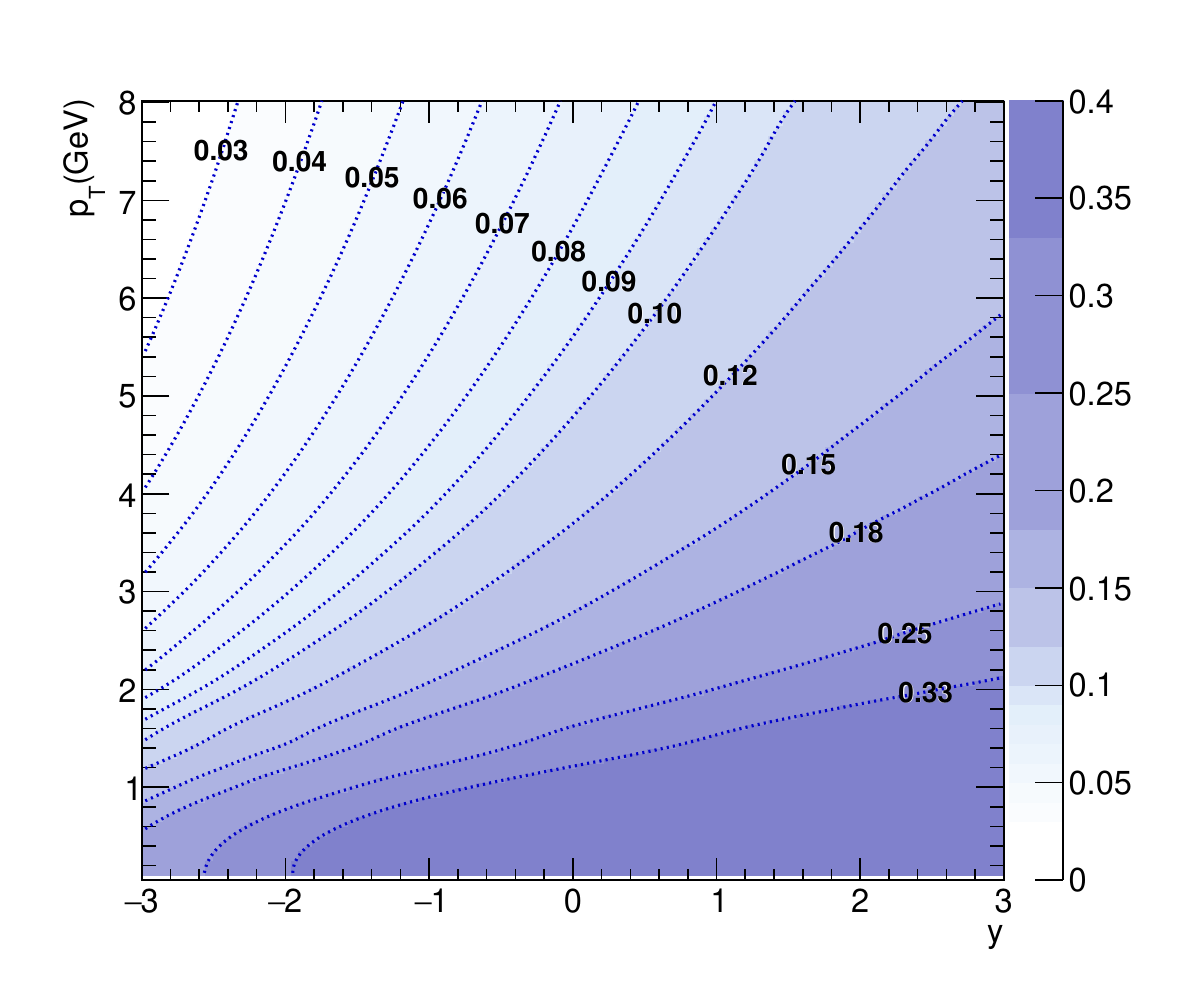}
    \end{subfigure}
   \begin{subfigure}{0.49\textwidth}
    \centering
    \includegraphics[width=\textwidth]{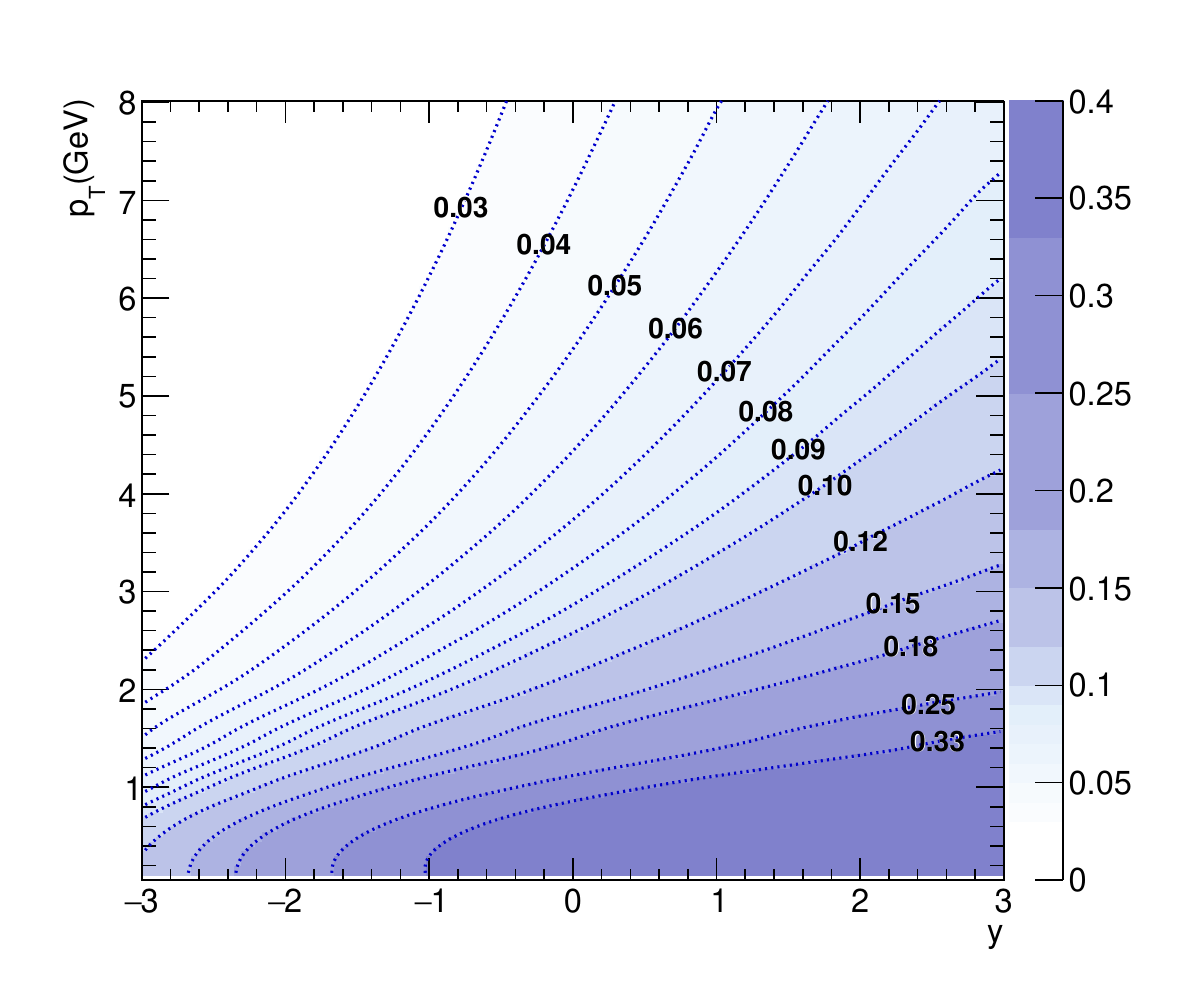}
    \end{subfigure}
    \caption{Ratio of the diffractive to inclusive \(D^0\) photoproduction cross sections in \(p\)Pb UPCs
    in the \((y,p_T)\) plane. The diffractive component is integrated over \(x_{\Pomeron}\) up to \(0.1\) in the left panel and up to \(0.04\) in the right panel.}
    \label{fig:diff_to_inclusive_inpA}
\end{figure}

Next, we calculate the diffractive component of heavy charm-meson photoproduction in proton--nucleus UPCs. For this calculation, we use the ZEUS SJ diffractive proton PDFs~\cite{ZEUS:2009uxs}. This set was extracted from fits to diffractive structure-function data, with diffractive dijet data also included to better constrain the gluon density. The diffractive component is integrated over the longitudinal momentum fraction \(x_{\Pomeron}\) carried by the diffractive exchange. For this calculation, we present results for two benchmark choices, \(x_{\Pomeron}\le 0.1\) and \(x_{\Pomeron}\le 0.04\), motivated by the cuts used in HERA measurements of diffractive charm photoproduction; see, for example, Ref.~\cite{H1:2006zxb}. In an LHC UPC measurement, the effective \(x_{\Pomeron}\) range would depend on the experimental conditions. Without proton tagging, it would be determined indirectly by the rapidity-gap requirement and detector acceptance. With proton tagging, it would instead be set by the kinematic acceptance of the forward proton spectrometer. 
\\[4pt]
Figure~\ref{fig:ydistr1diff} shows the predicted double-differential diffractive cross section for \(D^0\) production as a function of rapidity, in the same six \(p_T\) bins used in Fig.~\ref{fig:ydistr1}. The bands show the renormalization- and factorization-scale variation, with the constraint \(1/2\le \mu_R/\mu_F \le 2\). The blue bands correspond to \(x_{\Pomeron}\le 0.1\), while the red bands correspond to \(x_{\Pomeron}\le 0.04\). The prediction shows a sizable dependence on the upper limit of the \(x_{\Pomeron}\) integration, with the larger \(x_{\Pomeron}\) range giving the larger cross section. The cross section also shows a significant reduction in the most negative rapidity bin, where the available phase space for diffractive production is limited.
\\[4pt]
 To quantify the size of the diffractive contribution, Fig.~\ref{fig:diff_to_inclusive_inpA} shows the ratio of the diffractive to inclusive cross section in the \((y,p_T)\) plane. The left panel corresponds to \(x_{\Pomeron}\le 0.1\), while the right panel corresponds to \(x_{\Pomeron}\le 0.04\). The diffractive fraction is largest at low \(p_T\) and forward rapidity, and decreases rapidly with increasing \(p_T\). For \(x_{\Pomeron}\le 0.1\), it reaches about \(20\%\) at the largest rapidities and \(p_T\simeq 2~\mathrm{GeV}\). At \(p_T\simeq 4~\mathrm{GeV}\), the fraction is about \(10\%\) at \(y\simeq 2\). 

\clearpage
\section{Conclusions}
\label{sec:conclu}

In this work we have presented calculations for the 
diffractive photoproduction of 
charm \(D^0\) mesons in AA and pA UPCs at the LHC. 
For the UPC AA case, we have extended the 
 G$\gamma$A--FONLL
 framework set up in the previous work~\cite{Cacciari:2025tgr} to the case of diffraction. The diffractive cross section was computed in bins of the $D^0$ rapidity $y$ and the
 transverse momentum $p_T$ using the nuclear diffractive parton distributions obtained within the leading twist shadowing model. We found that the LTA model leads to the modest diffractive fraction in AA UPCs of up to about 7-15\% for 
 $p_T \sim 2$ GeV
 and forward rapidities $y \sim 1-2$,
 corresponding to the small values of \(x\) in the Pomeron-emitting target nucleus. The diffractive fraction decreases for negative rapidities, as well as for increasing values of the transverse momenta. 
 \\[4pt]
 Next, we have used the calculation of diffraction within the LTA model to correct the non-diffractive, inclusive calculation to reflect the $Xn0n$ selection used in the experiment. To this aim, we have subtracted the diffractive component convoluted  with the $An0n$ photon flux, and added back the diffractive part convoluted   with the  $Xn0n$ photon flux. We find that the modification of the inclusive calculation due to this subtraction is very small, much smaller as compared to uncertainties due to the factorization and renormalization scale variations. We also used an improved photon flux, as opposed to the pointlike flux used in the previous calculations, and found that the differences and impact on the calculation is also very small. 
\\[4pt]
In the second part of this work, we have performed calculations and made predictions for the inclusive and diffractive 
$D^0$ photoproduction
in the case of the proton-nucleus UPC at \(\sqrt{s_{\scriptscriptstyle{NN}}}=8.16 \rm \; TeV\), which is one possible future running scenario at the LHC. We assumed that the dominant source of the photons is the nucleus in $pA$ collisions, and thus such process effectively becomes a photoproduction of inclusive charm on the proton. We computed distributions in the $D^0$ transverse momentum and rapidity,
evaluating the diffractive contribution 
using the ZEUS SJ parametrization of the proton diffractive PDFs.


\section*{Acknowledgments}

GMI is supported by the U.S. Department of Energy, Office of Science, under Grant No. DOE-SC0011088, and by the DOE Early Career Research Program, under Grant No. 036923-00001. AMS is  supported by the U.S. Department of Energy grant No. DE-SC-0002145 and within the framework of the of the Saturated Glue (SURGE) Topical Theory Collaboration. The research of VG was funded by the Center of Excellence in Quark Matter of Research Council of Finland (projects 346325 and 346326).

\appendix
\section*{Appendix A: Typical $x$  in gluon density }

In this appendix we show the distributions in longitudinal momentum fraction \(x\) of the gluon density that are in different bins of \(p_T\) and rapidity \(y\) of the meson for the UPC AA.
That is we show the quantity
\begin{equation}
    \frac{d^3 \sigma}{dy dp_T dx} \Delta x \, ,
\end{equation}
as histograms, which when summed over the bins  give cross sections shown in Fig.~\ref{fig:ydpt_epps21bcfy01_subtr2_cms}.  We observe that the lowest values of $x$ are of course probed for highest values of rapidities and lowest values of transverse momenta. We observe however, that the distributions are quite wide. For example, for the \(2<p_T<5\) GeV and \(1<y<2\) the typical values of \(x\sim 10^{-4}-10^{3}\), about 30\% of the cross section in this bin comes from larger values of \(x>0.01\).

\begin{figure}[htbp]
    \centering
    \begin{subfigure}[b]{0.3\textwidth}
        \centering
        \includegraphics[width=\textwidth]{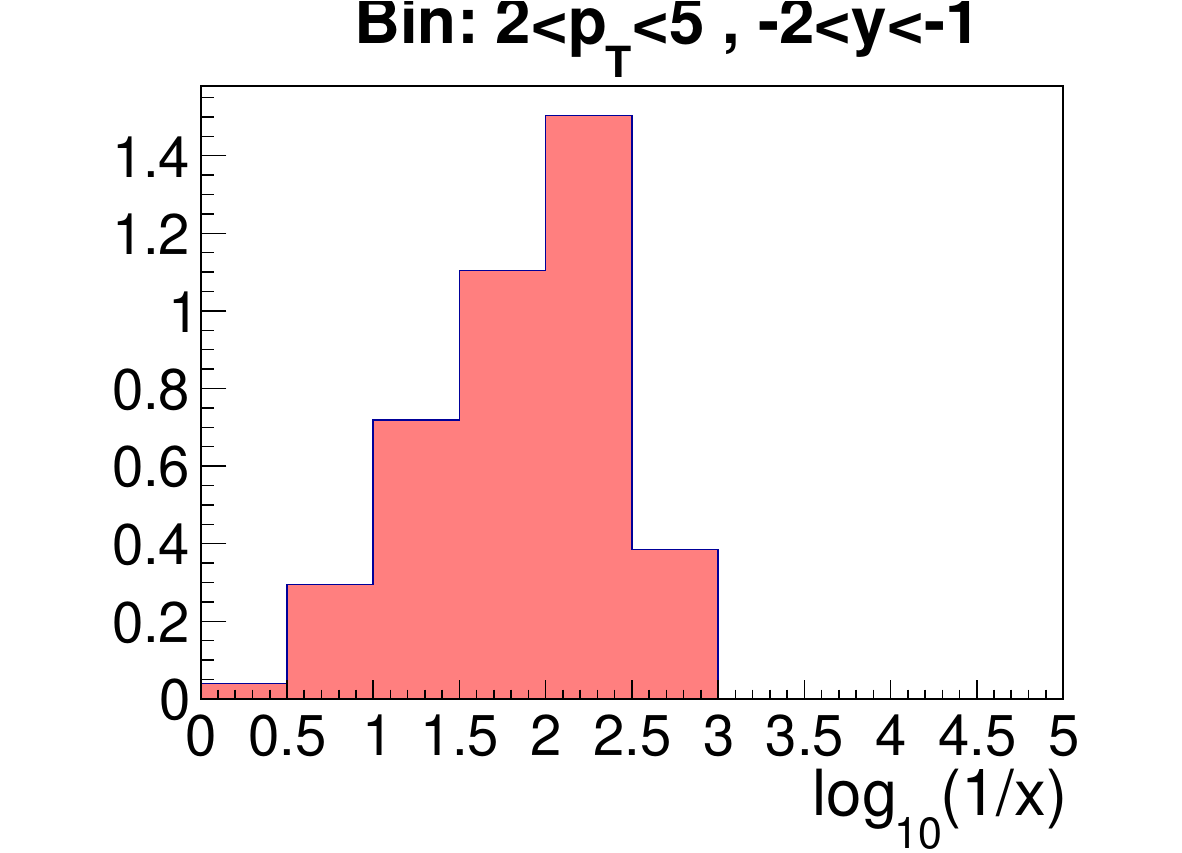}
    \end{subfigure}
    \hfill
    \begin{subfigure}[b]{0.3\textwidth}
        \centering
        \includegraphics[width=\textwidth]{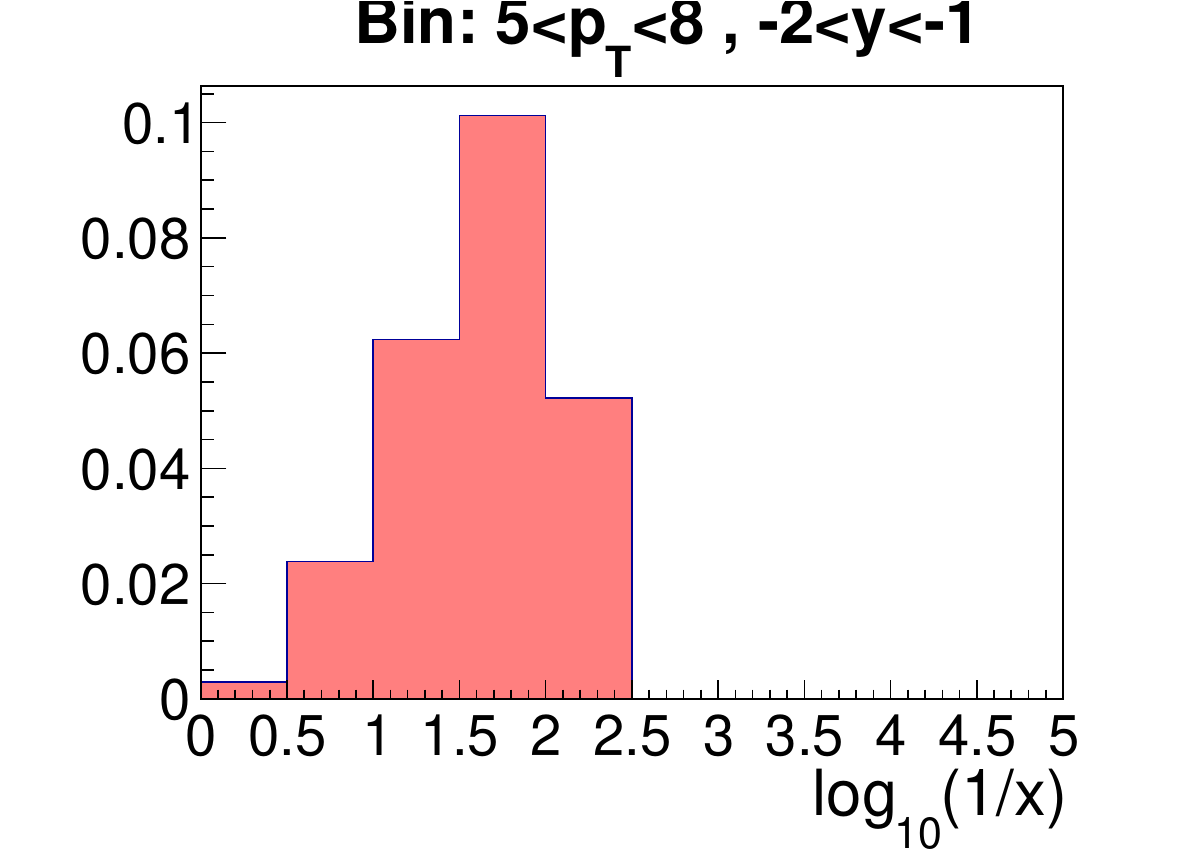}
    \end{subfigure}
    \hfill
    \begin{subfigure}[b]{0.3\textwidth}
        \centering
        \includegraphics[width=\textwidth]{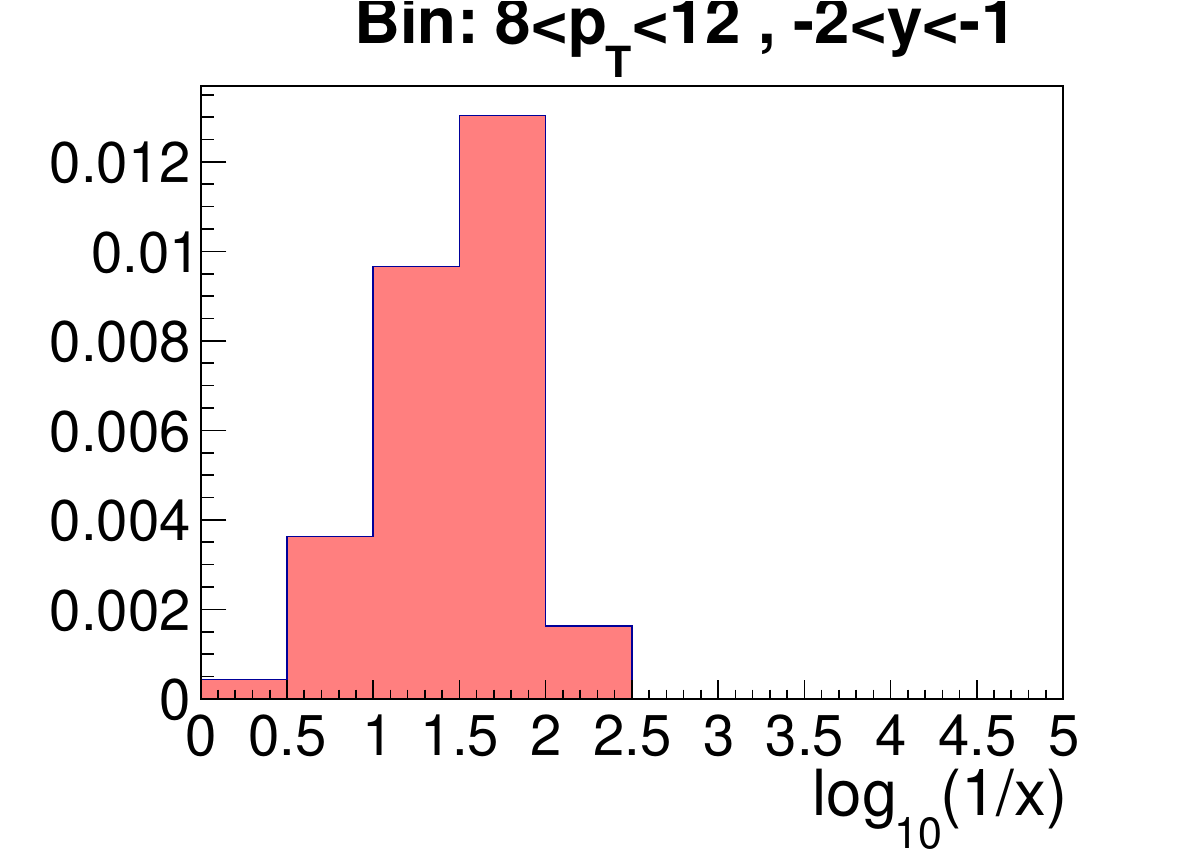}
    \end{subfigure}

 \vspace{1.0em}  
    \begin{subfigure}[b]{0.3\textwidth}
        \centering
        \includegraphics[width=\textwidth]{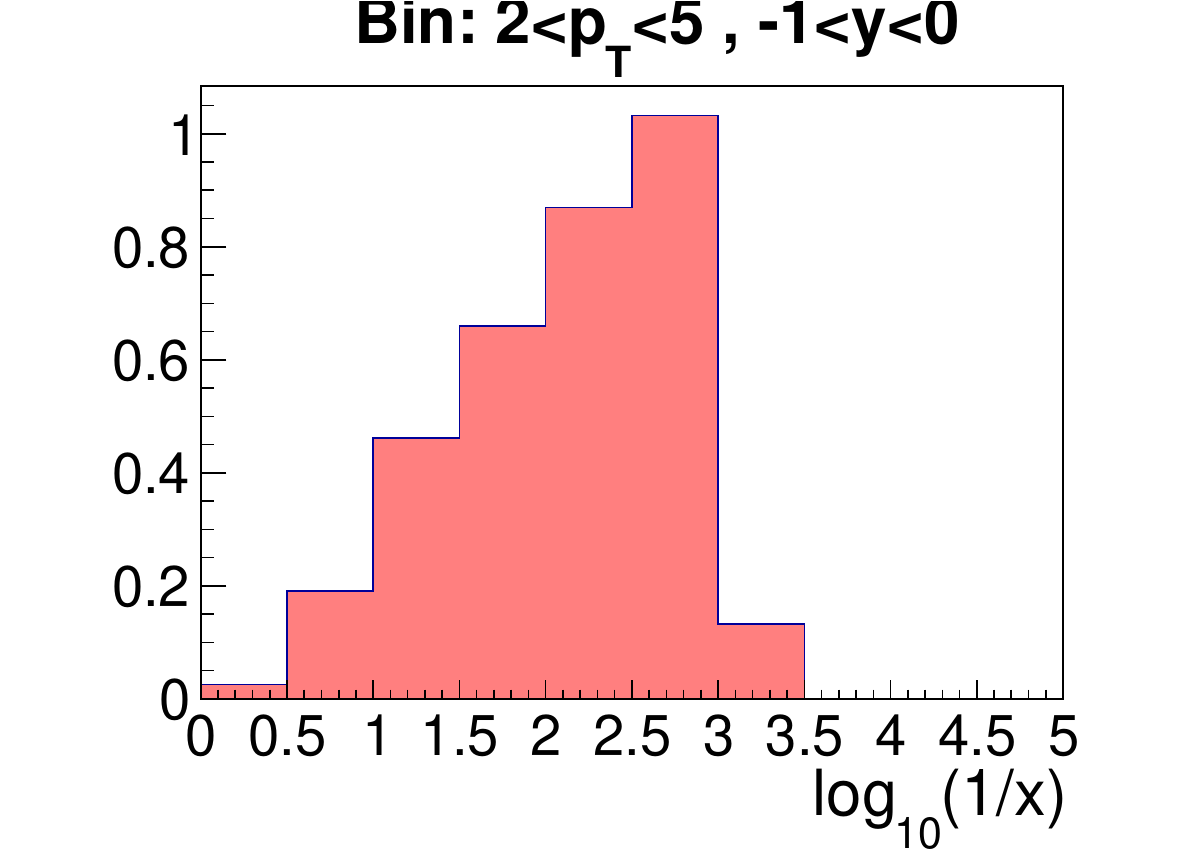}
    \end{subfigure}
\hfill
    \begin{subfigure}[b]{0.3\textwidth}
        \centering
        \includegraphics[width=\textwidth]{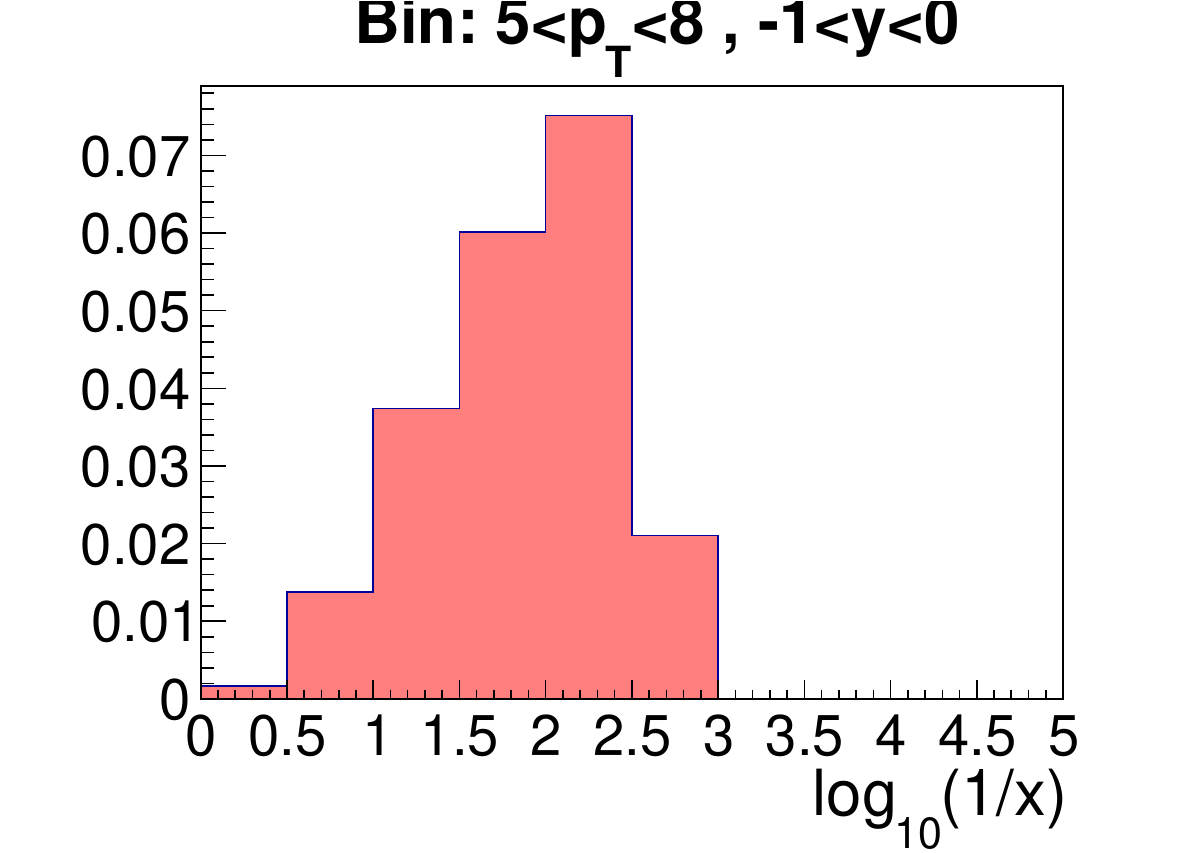}
    \end{subfigure}
    \hfill
    \begin{subfigure}[b]{0.3\textwidth}
        \centering
        \includegraphics[width=\textwidth]{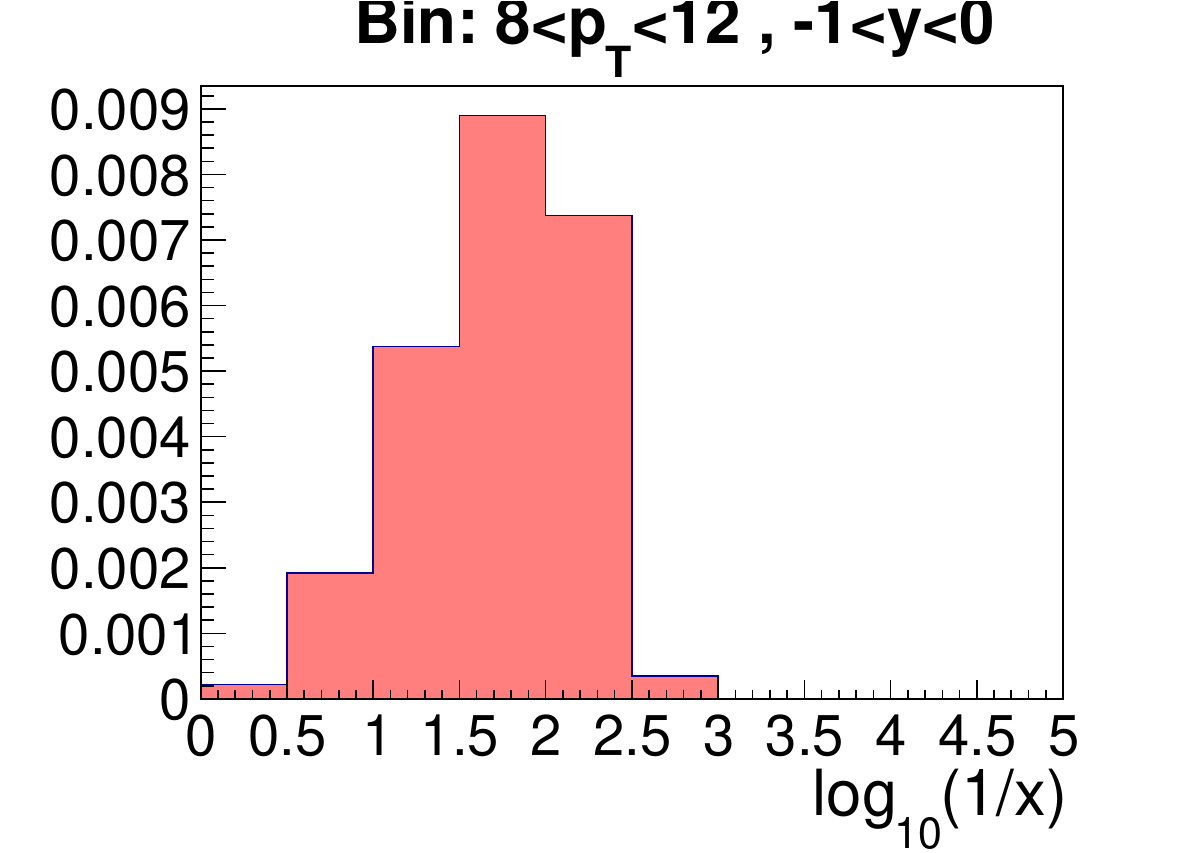}
    \end{subfigure}

 \vspace{1.0em}
 
    \begin{subfigure}[b]{0.3\textwidth}
        \centering
        \includegraphics[width=\textwidth]{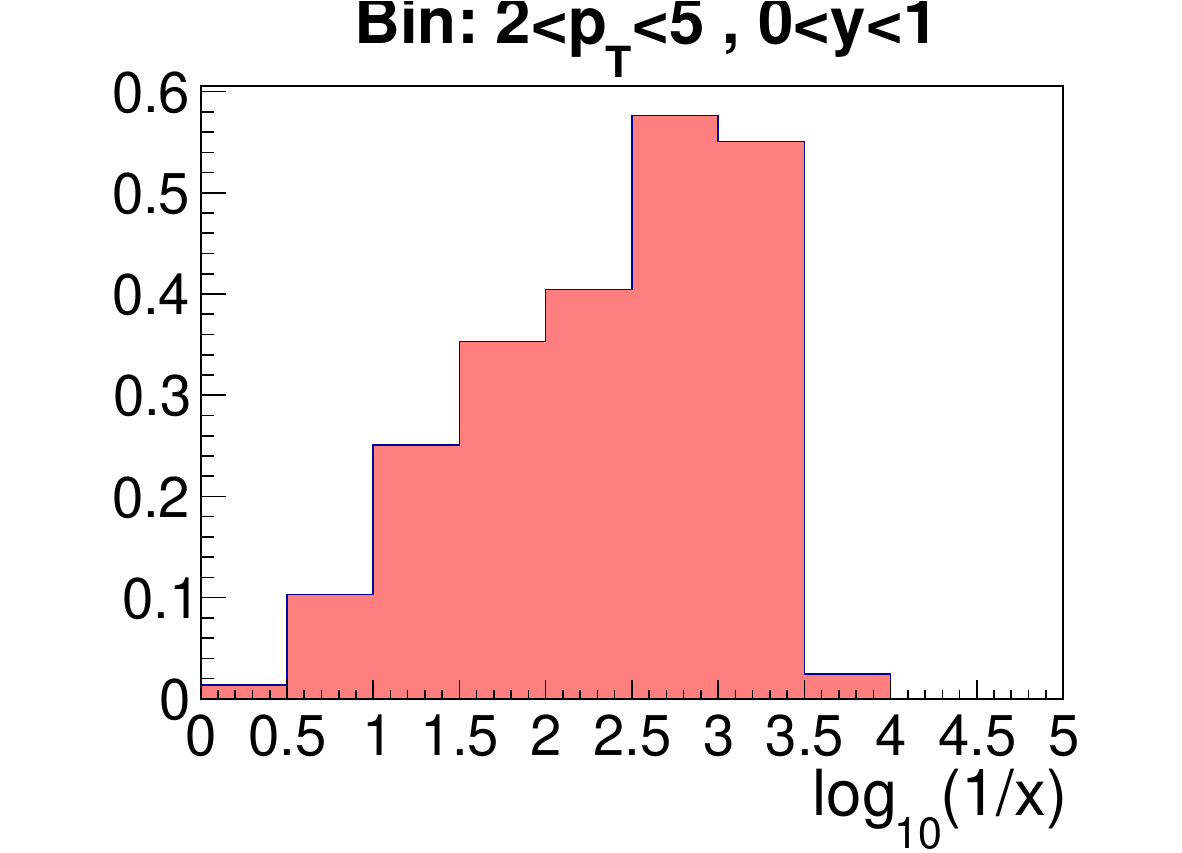}

    \end{subfigure}
  \hfill
    \begin{subfigure}[b]{0.3\textwidth}
        \centering
        \includegraphics[width=\textwidth]{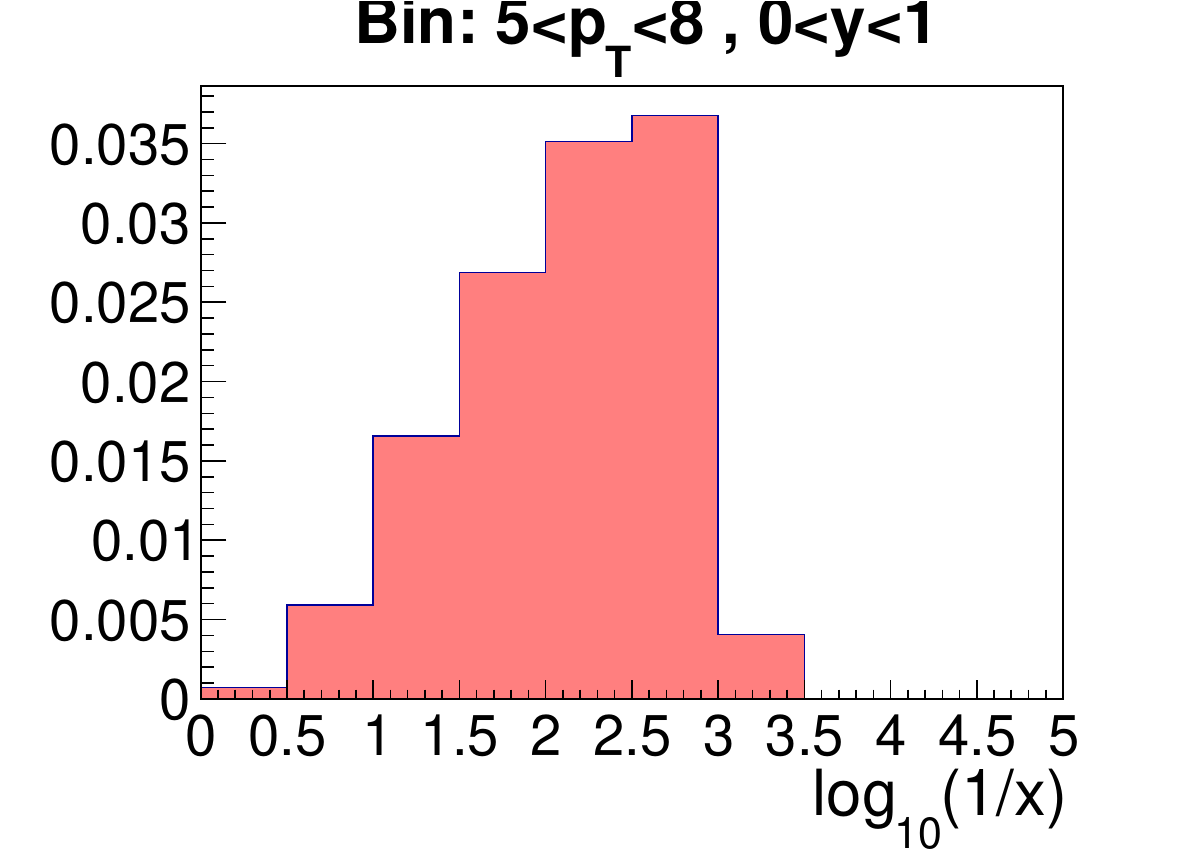}
    \end{subfigure}
\hfill
    \begin{subfigure}[b]{0.3\textwidth}
        \centering
        \includegraphics[width=\textwidth]{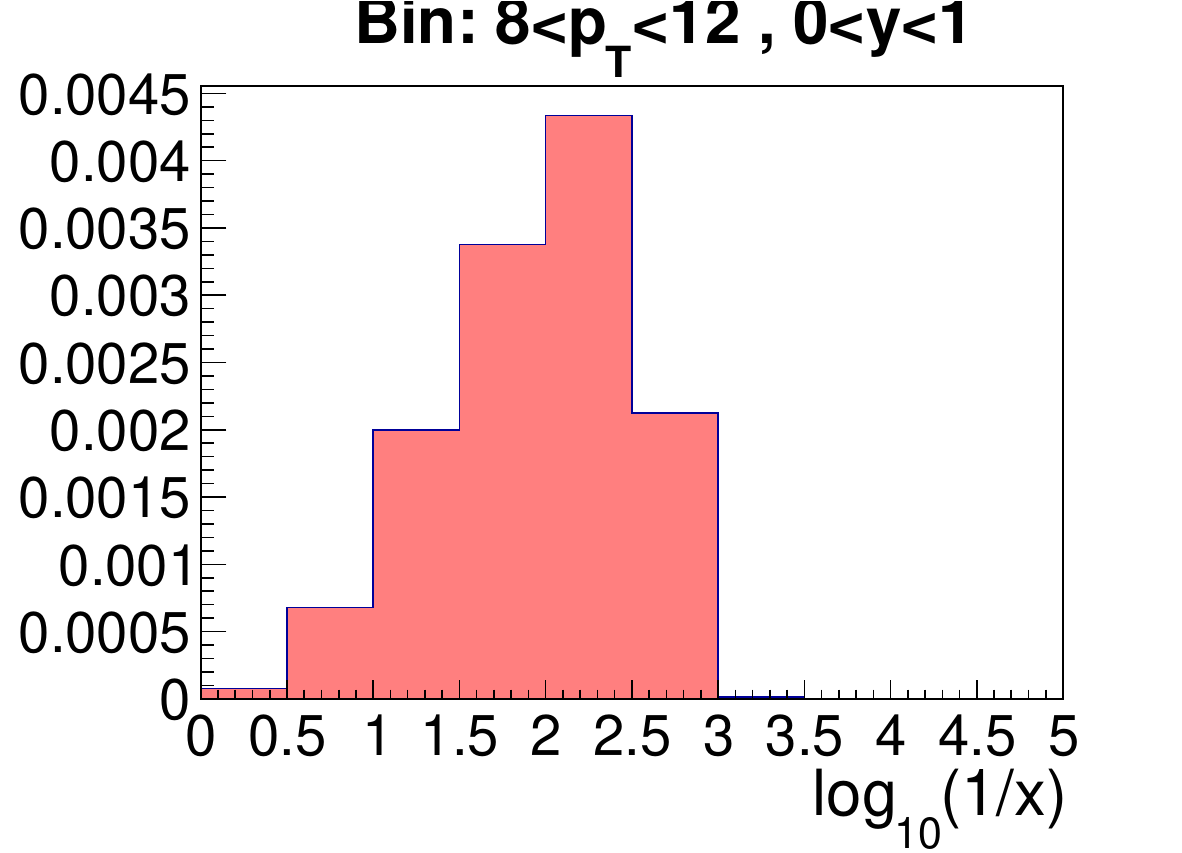}
    \end{subfigure}
    
    \vspace{1.0em}
    
    \begin{subfigure}[b]{0.3\textwidth}
        \centering
        \includegraphics[width=\textwidth]{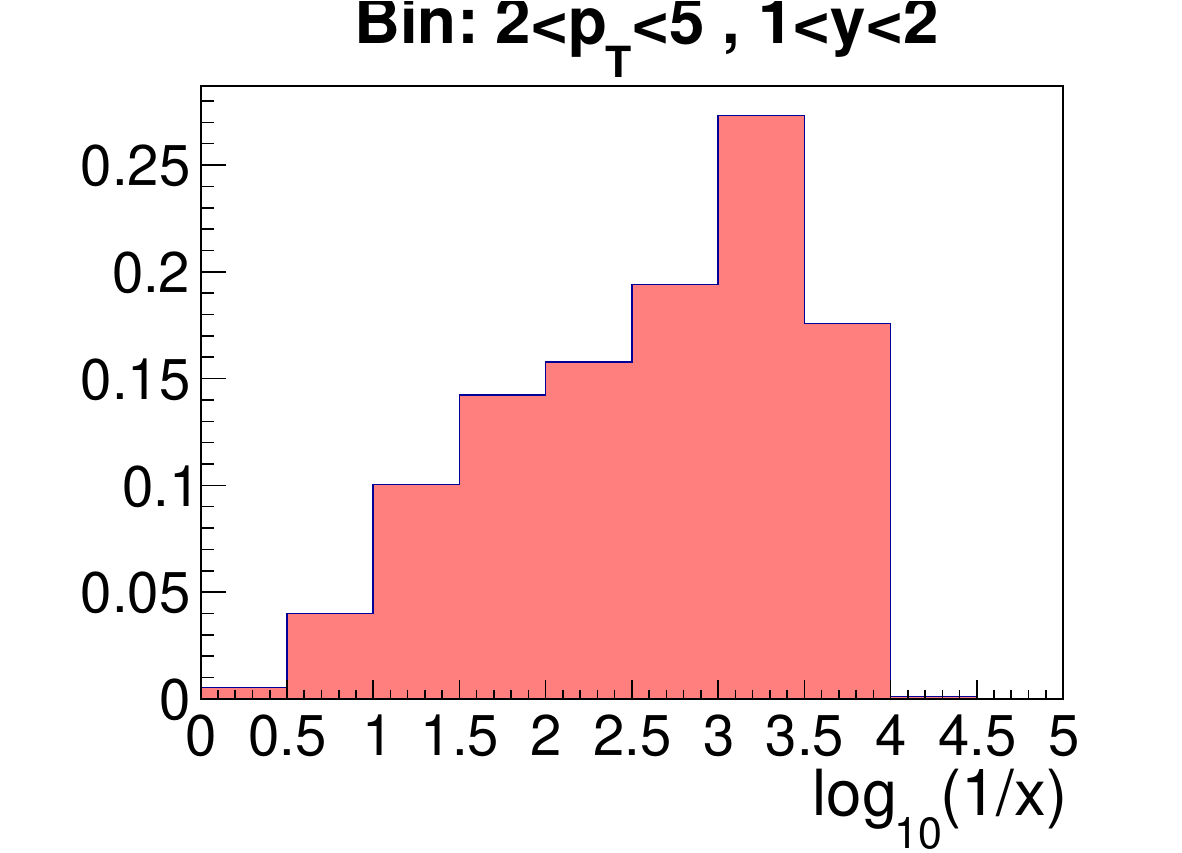}
    \end{subfigure}
    \hfill
    \begin{subfigure}[b]{0.3\textwidth}
        \centering
        \includegraphics[width=\textwidth]{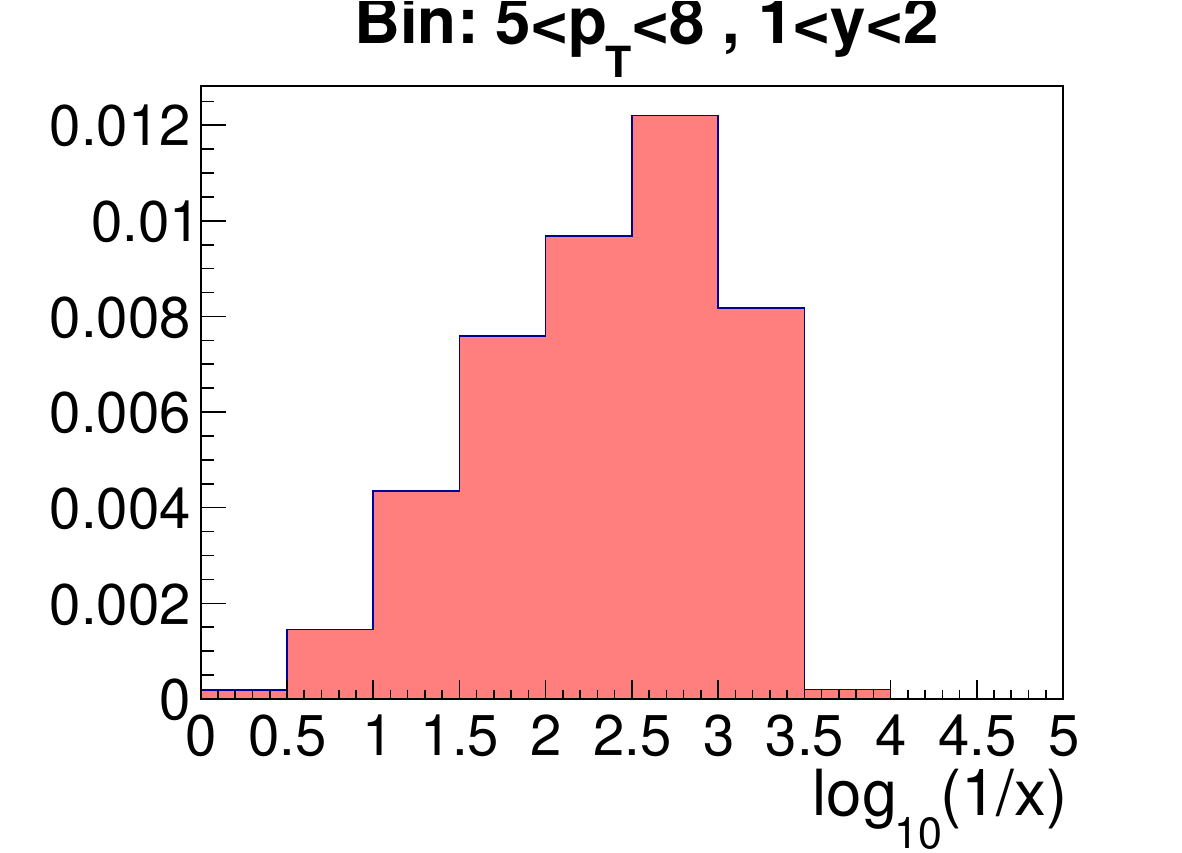}
    \end{subfigure}
    \hfill
    \begin{subfigure}[b]{0.3\textwidth}
        \centering
        \includegraphics[width=\textwidth]{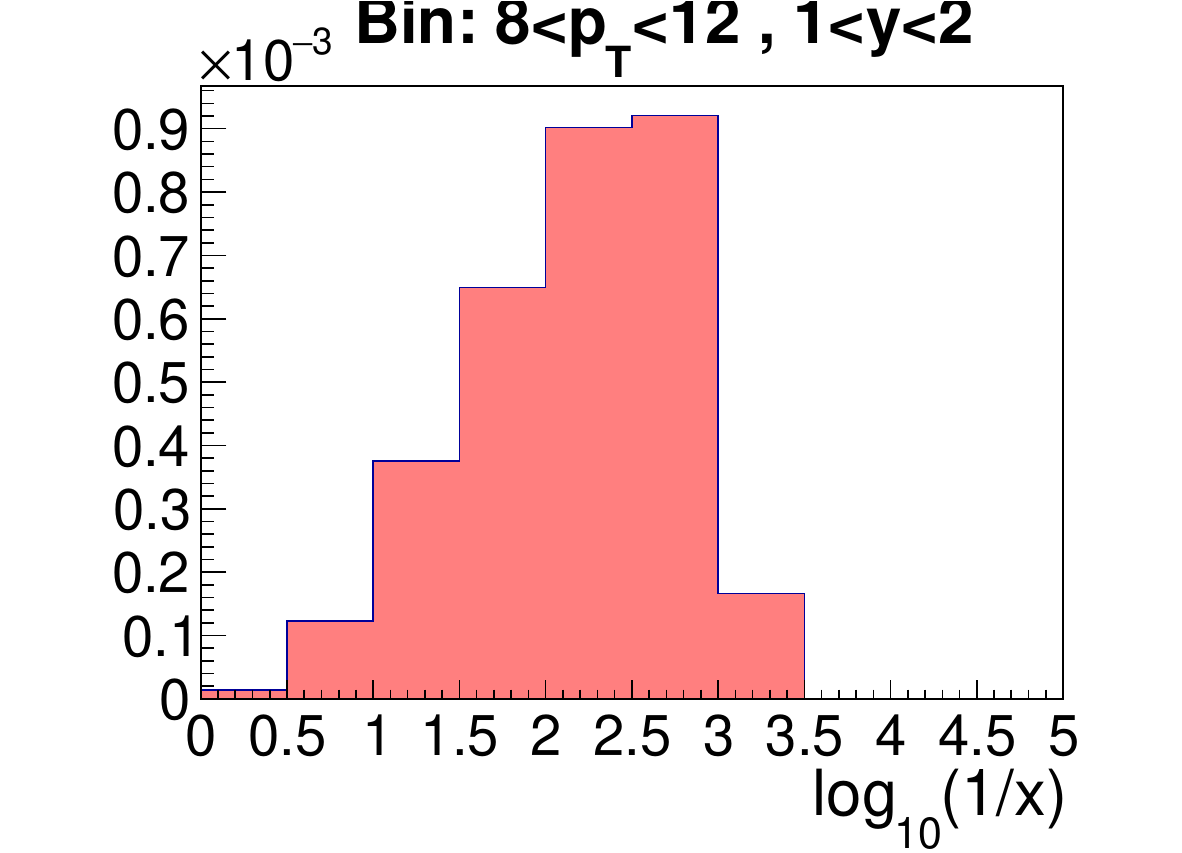}
    \end{subfigure}

    \caption{Plot of the inclusive cross section \(\frac{d^3 \sigma}{dy dp_T dx} \Delta x\) binned in \(x\), the longitudinal momentum fraction carried by the gluon. From left to right column: bins in \(p_T\): (2,5), (5,8), (8,12) GeV. Rows from top to bottom: bins in \(y\): (-2,-1), (-1,0), (0,1), (1,2).}
    \label{fig:grid}
\end{figure}

\bibliography{mybib}

\end{document}